# The PLATO 2.0 Mission

28.02.2014

H. Rauer[1,2], C. Catala[3], C. Aerts[4], T. Appourchaux[5], W. Benz[6], A. Brandeker[7], J. Christensen-Dalsgaard[8], M. Deleuil[9], L. Gizon[10,25], M.-J. Goupil[3], M. Güdel[11], E. Janot-Pacheco[12], M. Mas-Hesse[13], I. Pagano[14], G. Piotto[15], D. Pollacco[16], N.C. Santos[17], A. Smith[18], J.-C., Suárez[19], R. Szabó[20], S. Udry[21], V. Adibekyan[17], Y. Alibert[6], J.-M. Almenara[9], P. Amaro-Seoane[22], M. Ammler-von Eiff[23], M. Asplund[42], E. Antonello[24], W. Ball[25], S. Barnes[26], F. Baudin[5], K. Belkacem[3], M. Bergemann[27], G. Bihain[26], A.C. Birch[10], X. Bonfils[40], I. Boisse[17], A.S. Bonomo[28], F. Borsa[24], I.M. Brandão[17], E. Brocato[29], S. Brun[30], M. Burleigh[31], R. Burston[10], J. Cabrera[1], S. Cassisi[33], W. Chaplin[34], S. Charpinet[35], C. Chiappini[26], R.P. Church[36], Sz. Csizmadia[1], M. Cunha[17], M. Damasso[15,28], M.B. Davies[36], H.J. Deeg[37], R. F. Díaz[21], S. Dreizler[25], C. Dreyer[1,2], P. Eggenberger[21], D. Ehrenreich[21], P. Eigmüller[1], A. Erikson[1], R. Farmer[39], S. Feltzing[36], F. de Oliveira Fialho[38], P. Figueira[17], T. Forveille[40], M. Fridlund[1], R.A. García[30], P. Giommi[41], G. Giuffrida[29,41], M. Godolt[1], J. Gomes da Silva[17], T. Granzer[26], J.L. Grenfell[1], A. Grotsch-Noels[43], E. Günther[23], C.A. Haswell[39], A.P. Hatzes[23], G. Hébrard[44], S. Hekker[10,45], R. Helled[46], K. Heng[6], J.M. Jenkins[47], A. Johansen[36], M.L. Khodachenko[48], K.G. Kislyakova[48], W. Kley[49], U. Kolb[39], N. Krivova[10], F. Kupka[50], H. Lammer[48], A.F. Lanza[14], Y. Lebreton[3], D. Magrin[51], P. Marcos-Arenal[4], P.M. Marrese[29,41], J.P. Marques[25], J. Martins[17], S. Mathis[30], S. Mathur[52], S. Messina[14], A. Miglio[34], J. Montalban[43], M. Montalto[17], M.J.P.F.G. Monteiro[17], H. Moradi[53], E. Moravveji[4], C. Mordasini[54], T. Morel[43], A. Mortier[17], V. Nascimbeni[15], R.P. Nelson[60], M.B. Nielsen[25], L. Noack[32], A.J. Norton[39], A. Ofir[25], M. Oshagh[17], R.-M. Ouazzani[3], P. Pápics[4], V. C. Parro[55], P. Petit[56], B. Plez[57], E. Poretti[24], A. Quirrenbach[58], R. Ragazzoni[51], G. Raimondo[33], M. Rainer[24], D.R. Reese[43], R. Redmer[59], S. Reffert[58], B. Rojas-Ayala[17], I.W. Roxburgh[60], S. Salmon[43], A. Santerne[17], J. Schneider[3], J. Schou[10], S. Schuh[10], H. Schunker[10], A. Silva-Valio[61], R. Silvotti[28], I. Skillen[62], I. Snellen[63], F. Sohl[1], S.G. Sousa[17], A. Sozzetti[28], D. Stello[69], K. G. Strassmeier[26], M. Švanda[64], Gy. M. Szabó[20,65], A. Tkachenko[4], D. Valencia[66], V. Van Grootel[43], S.D. Vauclair[35], P. Ventura[29], F.W. Wagner[1], N.A. Walton[67], J. Weingrill[26], S.C. Werner[68], P.J. Wheatley[16], K. Zwintz[4]

Corresponding author: Heike Rauer (Email: heike.rauer@dlr.de)
[1]Institute of Planetary Research, German Aerospace Center, Rutherfordstrasse 2, 12489 Berlin, Germany
[2]Department of Astronomy and Astrophysics, Berlin Institute of Technology, Hardenbergstrasse 36, 10623 Berlin, Germany
[3] LESIA, Observatoire de Paris, PSL Research University, CNRS, UPMC, University Paris-Diderot , 5 Pl Jules Janssen, 92195 Meudon Cedex, France
[4]Instituut voor Sterrenkunde, KU Leuven, Celestijnenlaan 200D, 3001 Leuven, Belgium
[5]Institut d'Astrophysique Spatiale, Université Paris 11, Batiment 121, 91405 ORSAY Cedex, France
[6]Center for Space and Habitability, University of Bern, Physikalisches Institut, Sidlerstrasse 5, 3012 Bern, Switzerland
[7]Stockholm University, Department of Astronomy, AlbaNova University Center, 106 91 Stockholm, Sweden
[8]Stellar Astrophysics Centre, Department of Physics and Astronomy, Aarhus University, Ny Munkegade 120, 8000 Aarhus C, Denmark
[9]Laboratoire d'Astrophysique de Marseille, 38 rue Frédéric Joliot-Curie, 13388 Marseille Cedex 13, France
[10]Max-Planck-Institut für Sonnensystemforschung, Justus-von-Liebig-Weg 3, 37077 Göttingen, Germany
[11]Institute for Astronomy, University of Vienna, Türkenschanzstrasse 17, 1180 Vienna, Austria
[12]Instituto de Astronomia, Geofísica e Ciências Atmosféricas - IAG/USP, Brazil
[13]Centro de Astrobiología (CSIC–INTA), Madrid, Spain
[14]INAF- Osservatorio Astrofisico di Catania, Via S. Sofia 78, 95123 Catania, Italy
[15]Università di Padova, Dipartimento di Fisica e Astronomia "Galileo Galilei", Vicolo dell'Osservatorio 3, 35122 Padova, Italy
[16]Warwick University, Department of Physics, Gibbet Hill Rd, Coventry CV4 7AL, United Kingdom
[17]Centro de Astrofísica, Universidade do Porto, Rua das Estrelas, 4150-762 Porto, Portugal
[18]Mullard Space Science Laboratory, University College London, London, United Kingdom
[19]Instituto de Astrofísica de Andalucía – CSIC, Glorieta de la Astronomía, s/n. 18008, Granada, Spain
[20]Konkoly Observatory of the Hungarian Academy of Science, Konkoly Thege Miklos út. 15-17, 1121 Budapest, Hungary
[21]Observatoire de Genève, 51 chemin des Maillettes, 1290 Sauverny, Switzerland
[22]Max Planck Institut für Gravitationsphysik (Albert-Einstein-Institut), 14476 Potsdam, Germany
[23]Thüringer Landessternwarte Tautenburg, Sternwarte 5, 07778 Tautenburg, Germany
[24]INAF-Osservatorio Astronomico di Brera, Via E. Bianchi 46, 23807 Merate (LC), Italy
[25]Institute for Astrophysics, Georg-August-Universität Göttingen, Friedrich-Hund-Platz 1, 37077 Göttingen, Germany
[26]Leibniz Institute for Astrophysics, An der Sternwarte 16, 14482 Potsdam, Germany
[27] Institute of Astronomy, University of Cambridge, CB3 OHA, Cambridge, UK
[28]INAF - Osservatorio Aastrofisico di Torino, Via Osservatorio 20, 10025 Pino Torinese, Italy
[29] INAF - Osservatorio Astronomico di Roma, Via Frascati 33, 00040 Monte Porzio Catone (RM), Italy
[30]CEA, DSM/IRFU/Service d'Astrophysique, Bat 709, CEA-Saclay, 91191 Gif-sur-Yvette, France
[31]Department of Physics and Astronomy, University of Leicester, Leicester LE1 7RH, United Kingdom
[32]Royal Observatory of Belgium, Ringlaan 3, 1180 Brussels, Belgium
[33]INAF - Osservatorio Astronomico di Teramo, Via Mentore Maggini s.n.c., 64100 Teramo, Italy
[34]School of Physics and Astronomy, University of Birmingham, Edgbaston, Birmingham B152TT, United Kingdom

[35]Université de Toulouse, UPS-OMP, IRAP, 14 Av. E. Belin, 31400 Toulouse, France
[36]Lund Observatory, Box 43, 221 00 Lund, Sweden
[37]Instituto de Astrofısica de Canarias, Tenerife, Spain
[38]Laboratory of Automation and Control, Polytechnic School of the University of São Paulo, Brazil
[39] Dept of Physical Sciences, The Open University, Walton Hall, Milton Keynes MK7 6AA, United Kingdom
[40]Observatoire de Grenoble, Institut de Planetologie et d'Astrophysique de Grenoble, 38041 Grenoble Cedex 9, France
[41]ASI Science Data Center, Via del Politecnico snc, 00133 Rome, Italy
[42] Australian National University, Research School of Astronomy and Astrophysics, Cotter Road, Weston ACT 2611, Australia
[43]Institut d'Astrophysique et Géophysique de l'Université de Liège, Allée du 6 Août 17, 4000 Liège, Belgium
[44]Institut d'Astrophysique de Paris, UMR7095 CNRS, Université Pierre & Marie Curie, 98bis Bd Arago, 75014 Paris, France
[45]Sterrenkundig Instituut Anton Pannekoek, Universiteit van Amsterdam, Postbus 94249, 1090 GE Amsterdam, The Netherlands
[46]Department of Geophysics, Atmospheric and Planetary Sciences, Tel-Aviv University, Tel-Aviv, Israel
[47]SETI Institute, NASA Ames Research Center, USA
[48]Austrian Academy of Sciences, Schmiedlstrasse 6, 8042 Graz, Austria
[49]Universität Tübingen, Institut für Astronomie and Astrophysik, Morgenstelle 10, 72076 Tübingen, Germany
[50]Fakultät für Mathematik,Universität Wien, Oskar-Morgenstern-Platz 1, 1090 Wien, Austria
[51]INAF - Astronomical Observatory of Padova, Italy
[52]Space Science Institute, 4750 Walnut Street, Suite#205, Boulder, CO, 80301, USA
[53]Monash Centre for Astrophysics, School of Mathematical Sciences, Monash University, Victoria 3800, Australia
[54]Max Planck Institute for Astronomy, Planet and Star Formation Department, Königstuhl 17, 69117 Heidelberg, Germany
[55]Instituto Mauá de Tecnologia, Brazil
[56]Observatoire Midi-Pyrenées, 14 Av E Belin, 31400 Toulouse Cedex, France
[57]Laboratoire Univers et Particules de Montpellier, Université Montpellier 2, CNRS, 34095 Montpellier cedex 5, France
[58]ZAH, Landessternwarte, Universität Heidelberg, Königstuhl 12, 69117 Heidelberg, Germany
[59]University of Rostock, Institute of Physics, 18051 Rostock, Germany
[60]Queen Mary University of London, Astronomy Unit, School of Physics and Astronomy, Mile End Rd, London E1 4NS, United Kingdom
[61]CRAAM - Mackenzie University, Sao Paulo, Brazil
[62]Isaac Newton Group of Telescopes, Astronomy Group, Apartado 321, 38700 Santa Cruz de La Palma, Spain
[63]Leiden Observatory, Leiden University, Postbus 9513, 2300 RA Leiden, The Netherlands
[64]Astronomical Institute, Charles University in Prague, Faculty of Mathematics and Physics, V Holešovičkách 2, 18000 Prague 8, Czech Republic
[65]ELTE Gothard Astrophysical Observatory, 9704 Szombathely, Szent Imre herceg út 112, Hungary
[66]Atmosphere and Planetary Sciences Department, Massachusetts Institute of Technology, 77 Massachusetts Avenue, Cambridge, MA 02139, USA
[67]Institute of Astronomy, University of Cambridge, Madingley Road, Cambridge, CB3 0HA, United Kingdom
[68]Comparative Planetology, Centre for Earth Evolution and Dynamics, University of Oslo, PO 1048 Blindern, 0316 Oslo, Norway
[69]Sydney Institute for Astronomy (SIfA), School of Physics, University of Sydney, NSW 2006, Australia

# Abstract

PLATO 2.0 has recently been selected for ESA's M3 launch opportunity (2022/24). Providing accurate key planet parameters (radius, mass, density and age) in statistical numbers, it addresses fundamental questions such as: How do planetary systems form and evolve? Are there other systems with planets like ours, including potentially habitable planets? The PLATO 2.0 instrument consists of 34 small aperture telescopes (32 with 25 sec readout cadence and 2 with 2.5 sec candence) providing a wide field-of-view (2232 deg$^2$) and a large photometric magnitude range (4-16 mag). It focusses on bright (4-11 mag) stars in wide fields to detect and characterize planets down to Earth-size by photometric transits, whose masses can then be determined by ground-based radial-velocity follow-up measurements. Asteroseismology will be performed for these bright stars to obtain highly accurate stellar parameters, including masses and ages. The combination of bright targets and asteroseismology results in high accuracy for the bulk planet parameters: 2%, 4-10% and 10% for planet radii, masses and ages, respectively. The planned baseline observing strategy includes two long pointings (2-3 years) to detect and bulk characterize planets reaching into the habitable zone (HZ) of solar-like stars and an additional step-and-stare phase to cover in total about 50% of the sky. PLATO 2.0 will observe up to 1,000,000 stars and detect and characterize hundreds of small planets, and thousands of planets in the Neptune to gas giant regime out to the HZ. It will therefore provide the first large-scale catalogue of bulk characterized planets with accurate radii, masses, mean densities and ages. This catalogue will include terrestrial planets at intermediate orbital distances, where surface temperatures are moderate. Coverage of this parameter range with statistical numbers of bulk characterized planets is unique to PLATO 2.0. The PLATO 2.0 catalogue allows us to e.g.: - complete our knowledge of planet diversity for low-mass objects, - correlate the planet mean density-orbital distance distribution with predictions from planet formation theories, - constrain the influence of planet migration and scattering on the architecture of multiple systems, and - specify how planet and system parameters change with host star characteristics, such as type, metallicity and age. The catalogue will allow us to study planets and planetary systems at different evolutionary phases. It will further provide a census for small, low-mass planets. This will serve to identify objects which retained their primordial hydrogen atmosphere and in general the typical characteristics of planets in such low-mass, low-density range. Planets detected by PLATO 2.0 will orbit bright stars and many of them will be targets for future atmosphere spectroscopy exploring their atmosphere. Furthermore, the mission has the potential to detect exomoons, planetary rings, binary and Trojan planets. The planetary science possible with PLATO 2.0 is complemented by its impact on stellar and galactic science via asteroseismology as well as light curves of all kinds of variable stars, together with observations of stellar clusters of different ages. This will allow us to improve stellar models and study stellar activity. A large number of well-known ages from red giant stars will probe the structure and evolution of our Galaxy. Asteroseismic ages of bright stars for different phases of stellar evolution allow calibrating stellar age-rotation relationships. Together with the results of ESA's Gaia mission, the results of PLATO 2.0 will provide a huge legacy to planetary, stellar and galactic science.

# 1 Introduction

In the last 20 years, mankind has embarked on a quest which previously was only the subject of science fiction – the search for worlds similar to our own beyond the Solar System. This quest is ultimately motivated by mankind's desire to know its place in the Universe: Is our Solar System special? How did it form? And especially: Is there life elsewhere in the Universe?

Today, we know of about 1000 exoplanets with secure identifications (see e.g. list on exoplanet.eu, Schneider et al. 2011) and a few thousand as yet unconfirmed planet candidates, indicating that every second dwarf star might host a planet (e.g. Mayor et al. 2011, Fressin et al. 2013). Many confirmed exoplanets fall into new classes unlike any of the planets of the Solar System, e.g., 'hot Jupiters', 'mini-Neptunes' and 'super-Earths' (planets <10 MEarth). However this sample currently lacks exoplanets resembling the terrestrial planets of our own Solar System.

It is the goal of PLATO 2.0 (PLAnetary Transits and Oscillation of stars) to find these planets and provide the first catalogue of potentially habitable planets with known mean densities and ages. Mean densities will be obtained from planet radius measurements via the photometric transit method applied to stellar lightcurves obtained with the satellite, in combination with mass determinations from radial velocity (RV) follow-up measurements with ground-based telescopes. The PLATO 2.0 consortium will coordinate the world-wide observational effort needed to obtain the required RV follow-up data (see section 7), and we therefore consider this as part of the PLATO 2.0 mission activities and results. Stellar masses, radii and ages are derived by asteroseismic analyses of the photometric lightcurves.

PLATO 2.0 alone can systemically detect and characterize the bulk properties of Earth-like planets on Earth-like orbits around Sun-like stars. Crucially, this catalogue will contain targets accessible by future ground-based observatories, including the European Extremely Large Telescope (E-ELT), and by space missions, including the James Webb Space Telescope (JWST) and future M and L class missions designed to study exoplanet atmospheres.

While we have a good understanding of the structure and mass distribution of our Solar System and its planets and moons, we only have indirect and incomplete knowledge of how our system formed and evolved. This gap can be closed by the characterization of extrasolar planetary systems. Although each of the systems is itself a snapshot, obtaining an assembly of planetary systems, characterized by structure and age, will increase our understanding of planetary system evolution and formation significantly. In the same way as we understand stellar evolution from the diversity of ages present in the stellar population, PLATO 2.0 will enable us to understand the evolution of planetary systems. The acquisition of accurate masses and radii for planets around host stars with different chemical composition, ages, stellar activity, and in different planet system architectures, will form a major result from PLATO 2.0.

The PLATO 2.0 mission is technically identical to the PLATO mission concept proposed as M1/M2 candidate to ESA (ESA/SRE(2011)13). However, its science case takes into account the enormous developments in exoplanet science in recent years. To reflect this strongly updated science case and organizational changes in the consortium, the mission has been called PLATO 2.0 for ESA's M3 mission selection phase. PLATO 2.0 consists of 32 so-called 'normal' telescopes operating in white light and providing a very wide field-of-view (FoV) and two additional 'fast' cameras with high read-out cadence and colour filters (see Section 5). The unusual multi-telescope design allows for a large photometric dynamic range (4-16 mag). The current baseline observing plan foresees two long target field pointings (2-3 years each) and a step-and-stare phase of up to five months per field. The total mission lifetime is 6 years, and 2 years possible extension. In total, the mission will cover about 50% of the sky over its lifetime.

PLATO 2.0 promises precise planetary radii, masses and ages by utilizing precise stellar parameters and RV follow-up. Stellar radii will be available from the Gaia mission, and stellar masses and ages will be tightly constrained by the systematic use of asteroseismology for about 85,000 stars. The mission's main planet hunting target range is $4 \leq m_v \leq 11$. In this range asteroseismology can be performed and accurate planet parameters can be derived, including planet masses from radial velocity (RV) follow-up spectroscopy. Planetary radii and masses will be constrained to a few percent, and potentially even to 2% or below. Ages will be known to 10% for solar-like stars. PLATO 2.0 can in addition detect several thousands of terrestrial planets down to 13 mag, but with performances comparable to CoRoT and *Kepler*. Larger planets can be detected down to 16 mag, which will still be interesting discoveries for statistical studies. PLATO 2.0 is an unbiased, magnitude-limited survey that will observe stars throughout the Herzsprung-Russell Diagram, including the important main sequence F, G, K, as well as the brightest M dwarfs. The recent space missions CoRoT (Baglin et al., 2006) and *Kepler* (NASA, Koch et al. 2010), have provided more than 100 confirmed planets with known radii and masses, from hot gas giants to a limited number of hot super-Earths. *Kepler* furthermore provided planet frequency (or number of planet candidates per star). For the cool terrestrial planets, however, the faintness of the *Kepler* target stars does not allow for RV follow-up. Recently, Transit Timing Variations (see section 6.4) have been used to derive planet masses for 163 *Kepler* planets (Hadden & Lithwick 2013), providing at least constraints on mean planet densities.

Unfortunately, the upcoming future projects in this decade have limitations for the detection and characterization of terrestrial planets with long orbital periods (see Section 6.3 for a mission comparison). Transiting planets expected from Gaia mission photometry (Dzigan & Zucker 2012) will be large and mainly orbit stars fainter than 11mag, too faint for accurate follow-up spectroscopy on large scale, whereas astrometric detections are made for giant planets only (Casertano et al., 2008). From the ground, searches for small planets in the habitable zone of solar-like stars by radial velocity (RV) techniques, e.g., via ESO's ESPRESSO project, will help to unveil the presence of Earth-like planets orbiting other Suns, however not in large numbers. It is thus unclear whether our Solar System is typical or special, and this will remain so until we can reliably detect and characterize Earth-like planets in Earth-like orbits around all kinds of bright host stars, which is a primary objective of PLATO 2.0.

Two space missions targeting transits of bright host stars have recently been selected for launch in 2017. ESA's Small Mission CHEOPS (Broeg et al. 2013, launch 2017) has the goal of increasing the number of planets with known radii and masses by observing transits of planets detected previously by RV or by transit detection surveys. CHEOPS will therefore be an important first step towards the goal of a planetary bulk parameter survey. The TESS mission (NASA, launch 2017) will search over the whole sky for small planets down to Earth size around bright stars, which will be interesting targets for atmospheric follow-up by e.g., JWST (1 $R_{Earth}$ planets will be detected at approx. ≤7 mag for solar-like stars, and larger planets can be detected at much fainter hosts). However, TESS will focus mainly on planets in short period orbits (up to about 20 days) because of its pointing strategy, which covers most fields for 27 days only. At the ecliptic poles, about 2% of the sky will be covered for a whole year; this provides some potential for the detection of longer-period planets. TESS will detect the first small planets around stars close to our Solar System. It will, however, not address the science case of characterizing rocky planets at intermediate orbital distances (a>0.3au, including the HZ) around solar-like stars, which remains unique for PLATO 2.0. PLATO 2.0 will outnumber the detection of small, characterized planets by 1-3 orders of magnitude compared to *Kepler* and TESS (see Section 6.3). On the other hand, TESS, being the first all-sky survey, will identify interesting targets, defining science cases that PLATO 2.0 could address during its step-and-stare phase. In short, TESS and CHEOPS perform important first steps that will provide a glimpse of the planet bulk density parameter space. A complete picture of the planet population, however, including planets on Earth-like orbits, requires PLATO 2.0. This objective remains unique to this mission for the next decade and will be crucial for answering the question: how unique is our Solar System?

PLATO 2.0 will address a number of additional science goals, including stellar-planet interactions, exo-moons, rings, dynamical interactions in planetary systems, and asteroseismology of a wide variety of stars. Furthermore, the planets detected by PLATO 2.0 will orbit nearby stars, making their angular separation sufficiently wide to permit direct imaging. This offers the unique opportunity to investigate the planet atmospheres by transit spectroscopy and by direct imaging spectroscopy for a very large number of planets

Furthermore, the process of obtaining precise stellar parameters will validate the stellar models and allow us to constrain poorly-understood physical processes that introduce uncertainty in stellar parameters, especially age. Such models, improved by PLATO 2.0's constraints, can then even better characterize the orbiting planets and concurrently contribute to our understanding on a variety of topics that benefit from accurate stellar models. In total, the baseline observing strategy of PLATO 2.0 provides about 1,000,000 highly accurate stellar light curves.

The description of the mission in this paper is divided into two parts: the scientific goals (§2-4) and technical details (§5-8). The scientific goals are split into planetary science (§2), stellar science (§3), and complementary/legacy science (§4). The technical information is separated into the mission and instrument concept (§5), planet detection performance (§6), follow-up (§7), and data products and policy (§8). Since PLATO 2.0 addresses a wide science community, a basic overview of the methods used (photometric transits and asteroseismology) is given in Appendix A.

## 2 Science Goals I: Planetary Science

PLATO 2.0 is a transit survey mission with the goal of detecting and bulk characterising new planets and planetary systems around bright stars, including planet parameter ranges which will otherwise not be explored in the next decade. PLATO 2.0's design is optimized to answer conclusively the following key questions:

- What are the bulk properties (mass, radius, mean density, age) of planets in a wide range of systems, including terrestrial planets in the habitable zone of solar-like stars?

- What is the planet orbital separation-mass function for low-mass terrestrial planets?

- How do planets and planet systems evolve with age?

- How often are planetary systems co-planar, rather than having been dynamically excited by more massive planets?

- How do planet properties and their frequencies correlate with factors relevant for planet formation (e.g., stellar metallicity, stellar type, orbital distance, disk properties)?

- Does the frequency of terrestrial planets depend on the environment in which they formed?

Answering these questions requires the detection and determination of accurate bulk properties for a large number of planets.

Due to their brightness, PLATO 2.0 targets are more amenable to RV follow-up than Kepler targets. Furthermore, they will provide prime targets for spectroscopic follow-up observations investigating their atmospheres, e.g., by JWST, E-ELT or future L class missions dedicated to exoplanet spectroscopy.

## 2.1 Planet detection and bulk characterization – status and PLATO 2.0 prospects

Today, about 1000 extrasolar planets have been discovered (exoplanet.eu; Scheider et al. 2011). For most of these planets we could determine only one of their fundamental parameters directly: radius or mass. In those cases where planets have been observed by both the transit and RV methods, their mass, radius, and thus mean density have been accurately measured. This has led to exciting discoveries, including new classes of intermediate planets called 'super-Earths' and 'mini-Neptunes'. In addition to the confirmed planets, NASA's *Kepler* mission has published results on several thousands of planet candidates. Together with RV and microlensing survey detections, these results show that small planets are very numerous. Even though the precise frequency of planets in the Galaxy is a matter of debate, the community presently agrees that planets, in particular rocky planets like our Earth, are very common around solar-type stars (FGK and M dwarfs -- see e.g., Udry & Santos 2007; Mayor et al. 2011; Howard et al. 2012; Cassan et al. 2012; Bonfils et al. 2013; Fressin et al. 2013; Petigura et al. 2013; Dong & Zhu 2013). This idea is fully supported by state-of-the-art planet formation models based on the core-accretion paradigm, which predict small rocky planets to greatly outnumber their Jovian or Neptune-like counterparts (e.g., Ida & Lin 2004; Mordasini et al. 2009, 2012a).

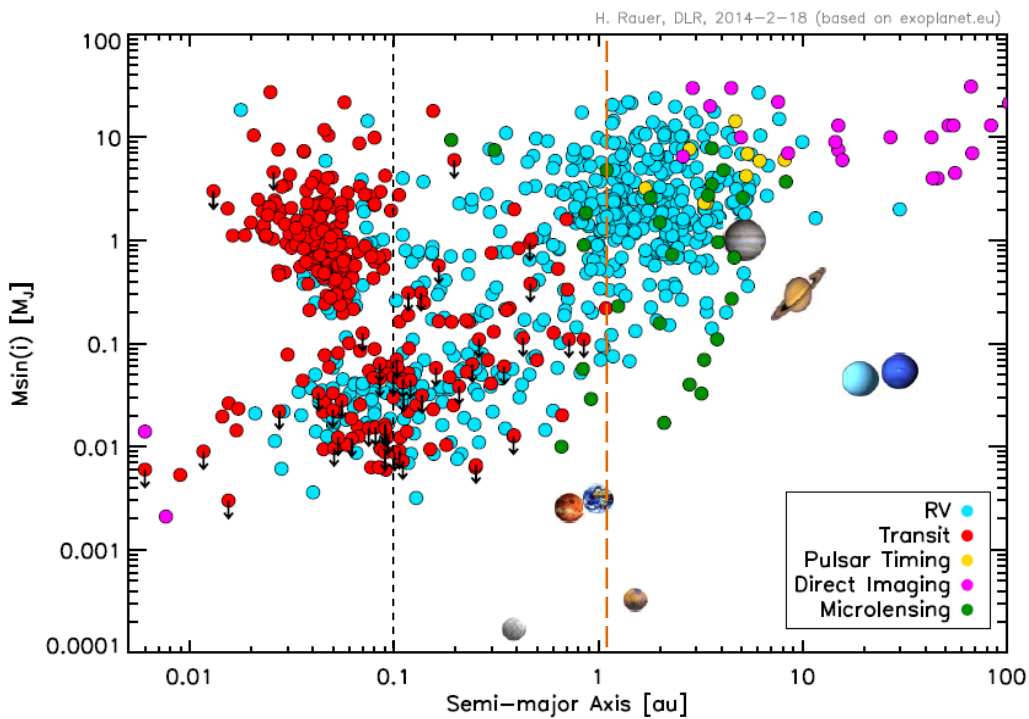

**Figure 2.1:** Current status of planet detections. Blue dots indicate RV detections with *m*sin*i* limits. Red dots are transit detections with known radii and masses. Red dots with downward arrows indicate transit detections with only upper mass limits available (update from Rauer et al. 2013). For illustration, the vertical black dashed line indicates the orbit of the widest transiting planet detected from ground today (HAT-P-15b (Kovacs et al. 2010)). The orange dashed line shows the envisaged orbital separation limit for PLATO 2.0 detections and characterization of super-Earths.

Figure **2.1** provides an overview of the confirmed exoplanet detections today. Jupiter-sized planets are well represented out to several au. Detections beyond ~0.1 au are dominated by the radial velocity technique which provides lower mass (m sin*i*) limits (blue dots). Masses <u>and</u> radii are known mainly for close-in planets, where data from both transits and RV are available (red dots). Transit

detections beyond about 0.1 au are very difficult from the ground due to the limited duty cycle of observations caused by the Earth's rotation (the most distant ground-based transit detection is HAT-P-15b at 0.095 au (Kovacs et al. 2010)). The known transits at intermediate orbital separations result from CoRoT and *Kepler*, showing that transit detections of planets at larger orbital separations are feasible from space. However, *Kepler* did not provide us with planet bulk parameters for the vast majority of its discoveries since most *Kepler* targets are too faint to allow for a direct measurement of terrestrial planet masses. We point out that the range of terrestrial planets as found in our Solar System, with masses from Earth- down to Mercury-sized objects beyond 0.3au, is still basically unexplored today.

Figure 2.2 shows the current status of Super-Earth planet detections in comparison to the position of the HZ, defined as the region around a star where liquid water can exist on a planetary surface (scaling based on Kasting et al., 1993). Most super-Earths have been found at orbital distances to the star closer than the HZ. Detections in the HZ have been made by RV or transit measurements (red and blue dots). However, only a small number of super-Earths have both mass and radius determined (purple dots), and these do not lie in the HZ. A recent example is the system around Kepler-62 with two planets orbiting in the HZ; no masses could be derived due to the faintness of the host star (Borucki et al. 2013). The black dashed line indicates the most distant super-Earths for which radii and masses could be directly measured by transits and RVs. Transit Time Variations (TTV) are capable expanding the distance limit (dotted line) for which masses of transiting planets are available, but we recall that TTV determinations of masses can have relatively large uncertainties, unless we observe co-planar transiting systems (see Section 6.4 for a more detailed discussion of TTV detections). The goal of PLATO 2.0 therefore focusses on providing terrestrial planets in the HZ of solar-like stars (up to about 1 au, orange dashed line) with accurately determined bulk parameters, which necessitates direct transit and RV measurements, hence planets orbiting bright host stars. TTVs will extend this distance range further, as they do for *Kepler*. In addition, PLATO 2.0's bright target stars allow for asteroseismology studies increasing not only the accuracy of stellar, and thus planet parameters (see Appendix A), but also providing the age of the systems detected (see Section 3.1). For M and K dwarfs, PLATO 2.0 will be able to detect planets beyond the snow lines (the distance to the snow line roughly scales as $2.7(M/M_{Sun})^2$), providing targets that could be further studied (e.g., by the JWST) for atmosphere signatures, giving PLATO the unique possibility to build a sequence of planets in largely different temperature and irradiation conditions, and test the fraction of volatiles incorporated in planets from beyond the snow line to very short stellar distances.

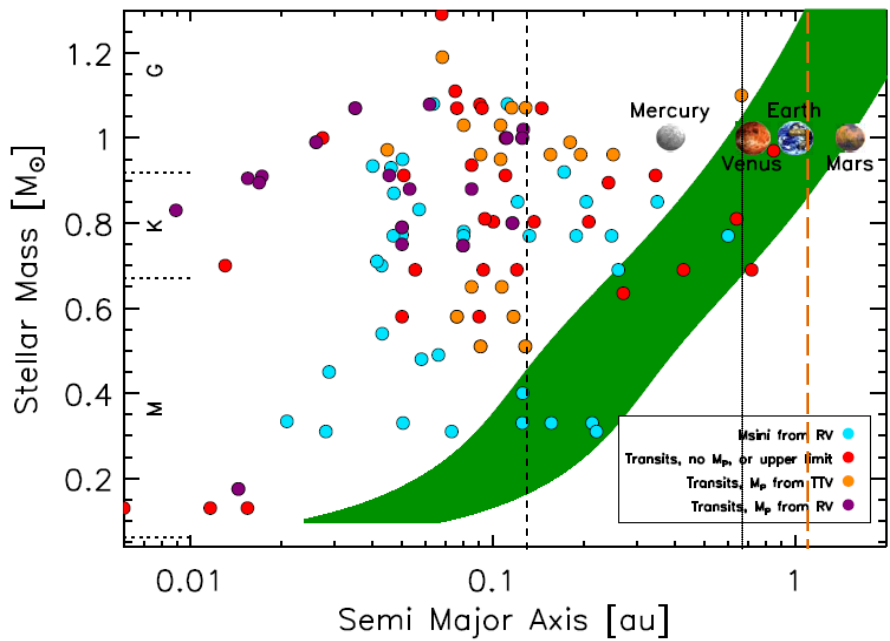

**Figure 2.2:** Super-Earth exoplanets (1 < $m_{planet}$ ≤ 10 $M_{Earth}$ or $r_{planet}$ ≤ 3 $R_{Earth}$) for different host star masses in comparison to the position of the habitable zone (green). Black dashed line: current max. distance of super-Earths with RV and transit measurements; Dotted line: most distant planet with transits and TTVs. Orange dashed line: distance goal of PLATO for fully characterized (transit+RV) super-Earths.

PLATO 2.0 therefore aims not only at a statistical approach studying the frequency of planet occurrence, but also asks about the nature of these planets: their bulk properties, atmospheres (gas versus terrestrial planets), and ultimately whether they could harbor life (hence orbit in the habitable zone). These new questions impose new requirements on planet detection surveys, because they need detailed follow-up observations which require high signal-to-noise ratios (SNRs). To address these new science questions, we need to

•	detect planets around bright stars (≤11 mag) to determine accurate mean densities and ages, and allow for follow-up spectroscopy of planetary atmospheres;

•	detect and characterize terrestrial planets at intermediate orbital distances up to the HZ around solar-like stars to place our Solar System into context;

•	detect and characterize planets in statistically significant numbers for a broad range of planet and planetary system classes to constrain planet formation scenarios.

These requirements define the design of PLATO 2.0 and its prime target range.

Figure 2.3 shows that past and existing transit surveys, including CoRoT and *Kepler*, have target stars which are too faint to fully characterize most detected planets. PLATO 2.0's main detection range is however <11 mag and will provide large numbers of targets for follow-up spectral characterization.

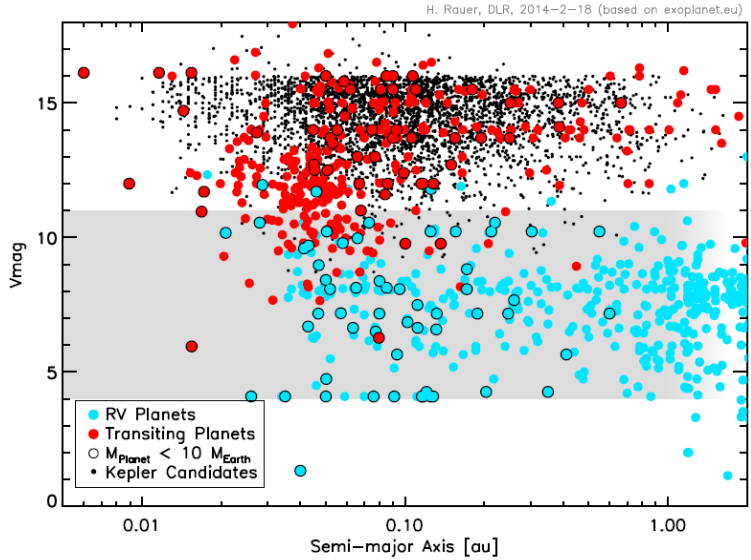

**Figure 2.3** Magnitude of known planet hosting stars versus planet orbital distance. The grey shaded band indicates the prime detection range of PLATO 2.0 (4-11 mag) for accurate bulk planet parameters and asteroseismology down to Earth-sized planets. Detection of Earth-sized planets is still possible down to 13 mag, and of larger planets down to 16 mag, but with lower bulk parameter accuracy.

The detection of Earth-like planets at intermediate orbital separations as in the Solar System is very time consuming, for RV as well as transit techniques. Furthermore, the geometrical transit probability decreases to 0.5% for Earth-like orbits around solar-like stars, affecting transit searches and also decreasing chances of finding transits for planets previously detected by RV. Fortunately, a large scale, wide-angle space transit survey like PLATO 2.0 can be optimized by observing a very large number of stars at the same time continuously and by adopting an appropriate observing strategy (see Section 5.2). PLATO 2.0 will detect thousands of new planetary systems of all kinds and hundreds of small/low-mass planets in the habitable zone of bright solar-like stars for which accurate radii, masses, mean densities, and ages can be derived. This goal is unique to PLATO 2.0. In addition, PLATO 2.0 will be able to detect exomoons, planetary ring systems, Trojan-planets, exo-comets, etc., thereby expanding our knowledge about the diversity of planetary systems.

## 2.2 Constraints on planet formation from statistics

A prime goal of PLATO 2.0 will be to detect a large number of planets, down to the terrestrial regime, with well-determined masses, radii and hence mean densities. Mean density is a testable quantity from theoretical planet formation models and is the key parameter to evaluate simulated planet population distributions and their input physics. This requires a large statistical sample covering the complete parameter space.

Figure 2.4 shows the mean density of planets versus planetary mass (a: for all planets, b: planets with P>50 days). We point out that unconfirmed planets cannot be used to derive reliable mean densities. In the figure, we note again that few planets in the mass range of Earth, Venus and below with measured densities and masses are available to date. Generally the mass range below 0.1 $M_{Jup}$ is sparsely populated. This is the highest priority detection space for PLATO 2.0. The PLATO 2.0 mission can provide thousands of rocky and icy planets with well-known radii (2%), masses (10%) and ages (10%) around ≤11mag stars, filling the left branch of Figure 2.4 a with a high number of planets.

Dashed lines in Figure 2.4 indicate modeled densities for planets with different bulk compositions (following Wagner et al., 2012). The right branch in each figure contains gas giant planets which follow roughly the green dashed line computed for planets with a Jupiter-like H-He bulk composition. The left branch of the roughly V-shaped density- mass distribution in Figure 2.4a is composed of planets with bulk densities from silicate to ice, some with extended atmospheres.

The formation of planets is presently believed to result from two different scenarios, which may or may not be mutually exclusive. In the core-accretion scenario, a planetary core is first formed by the collision of solid planetesimals. During this phase, the growing planet is in quasi-static equilibrium, the energy loss at the surface of the planet being compensated for by the energy resulting from the accretion of planetesimals. When the mass of the core reaches a so-called critical mass however, this compensation is no longer possible, and the planet envelope starts to contract, the contraction energy being radiated away at the surface. This contraction triggers a very rapid accretion of gas, which is limited by the amount of gas that can be delivered by the protoplanetary disk surrounding the forming planet. This scenario has been studied by many authors, accounting for different physical effects, like protoplanet migration (Alibert et al. 2004, 2005), opacity reduction in the planetary envelope (e.g., Hubickyj et al. 2005), excitation of accreted planetesimals by forming planets (e.g., Fortier et al. 2007, 2013), competition between different planets (e.g., Guilera et al. 2011).

In the second scenario, the disk instability model, the formation of a giant planet clump results from the presence of a gravitational instability in a cold and massive protoplanetary disk (e.g., Boss 1997, Mayer et al. 2005, Boley et al. 2010). After its formation, a giant planet clump is believed to cool and contract, and eventually accrete some planetesimals, forming a planetary core (e.g., Helled & Bodenheimer 2011, Vazan & Helled 2012).

In the framework of the core-accretion model one of the central issues is the possibility to build a core larger than the critical mass. The critical mass, in turn, depends on a number of processes which are poorly known. The critical mass depends strongly on the core luminosity (which results mainly from the accretion of planetesimals), and decreases for low luminosity (e.g., Ikoma et al. 2000). Moreover the critical mass depends on the opacity inside the planet envelope. Indeed, low opacity envelopes lead to a reduced critical mass, and a larger envelope mass (for a given core mass). Finally, the critical mass depends on the mean molecular weight inside the planetary envelope, which again depends on the planetesimals´ characteristics (size, strength, composition, see e.g., Hori and Ikoma 2011).

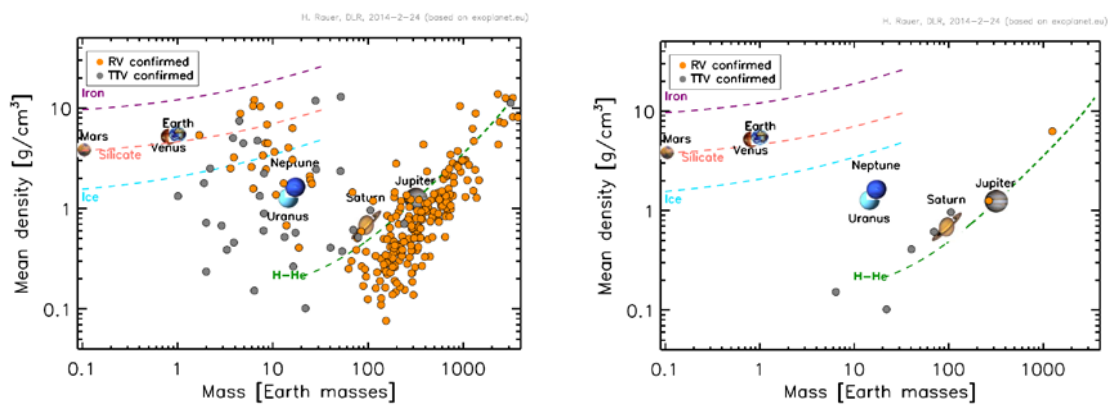

**Figure 2.4** Mean planet density versus mass with density lines for different bulk compositions. **Left**: All currently known planets with measured radius and mass (hence mean density). **Right**: As for (a), but only planets with orbital periods >80 days.

Determining observationally the critical core mass as a function of distance to the star, stellar metallicity, and other parameters would therefore place constraints on the characteristics of planetesimals (e.g., mass function, excitation state, internal strength). It would moreover determine up to which mass planets may be potentially habitable, since the presence of a massive $H_2$-He envelope in a super-critical planet probably prevents habitability.

The core accretion model scenario can be tested in particular by increasing the sample of high density, low-mass rocky exoplanets, since those planets define the critical mass limit beyond which efficient gas accretion starts. This is most likely the reason why basically no planets appear at high densities in the (approximately V-shaped) density-mass distribution in Figure 2.4, left. *For example, following the silicate composition line with increasing mass in* Figure 2.4*, left, we find no planets beyond about 0.03 $M_{Jup}$ (about 6-9 $M_{Earth}$).* This is consistent with the core accretion scenario where higher-mass silicate planets would quickly accrete significant H-He envelopes and end up as high mass but lower density planets in Figure 2.4, left, e.g., as ice planets or even growing to gas giant planets.

Interesting are two planets, Kepler-24b and c (Fabrycky et al. 2012, Ford et al. 2012; Wu & Lithwick 2013) with silicate mean density but masses beyond Uranus and Neptune, hence little gas envelope (see Figure 2.4, left). Such massive rocky planets challenge commonly accepted planet formation models. How can such planets form? Are they formed after the gaseous disc disappeared? These planets, were detected by Transit-Timing Variations (TTVs, see also Section 6.4) in *Kepler* data. Hence their masses have not been measured directly by RV but inferred from gravitational perturbations with respect to their Keplerian orbits, leading to potentially large mass uncertainties. This example illustrates that well-determined bulk densities are necessary to securely identify exciting new planet types that challenge formation theories. This is particularly true for low mass, high density planets, which are of central interest. Multi-planet, co-planar systems can be supplemented by TTV measurements. All of these discoveries will be facilitated by the large number of planets detected around bright stars which will on one hand allow us to obtain sufficient objects for statistical studies, and on the other hand uncover 'Rosetta Stone' objects with the potential to resolve some of the outstanding key questions.

The low-mass planets in Figure 2.4, left, show a wide range of densities: more than an order-of-magnitude. Planets of low mass and density below the (blue) pure ice line are indicative of large H-atmosphere envelopes. By filling the parameter space in Figure 2.4, PLATO 2.0 will identify a large sample of low-mass planets which likely have H envelopes, around different types of stars with different ages. Planet population synthesis models (e.g., Mordasini et al., 2012, right,) predict a large number of low-mass planets (super-Earths and below) with large hydrogen envelopes. Such predictions can be validated by PLATO 2.0, testing our planet formation theories. The situation becomes even more interesting if one considers also atmospheric loss processes (see also Section 2.11) which can remove a primordial H-atmosphere over time. These processes will be stronger closer to the host star. It will be interesting to study these effects observationally by correlating planetary mass and mean density of low-mass objects with e.g., orbital distance and age of the system (Lopez et al. 2012). We also expect the lowest-mass planets to lose their H atmospheres completely (e.g., like Earth, Venus, Mars). PLATO 2.0 will determine for which planets primordial atmospheres are unlikely to survive after a given time, and it will determine which planets have likely developed secondary atmospheres resulting in smaller apparent radii and higher mean densities.

Figure 2.4, right, shows the current situation for bulk characterization of planets with orbital periods >80 days. Only two exoplanets with measured transits and RV signals are currently known in this parameter range (orange dots); an additional five (red dots) arise from TTV mass determinations. Furthermore, few additional planets are expected to fill this diagram from the future space missions CHEOPS and TESS. Thus, while we will be able to compare planet population synthesis models with observations for planets at small orbital separations, the picture will be very limited for planets on larger orbits, i.e., orbits where planets are undisturbed by their host star and with potentially temperate surface conditions. PLATO 2.0 will be crucial to probe these orbital distances.

PLATO 2.0 will be the first mission to cover the parameter range of small, characterized (mass, radius, mean density, ages) planets with sufficiently large detection statistics to provide direct observational constraints to formation models.

**PLATO 2.0 will answer fundamental questions about planetary formation such as:**

- What is the bulk density distribution of low-mass, terrestrial planets?
- What is the observed critical core mass for giant planet formation?
- Can super-massive rocky planets exist and how are they formed?
- When and where do planets stop accreting gas?
- Which planets likely have extended, primordial H-envelopes?
- How do these parameters depend on stellar type, metallicity, chemical composition or age?

All of these questions have to be studied as a function of planet orbital separations, stellar metallicities and spectral type. They can only be addressed with a sufficiently large sample of planets of all sizes, from rocky to giant, with well determined masses, radii and bulk densities, around stars of different types and ages.

## 2.3 Terrestrial planets

This section discusses in more detail what can be learned from accurate radii and masses of terrestrial exoplanets, despite the limitations in observables for such distant planetary systems compared to our Solar System.

Terrestrial exoplanets up to about ten Earth masses are thought to have similar interior structures and bulk compositions as the terrestrial bodies in the Solar System. Their interiors are thought to be composed of rock-forming elements and metals such as iron, the latter evenly distributed or concentrated in central cores (Elkins-Tanton & Seager 2008). Gravitational and magnetic field measurements indicate that terrestrial planet interiors are strongly differentiated and subdivided into distinct layers. The composition of the layers varies with depth in such a way that the heaviest materials are concentrated in the center (core). An example of a differentiated terrestrial planet is the Earth which is divided into a partly or entirely liquid metallic core, a silicate mantle, and an outermost magmatic crust derived from partial melting of the mantle below. Unlike for the Solar System inner planets, there are fewer constraints than unknowns in the case of solid terrestrial extrasolar planets, and even basic interior structure models that would involve only two or three chemically homogeneous layers of constant density suffer from inherent non-uniqueness (e.g., Sohl & Schubert 2007, and references therein). To address these degeneracies, assumptions are usually made about their composition and its depth dependence.

Numerical models of planetary interiors using laboratory data on material properties aim at improving the general understanding of their origin, internal evolution, and present thermal states. In the case of the rocky planets within the Solar System, the resulting radial profiles are required to be consistent with geophysical observations and cosmochemical evidence for the compositions of crust, mantle and core (e.g., Sohl & Schubert, 2007, and references therein). For rocky exoplanets, the numerical models have to be consistent with the observed planetary masses and radii. Such models have been used to derive mass-radius relationships for exoplanets assuming a range of different mineralogical compositions to gain insight into the interior structure and possible bulk compositions of these planets (Valencia et al. 2006, Fortney et al. 2007, Seager et al. 2007, Sotin et al. 2007, Valencia et al. 2007, Grasset et al. 2009, Figueira et al. 2009, Wagner et al. 2011, Zeng & Sasselov 2013). The principal uncertainties mainly arise from the extrapolation of an equation-of-state to high pressures owing to the lack of reliable experimental data in the pressure range of 200 GPa to 10 TPa, whereas the surface temperature and internal thermal state of a massive rocky exoplanet are less important for its radial density distribution (e.g., Seager et al. 2007). Nevertheless, the latter are expected to have severe consequences for geodynamical processes. Furthermore, scaling laws for key physical and chemical properties have been obtained (e.g., Wagner et al. 2012, and references therein), which are essential for a better understanding of the global planetary processes controlling the general evolution of a planetary body and its astrobiological potential to be life-sustaining.

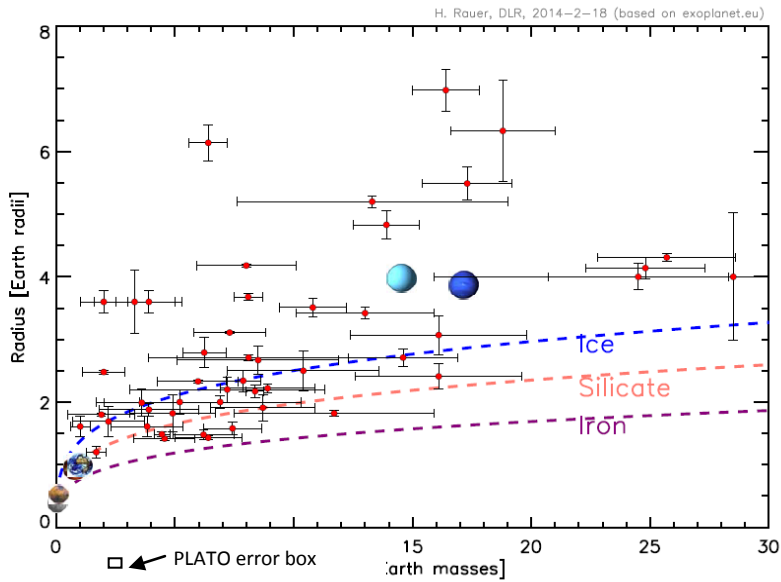

**Figure 2.5** Mass-radius diagram for planets with different bulk compositions. Water ice (blue line), silicate rock (orange line), iron (purple line) (see Wagner et al. 2011, for details) are compared to known low-mass planets (with 1 sigma error bars).

Figure 2.5 shows modeled mass-radius relationships in comparison to the relatively large (1 sigma) error bars obtained for low-mass planets to date. For the smallest planets, radii are better constrained than masses. These planets are usually detected by space missions (CoRoT and *Kepler*) providing photometrically accurate light curves, and hence radii, but the target objects are too faint to permit an accurate mass determination. In many cases, even a rocky or icy nature cannot be distinguished within the 1 sigma error bars shown. There is a need to reduce the error bars, as planned for PLATO 2.0, by providing highly accurate radii and masses with corresponding uncertainties of merely a few percent.

The knowledge of mean planet density is foremost dependent on the quality of the stellar mass and radius determinations that feed into the determinations of planetary mass and radius. One of the main goals of PLATO 2.0 is therefore to provide highly precise and accurate measurements of the planet host stars' characteristics, in particular their radii, masses and ages. Typical current uncertainties for radius and mass determinations of small planets are around ±6% and ±20%, respectively, leading to uncertainties of 30 to 50% in mean density. The observational accuracy envisaged for PLATO 2.0 will reduce the uncertainty in mean density to about 10%.

Provided the solid planet interior is fully differentiated into an iron core and silicate mantle, Figure 2.6 illustrates that the present detection limits are not sufficient to determine satisfactory the interior structure of an Earth-like planet (after Noack et al. 2013). Figure 2.6 (left) shows the iron core size for a radius of 1 Earth radius with 1σ uncertainty of +/- 6%, while the planet mass is taken constant at 1 Earth mass. To satisfy mass balance constraints, a larger planet radius is then compensated by a smaller iron core size. The dark-shaded band indicates the expected improvement in core size determination using PLATO 2.0 (radius ±2%). Figure 2.6 (right) shows the possible interior structure if the mass is determined as 1 Earth mass +/- 20% and the planet radius is held fixed at 1 Earth radius. The dark-shaded band again shows the improvement owing to the enhanced PLATO 2.0 accuracy. In summary, within the present observational limits, it is difficult to distinguish between an almost coreless planet and a Mercury-like planet interior with a large iron core. The situation will significantly improve with PLATO 2.0 accuracies.

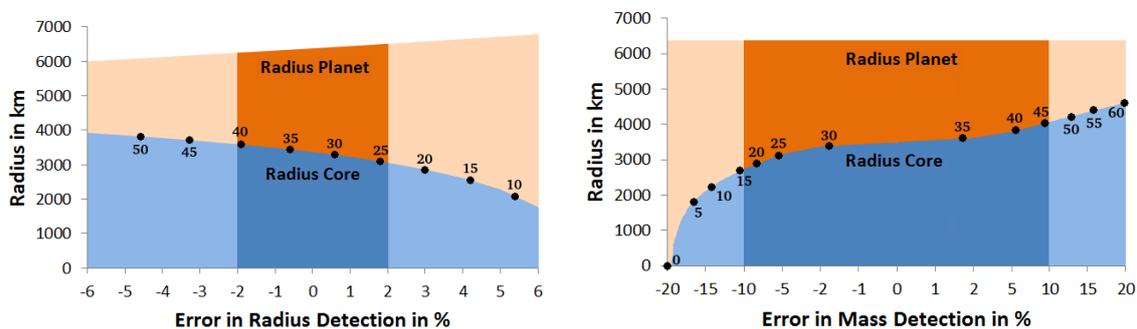

**Figure 2.6** *Left:* Radius of planet and its core depending on the uncertainty in radius $R_{planet}$ or *Right:* planetary mass $M_{planet}$. The calculations are based on mass-radius relationships reported in Wagner et al. (2011). Left, we assume a planet of 1 Earth-mass and vary the radius (0 corresponds to 1 $R_{Earth}$) within current uncertainties (±6% in radius). Right: same, but keeping the radius fixed at 1 $R_{Earth}$ and vary the planet mass within current uncertainties (±20%). Numbers at black dots provide the core mass fraction as percentage of total mass. The dark shaded regions illustrate the expected PLATO 2.0 accuracy (±2% and ±10% in radius and mass, respectively). See Noack et al. (2014) for details.

The ratio of core radius to planet radius is important for understanding the interior evolution of a terrestrial planet, which can also influence its surface habitability. For example, the volume of the silicate mantle and the hydrostatic pressure in the upper mantle both influence the amount of partial melting and hence the rate of volcanism at the surface. Greenhouse gases are trapped in the uprising melt and are released at the surface feeding the atmosphere. In view of the large uncertainties involved in the underlying exchange processes, important bounds on the present models must be expected from a large and diverse population of well-characterized low-mass planets. Accurate determinations of both mass and radius are therefore important to impose bounds on interior-surface-atmosphere interactions with possible consequences for surface habitability (e.g., Noack et al. 2014).

Current detection limits have prevented the discovery of more than a few rocky exoplanets, although low-mass planets around other stars are most likely abundant. The future detection of hot (super-)Earths by e.g., TESS, and their follow-up by CHEOPS, will provide the first fundamental information to better constrain the bulk compositions of these planets. PLATO 2.0 will then provide masses and radii of a large number of solid planets up to 1 au distance from their host star. Studying temperate planets at large orbital separations allows us to address the architecture of planetary systems and the connection to proto-planetary disk properties, and finally to study the relationship of interiors to atmospheres in planets up to the HZ. These will be complemented by the detection of giant planets at larger orbital separation expected from the Gaia mission, expanding our characterization of these planetary systems.

Constraining the mean composition and bulk interior structure of small planets, PLATO 2.0 will enable us to answer the following questions:

- Is there another planetary system including a terrestrial planet like Earth?

- What is the typical mean density distribution (and mass function) in planetary systems?

- How is the planet mean density distribution correlated with stellar parameters (e.g., metallicity, mass, age, etc.)?

## 2.4 Gas giants and icy planets

We discuss here the improvements for our understanding of gas and ice planet interior and formation mechanisms due to the PLATO 2.0 mission. Many gas and ice planets are already known, and more are expected from ground-based surveys (e.g., NGTS (Wheatley et al. 2013)) in the near future. The CHEOPS mission will provide significantly improved constraints on the radii of transiting planets discovered from the ground (Broeg et al. 2013). Thus, first steps to a deeper understanding of gas and ice planet interiors will be made in this decade. The main role of PLATO 2.0, following these activities, is to dramatically increase the mass-radius parameter space for exoplanet detection and to further extend it towards intermediate orbital separations. Again we note that this data set will be complemented at large separations by detections from the Gaia mission, which complement the PLATO observations at orbital separations where the transit probability is low.

We first discuss what is known about gas and ice planets from our Solar System. Giant planets are planetary bodies which primarily consist of hydrogen and helium and a small fraction of heavy elements (i.e., rocks, ices). The Solar System gas giants are Jupiter and Saturn. They orbit the Sun at distances of 5.2 and 9.6 au, respectively. The composition of a giant planet and its depth dependence are calculated by interior models, which are constrained by the observational properties of the planet, such as its mass, radius, rotation rate and gravitational field coefficients. For Jupiter and Saturn these physical parameters are well known from space missions.

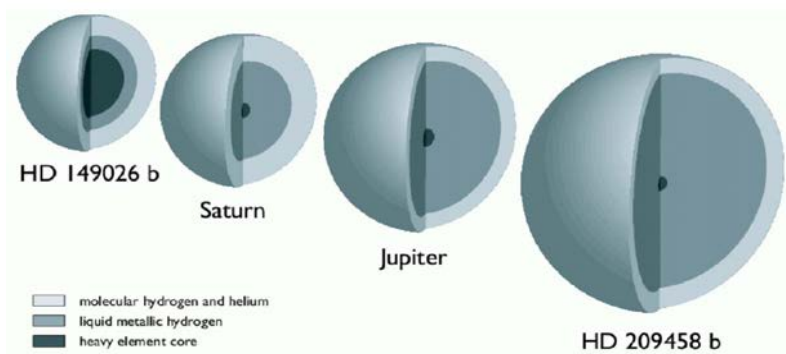

**Figure 2.7** Illustration of the interior structure of HD149026b and HD209458b in comparison to Jupiter and Saturn (from Charbonneau et al. 2007).

There is still uncertainty in the bulk composition of Jupiter and Saturn, in particular, the amount of high atomic number (Z>2) material and the presence of a central core. The uncertainty in giant planet interior models reflects the uncertainty in the equations of state (EOSs) and assumptions such as the number of layers, the distribution of the heavy elements within the planet, and the rotation profile/state. An additional uncertainty arises from the fact that the planets are assumed to be adiabatic, and therefore fully convective. However, if the planets have non-adiabatic structures, which is an outcome of double diffusive convection (Leconte & Chabrier 2012, 2013), the heavy element mass can be significantly higher due to the higher internal temperatures, and constraining the bulk composition of the planets becomes even more challenging.

Internal models of Jupiter and Saturn suggest that Jupiter's core mass ranges between 0 and 10 $M_{Earth}$ and that the mass of high-Z material in the envelope is about 30$M_{Earth}$. The total mass of heavy elements in Jupiter ranges from ~10 to ~30 $M_{Earth}$ (see e.g., Saumon & Guillot 2004; Nettelmann et al. 2008). Recently, Militzer et al. (2008) suggested that Jupiter's interior consists of a core of about 14 to 18 $M_{Earth}$ surrounded by a homogenous envelope composed mainly of hydrogen and helium. Determinations of Saturn's total enrichment in heavy elements typically range from ~10 to ~30 $M_{Earth}$, with core masses between ~0-15 $M_{Earth}$ (e.g., Saumon & Guillot 2004; Fortney & Nettelmann 2010; Helled & Guillot 2013).

The icy planets of the Solar System are Uranus and Neptune. Standard interior models suggest that they consist of three main layers: (1) an inner rocky core; (2) a water-rich envelope; (3) a thin atmosphere composed mostly of hydrogen and helium with some heavier elements (e.g., Podolak et al. 1995; Marley et al. 1995; Fortney & Nettelmann 2010). However, it should be noted that due to the uncertainties of the measurements it is still unclear whether Uranus and Neptune are truly 'icy planets', as their names suggest, or planetary bodies which primarily consist of silicates, with hydrogen and helium envelopes (e.g., Helled et al. 2011). In addition, calculations of Uranus' cooling history imply that the planet contracts ´too slowly´, i.e., simulations find that Uranus cannot cool to its measured intrinsic luminosity by the age of the Solar System assuming an adiabatic interior. This suggests that Uranus' interior may not be fully convective, and/or that it contains an additional energy source (e.g., compositional gradients) besides its gravitational contraction (e.g., Fortney & Nettelmann 2010). Neptune too, likely has a significant internal energy source. Another important open question regarding these planets is their formation process. It is still unclear what conditions and physical mechanisms lead to the formation of these fairly low-mass objects, especially at the large radial distances we find them today in the solar-system (e.g., Dodson-Robinson & Bodenheimer 2010). It was suggested by so called `Nice model' (Tsiganis et al. 2005) that the architecture of the Solar System changed over time thanks to dynamical interactions of the giant planets with the dense planetesimal disc. Due to the momentum exchanges the giant planets migrated (except for Jupiter all increased their orbital distances) and thus their current distances from the Sun do not have to reflect the distances where the giants formed. By capturing various planetary systems at different evolutionary states PLATO 2.0 may provide some verification of the planetary migration hypothesis.

The compositions and internal structures of extrasolar giant and Neptune-sized planets are less constrained than the planets in the Solar System, but they offer the opportunity to study giant planets as a class. The diversity of gas giant and 'icy' exoplanets is much broader than found in our Solar System, thus expanding the parameter range that can be studied.

Current technology still limits the detection of transits to planets that orbit fairly close to their host stars. Although the majority of transiting giant planets are composed mostly of hydrogen and helium (e.g., Guillot et al. 2006, Miller & Fortney 2011), their internal constitution is not necessarily similar to those of the gas giants in our Solar System. In fact, exoplanets show a large diversity of masses and radii, which has yet to be explained. Extrasolar giant planets can differ significantly from Jupiter and Saturn (e.g., Figure 2.7) since they formed in different environments. In addition, giant planets close to their parent stars are exposed to intense stellar radiation that prevents their atmospheres from cooling and therefore affects the contraction of their interiors (Koskinen et al. 2007). Although our understanding of 'hot Jupiters' is still incomplete, substantial progress in studying these objects has been made. Interior models including the effects of irradiation have been computed (e.g., Guillot et al. 1996; Bodenheimer et al. 2003, Batygin et al. 2011) and detailed models of the giant planets' atmospheres are now available, although which kind of mechanisms can inflate the radius of irradiated giant planets is so far still unclear (e.g. Laughlin et al. 2011). With its precise determinations of planet radii, PLATO 2.0 will also significantly contribute to understanding the mechanisms responsible for the inflation of gas giants. Indeed, it has been shown by Schneider et al. (2011) (their Fig. 9) and confirmed by Demory et al. (2011a) that inflation decreases with the planet illumination by the parent star. PLATO 2.0 will provide a broad statistics of this correlation and the influence of the stellar wind.

In addition, detailed studies of the interior structures of extrasolar giant planets suggest that these objects typically possess cores (Figure 2.8) of at least $10 M_{Earth}$ (Guillot et al. 2006, Miller & Fortney 2011). The heavy element mass is proportional to stellar metallicity ([Fe/H]) while the planetary enrichment is inversely proportional to the planetary mass (Miller & Fortney 2011). Recently, a class of planets has emerged that possess a large fraction of rocky material in their cores, see CoRoT-13b (Cabrera et al. 2010b) as an example (Figure 2.8).

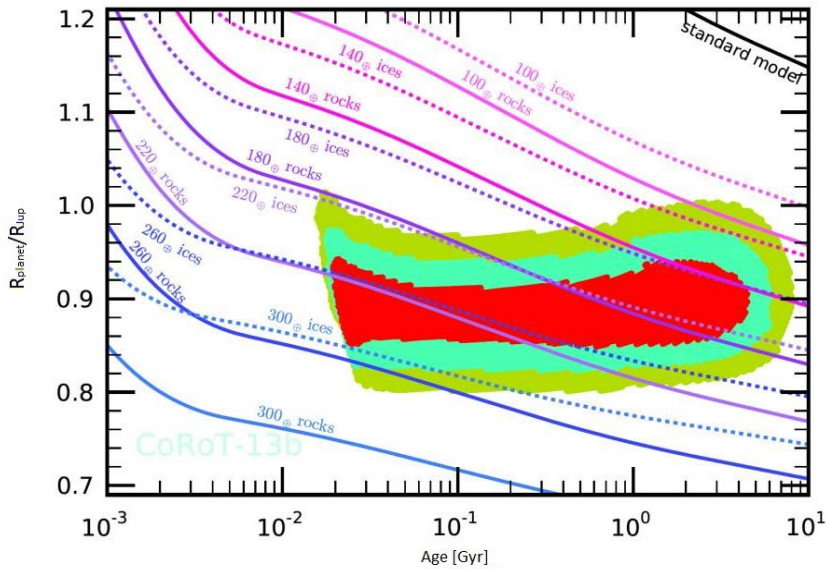

**Figure 2.8:** CoRoT-13b radius development over age (Cabrera et al. 2010b and references therein). The coloured areas provide the uncertainty in planet radius and in stellar age derived from stellar evolution models matching the stellar density and effective temperature (within 1 (red) to 3 (green) sigma uncertainty). The curves show evolution tracks for CoRoT-13b (assuming M = 1.308 $M_{Jup}$, $T_{eq}$ = 1700 K) for different amounts of heavy elements concentrated in a central core, surrounded by a solar-composition envelope.

As discussed in Section 2.2, there are two leading theories for giant planet formation: core accretion, the standard model, and disk instability. While both formation scenarios can lead to a large range of compositions and internal structures the core accretion model typically predicts a non-solar composition for giant planets, while giant planets formed by gravitational disk instability can have different compositions, including stellar, depending on stellar metallicity, planetary mass, the efficiency of planetesimal accretion etc. (Helled & Schubert 2008). More accurate determinations of the bulk compositions of giant extrasolar planets which are *not* strongly irradiated, as expected to be obtained by PLATO 2.0, can provide valuable constraints on giant planet formation and evolution models. For close-in gaseous and icy giant planets, the knowledge of their interior structure will furthermore allow us to better understand tidal dissipation processes in these planets (e.g. Remus et al., 2012; Ogilvie 2013).

PLATO 2.0 will improve our understanding of the composition and evolution of gas giant and Neptune-sized planets in two major ways. Firstly, the planets discovered by PLATO 2.0 around bright stars will have 3 times more accurate radius determinations and 5 times more accurate mass determinations than current results. This will allow us to classify detected planets as rocky, icy or gas giant with high accuracy. High precision measurements of planetary radii and masses will allow us to constrain core masses from interior modelling. These can be compared to the largest observed core sizes that failed to undergo gas accretion and help to constrain planet formation. Finally, PLATO 2.0 will provide the masses and radii of giant planets of various ages. This will allow us to address the contraction history of ice and gas giant planets (see Figure 2.8). Furthermore, it will allow us to address the possibility of compositional change with time and the connection between age, inflation, and atmospheric loss rate. In summary, PLATO 2.0 will address the following questions regarding gas and ice planets:

- How do gas giants with massive cores form?

- Up to which orbital separation do we find inflated gas giant planets?

- How does this correlate with stellar parameters (e.g., type, activity, age)?

- Are gas giants with massive cores frequent and how does their distribution depend on orbital distance and stellar type?

## 2.5 *Planets orbiting intermediate mass stars*

It is now well established that planets form within a few million years from the dusty, circumstellar disk of young stars. It is thus expected that the properties of the planets must be closely related to the structure, lifetime, mass and chemical composition of the disk, but how they relate to each other is not known. Finding such correlations would give us key information on how planets form. In order to find out how the properties of the planets relate to the properties of the disk, we have to take the statistical approach. Theoretical studies have shown that more massive stars should also have more massive planets, because they had also more massive disks (Kennedy & Kenyon 2008). This prediction is in fact confirmed observationally: Intermediate-mass stars (1.3-2.1 $M_{Sun}$) have twice as many massive planets as solar-like stars, and they can also have planets that are much more massive than solar-like stars (Johnson et al. 2010a; 2010b; Vigan et al. 2012). PLATO is the first mission that allows the study of how the planet population changes with the mass of the host star over a wide range of masses for a large number of targets. Since we now know, thanks to Spitzer and Herschel observations, how the average properties of disk change with the mass of the host star, we will for the first time be able statistically relate properties of the disks with the properties of the planets. A key question to be addressed is: How do the properties of planets change with the mass of the host stars?

## 2.6 *Planets around Subgiant and Giant Stars*

Several ground-based Doppler planet searches target subgiant and giant stars instead of main-sequence stars. The number of planets known to orbit giant stars (about 50) is still small compared to those known to orbit main-sequence stars, but their number has dramatically increased in recent years and is expected to do so in the near future. The discovery and characterization of planets orbiting subgiant and giant stars is of particular importance for the following reasons:

- Confirmation of a planet orbiting a giant star is in many cases almost impossible based on radial velocities alone, since the RV signal of an orbiting planet is hard to disentangle from the RV signature of radial and non-radial pulsations, unless in cases where their timescales are very different. Thus, independent confirmation of planets orbiting giant stars are most useful.

- Subgiants and giant stars can be more massive than solar-like main-sequence stars, so by finding more planets around giant stars we can disentangle the influence of the host star's mass and its disk on the forming planets and their properties.

- Subgiants and giant stars have undergone significant stellar evolution, which affected planetary orbits. Studying the planet population around subgiant and giant stars offers the opportunity to investigate the influence of stellar evolution on the properties of the planetary population.

- It is not still clear if evolved giant planet hosts are mostly metal-rich (Pasquini et al 2007, but see also Mortier 2013). This could be checked using the PLATO 2.0 planet sample in combination with its well-characterized host stars.

CoRoT and *Kepler* have made few detections of planets around giant stars. A recent example is Kepler-56 (Steffen et al. 2013) and further candidates can be found in, e.g., Huber et al. (2013a). Recently, the false-alarm rate of planets around giant stars in *Kepler* data (KOIs) was found high (Sliski

& Kipping, 2014). PLATO 2.0 is in a better position to find such planets. The depth of a transit of a Jupiter-sized planet in front of a giant star with a radius of 10 solar radii is 100 ppm, which is within the reach of PLATO 2.0. The planets found by RV surveys around this type of star have typically periods of several hundred of days. For example, at a 400 day orbital period, the transit probability is 3% and the transit duration almost 4 days. Its detection will be challenging since the photometric activity of the giant star must be well characterized, but the detection of such transits is within the detection capabilities of PLATO 2.0.

## *2.7 Planets around post-RGB stars*

To date not a single bona fide planet has been identified orbiting an isolated white dwarf (e.g., Hogan et al. 2009). Therefore, we remain ignorant about the final evolutionary configuration of >95% of planetary systems. Theoretical models (e.g., Nordhaus & Spiegel 2013) predict a gap in the final distribution of orbital periods, due to the opposite effects of stellar mass loss (planets pushed outwards) and tidal interactions (planets pushed inwards) during the red giant branch (RGB) and asymptotic giant branch (AGB) phases. If a planet enters the envelope of the expanding giant star, its survival depends a number of poorly constrained parameters, particularly its mass. Currently, the lowest mass brown dwarf companion known to have survived such "common envelope" evolution to the WD stage has 25-30 $M_{Jup}$ (Casewell et al. 2012), but theoretical models suggest much lower mass gas giants may survive.

Over its five year primary mission, Gaia is expected to astrometrically detecttens or hundreds of WD planets (M > ~1 $M_{Jup}$) in long period orbits (Silvotti et al. 2011), but the likelihood of planets surviving in close orbits around WDs will likely remain an open question for some years. Recently, more than 15 planets around post-RGB were detected, orbiting extreme horizontal branch subdwarf B (sdB) stars, or cataclysmic variables. Most of them, on long-period orbits, were discovered from eclipse or pulsation timing (e.g., Silvotti et al. 2007), while two sdB planetary systems with very short orbital periods of few hours were detected by *Kepler* through illumination effects (Charpinet et al. 2011, Silvotti et al. 2014). The *Kepler* discoveries suggest that ~10% of sdB stars could have close planets (or planetary remnants) and ~1/40 of sdB stars could show a transit. Although we expect that some new results may come in the next years from ground-based Doppler surveys, PLATO 2.0 can easily collect large-number statistics on these objects, allowing detecting sdB planets not only from illumination effects but also from the first transits, giving first estimates of their radii.

Even more importantly, PLATO 2.0 has the capabilities to detect the first WD planet transits, which require large statistics (Faedi et al. 2011). PLATO 2.0 can easily detect gas giants eclipsing WDs, placing limits on the masses of planets that can survive 'common envelope' evolution. In addition, since WDs are similar in radius to Earth, PLATO 2.0 can detect transiting bodies down to sub-lunar sizes. Such objects may exist in close orbits to WDs, possibly through perturbations with other planets in a complex and unstable post-main sequence system. Indeed, at periods of ~10-30 hours, these rocky bodies would exist in the WD's 'habitable zones' (Agol 2011), and their atmospheres would be detectable with JWST (Loeb & Maoz 2013).

Discovery and characterization of post-RGB planets is essential to study planetary system evolution and planet-star interaction during the most critical phases of stellar evolution: RGB and AGB expansion, thermal pulses, planetary nebula ejection. We note that sdB/WD asteroseismology allows a very good characterization of these stars and their planets.

## *2.8 Circumbinary Planets*

Planets that orbit around both components of a stellar binary were suggested as favorable targets for transit surveys (Borucki 1984) due to the expected alignment between the planetary and the stellar orbital planes, which strongly increases detection probabilities on eclipsing binaries with near edge-on orbits. Some early surveys (e.g. Deeg et al. 1998) subsequently centered on them, but it was not until the *Kepler* mission that the first transiting CBPs were found (Doyle et al. 2011). The discovery of 7 circumbinary planets (CBPs) in 6 systems has been announced to date. Their characteristics are rather distinct to those found by timing methods, with orbital periods on the order of several months and planet-masses that are relatively low, the heaviest one being Kepler-16b with 0.33 $M_{Jup}$. All CBP orbits have an inner limit to their stability (e.g. Dvorak et al., 1989, Chambers et al. 2002) and most of the transiting CBPs orbit rather close to that limit (Welsh et al. 2012). It is also notable that all planet-hosting binaries have orbital periods on the order of 10 days or longer. An additional photometric method to detect CBPs, based on the detection of the binaries' eclipses in the planet's reflected light has been presented by Deeg & Doyle (2011). In *Kepler* data, this method could detect CBPs that are not far from the inner stability limit around short-periodic binaries in a large range of orbital inclinations, but no discoveries have been reported yet.

Formation and evolution models predict in general the formation of circumbinary protoplanets in relatively distant disks and subsequent migration, combined with the further accretion of matter, to the planet's observed positions. In particular, the accumulation of CBPs near the inner stability limit has been foreseen by Pierens & Nelson (2007), who predicted that an inward drift of a protoplanet can be stopped near the edge of the cavity formed by the binary. In more general terms however, any generic theory on planet system formation and evolution needs to be compatible with planets found around binary stars, making this population of planets therefore an interesting test-bed for many theoretical advances.

For PLATO 2.0, this presents the following objectives:

- What are the properties of the circumbinary planetary systems? What are their masses, orbital periods and the types and ages of host stars? Can their special features be explained by existing planet formation theories, and/or do they need modifications?

- Do other classes of CBPs besides currently known ones exist? In particular, no CBPs around short-period binaries have been found to date, although these binaries are by far the most common ones and there are no special obstacles to the detection of their planets.

The number of CBP detections that was found to date in *Kepler* data is likely limited by the number of lightcurves sampled and not so much by its photometric precision, e.g. all known CBP transits can be identified 'by eye' in the lightcurves. This indicates that the discovery of these systems in *Kepler* data may be rather complete, although some efforts to detect shallow transit CBPs are still ongoing. With the sample size and observing duration of PLATO 2.0, we can expect that the sample of transiting CBPs of the types that are currently known will multiply several-fold. We can also expect a clear answer on the existence of short-periodic CBPs. For this population, long observing durations are not essential, and the PLATO 2.0 step-and-stare phase with its very large sample will be decisive to resolve their abundance.

## 2.9 Evolution of planetary systems

The ability to derive the ages of planetary systems is one of the key assets of PLATO 2.0. Unfortunately, the ages of stars are traditionally very poorly constrained (to within only a few Gyr for stars on the main sequence). Furthermore, young planets, that are the most important in order to decipher the conditions under which planetary systems are formed, orbit around active stars and the determination of their parameters has remained at best elusive (see e.g., Gillon et al. 2010, Czelsa et al. 2009, Guillot & Havel 2011). With relative ages of main sequence stars known to within 10%, PLATO 2.0 will essentially remove the age ambiguity. Being able to know planetary systems ages for a large sample will allow us to search for type cases of planet evolution and possible correlations with e.g., the host star parameters, and the planet interior composition and structure. Furthermore, once future large-scale missions are able to spectroscopically characterize nearby Earth-like planets for signatures of life, their host star has likely been characterized by PLATO, such that the age of the system will be well known from the PLATO catalogue.

Planets and planetary systems evolve with age in several aspects, which we briefly summarize here:

- Gas giant planets progressively cool and contract, a process that lasts up to several Gyrs (see Section 2.4). An accurate knowledge of age is therefore crucial for the interpretation of measured radii and a determination of interior structure (ESA/SRE(2011)13).

- Terrestrial planets also evolve with time. Planet formation theories predict rocky planets with primordial hydrogen atmospheres. The atmospheres of the terrestrial planets in our Solar System are secondary atmospheres produced by outgassing from the interior and impacts of (small) bodies, both processes being more intense in the young Solar System. In the case of e.g., Mars a possible denser young atmosphere has meanwhile been lost to space. In the case of the Earth its atmosphere has been modified by the development of oxygen-producing life (tertiary atmosphere) since about 2.5 Gyrs. The distinct evolution of the terrestrial planets in our Solar Systems is far from being fully understood. Exoplanets can complement our investigations of terrestrial planet evolution by contributing information not accessible in the Solar System: A large number of planets over a wide bulk parameter range and at different ages. This will allow us to search for type cases and possible correlations of planetary evolution processes with stellar and planetary system parameters, which will provide a breakthrough in our understanding of the evolution of atmospheric composition and habitability. PLATO 2.0 will provide the initial key steps towards this ultimate goal. Crucial here are planets at intermediate orbital distances, which are less affected by interactions with strong stellar radiation or winds.

- Host stars evolve with time and expose young planets with much higher UV and high-energy radiation levels than found on Earth today (see Section 2.11). This affects processes like atmospheric losses, but also radiation levels affecting life on the surface of terrestrial planets. Therefore, a good characterization and dating of the host stars is crucial to obtain an understanding of the evolution of planetary atmospheres and habitable conditions.

- The architecture of planetary systems is shaped through planet formation and subsequent dynamical processes that cover a wide range of timescales, up to billions of years. The comparison of planet system populations of different ages will allow us to investigate whether typical scenarios at different ages exist (e.g., hot Jupiters: disk or Kozai migration). Kepler has discovered several compact multiple systems which show significant dynamical interactions (with 7 planets Kepler-90 (Cabrera et al. 2014); with 6 planets Kepler-11 (Lissauer et al. 2011), or Kepler-154 (Ofir & Dreizler 2013, Rowe et al. 2014)) and which are very interesting for the understanding of planetary formation. PLATO 2.0 will be able to provide accurate masses for similar complex planetary systems. Furthermore, TTV observations over long time periods, e.g., by combining PLATO 2.0 with already available *Kepler* and TESS observations, constraining the Q-factors describing the internal tidal energy dissipation in stars and planets, is crucial to understanding the evolution of close-in planets (Goldreich & Soter 1966).

The accurate determination of planetary system ages for thousands of systems is therefore among the key features of PLATO 2.0. This crucial goal will not be achieved by any other ongoing or planned future transit mission. Key science questions PLATO 2.0 can answer are:

•        What are the ages of planetary systems?

- How do planet parameters (e.g., mean densities, radii of gas giants, planet star distance distributions, and (if combined with spectroscopic follow-up) atmospheres) correlate with age?

- How many super-Earths retain their primary atmosphere (inferred from low density)? Is there a correlation of these primary atmospheres with system age? What are the main parameters governing the presence of primary atmospheres (e.g., formation mechanism, stellar type, orbital distance, age, metallicity …)?

- How does the architecture of planetary systems vary and evolve with age?

## *2.10 Planetary atmospheres*

In the past decade, numerous studies have been published on the use of wavelength-dependent primary transits and secondary eclipses to characterise the atmospheres of exoplanets, including e.g. GJ 1214b (e.g., Charbonneau et al. 2009; Bean et al. 2010; Berta et al. 2012; de Mooij et al. 2012) and 55 Cancri e (e.g., Crossfield et al. 2012; Demory et al. 2012; Ehrenreich et al. 2012). Highlights include the claimed detections of molecular features in the infrared (e.g., Knutson et al. 2011) to the inferred presence of clouds/hazes in the visible (e.g., Pont et al. 2013) in the atmospheres of hot Jupiters, and even the detection of the exosphere (Vidal-Madjar et al. 2003; Lecavelier des Etangs et al. 2012). Visible data determine the albedo, the identity of the major, spectroscopically inert molecule and the relative abundance of clouds/hazes of the atmosphere. Clouds have long been an obstacle in our understanding of the atmospheres of Earth, Solar System objects and brown dwarfs, and are rapidly emerging as a major theme in the study of hot Jupiters, super Earths and directly imaged exoplanets. For small exoplanets, visible data help to determine if a thick, gaseous atmosphere is present and thus identify the exoplanet as a prime candidate for follow-up, atmospheric spectroscopy with JWST, E-ELT and future L-class missions.

The albedo measures the fraction of starlight reflected by an atmosphere and therefore its energy budget. It is of central importance in determining the thermal structure of the atmosphere. Measuring the secondary eclipse (occultation depth) in the visible directly yields the geometric albedo, which is the albedo of the atmosphere at full orbital phase (e.g., Demory et al. 2011b). Detecting reflected light over the planet orbit, the spherical albedo can be derived. For the hottest objects (~2000 to 3000 K) thermal emission from the exoplanet may contaminate the broadband visible data, thus confusing the measurement of reflected light versus thermal emission. In these situations, the two broadbands of the fast cameras of PLATO 2.0 will be useful in decontaminating the occultation depth measurements for the brightest stars.

The spectroscopically active molecules of an atmosphere typically contribute spectral features in the infrared, but these molecules are often minor constituents of an atmosphere (by mass). Of central importance in interpreting an exoplanetary atmosphere is knowledge of the pressure scale height, which is set by the mean molecular weight. This is determined by the dominant (by mass) inert molecule, and the gravity of the planet. On Earth, the dominant, inert molecule is nitrogen; in gas giants like Jupiter, it is believed to be molecular hydrogen. Analyses of the spectra of hot Jupiters often assume the atmosphere to be hydrogen dominated (Madhusudhan & Seager 2009). For rocky or terrestrial exoplanets with secondary atmospheres, the mean molecular weight cannot be assumed. First indications of the mean molecular weight can be obtained by measuring the primary transits at two visible wavelengths (Benneke & Seager 2012), which can be accomplished using the two broadbands of the fast cameras of PLATO. The method is complicated in the presence of clouds, but still provides first hints (strong/weak Rayleigh slope) on the nature of the atmosphere, to be followed on later with spectroscopic observations. If only one broadband measurement is made, then one may be able to distinguish between hypothesized atmospheres (e.g., hydrogen-dominated versus water-dominated models; de Mooij et al. 2012). Alternative methods include the detailed analysis of the line shape of a certain molecular species or the relative strength of its features at different wavelengths (Benneke & Seager 2012), but such an approach requires the line opacity list in question to be robust, which is not always the case. Visible data thus provides an important check on the analysis of infrared data of exoplanetary atmospheres. Identifying the dominant, inert molecule in an atmosphere has significant implications for inferring its thermal structure and spectrum, as the inert component often exerts an indirect influence on the spectroscopically active molecules via processes such as pressure broadening and collision-induced absorption.

Phase curves show the flux as a function of orbital phase, which may be deconvolved to obtain the flux versus longitude on the exoplanet, known as a "brightness map" (Cowan & Agol 2008). Infrared phase curves contain information about the efficiency of heat redistribution from the dayside to the nightside of an exoplanet (Showman & Guillot 2002; Cooper & Showman 2005; Showman et al. 2009; Cowan & Agol 2011; Heng et al. 2011), as previously demonstrated for hot Jupiters (e.g., Knutson et al. 2007, 2009). To a lesser extent, infrared phase curves constrain the atmospheric albedo and drag mechanisms (shocks, magnetic drag). By contrast, visible phase curves encode the reflectivity of the atmosphere versus longitude, which in turn constrains the relative abundance of clouds or hazes if they are present. The cloud/haze abundance depends on the size and mass density of the particles, as well as the local velocity, density, pressure and temperature of the flow, implying that a robust prediction of the cloud properties requires one to understand atmospheric chemistry and dynamics in tandem. Examples of exoplanets where clouds are likely to be present include Kepler-7b, which has a high albedo (~0.3) and a phase curve containing a surprising amount of structure (Demory et al. 2011). The feasibility of obtaining visible phase curves has already been demonstrated for the CoRoT (Alonso et al. 2009a,b; Snellen et al. 2009, 2010) and *Kepler* (Borucki et al. 2009; Batalha et al. 2011) missions.

A more ambitious goal is the use of the information from the phase light curve of the planet to constrain the temporal evolution of the temperature distribution of its upper atmosphere and set the first constraints on the dynamics of its atmosphere (e.g., Knutson et al. 2009). An interesting goal would be to establish the frequency of planets showing super-rotation on their atmospheres, a phenomenon which involves displacement of the hottest atmospheric spot of a tidally locked planet by an equatorial super-rotating jet stream (see Faigler et al. 2013 and references therein). PLATO 2.0 will provide bright targets for such investigations. High-accuracy photometry also allows for the measurement of the tidal distortion created by a transiting planet on its star (Welsh et al. 2010), which can provide a wealth of information about the star-planet interaction. Among the PLATO 2.0 detections will also be nearby giant planets on wide orbits for which transit spectroscopy and direct imaging spectroscopy will be possible. The comparison of these two approaches will then allow us to study the vertical structure of the planet atmosphere.

As the scientific community prepares for the launch of the JWST and also ground-based telescopes such as E-ELT, a central question to ask is: what are the best targets for follow-up, atmospheric spectroscopy of small exoplanets? Earth-like exoplanets with sizes of about 2 Earth radii are believed to be either composed predominantly of rock or scaled-down versions of Neptune with thick gaseous envelopes. If the bulk composition of an exoplanet cannot be made from a material lighter than water, then one can calculate the thickness of the atmosphere, relative to the measured radius, by utilising the mass-radius relation of pure water (Kipping et al. 2013). It was shown that such simple approach can be used to imply a mostly rocky composition (e.g., Earth, Kepler-36b; Kipping et al. 2013). By quantifying this metric for the entire PLATO 2.0 catalogue of small exoplanets, one can construct a valuable database of optimal follow-up targets. Knowledge of the fraction of small exoplanets with and without thick atmospheres, as a function of their other properties, provides a direct constraint on planet formation theories (see Section 2.2).

In summary, the key science questions that PLATO 2.0 can answer about the atmospheres of exoplanets are:

- What is the diversity of albedos present in exoplanetary atmospheres? How does the albedo correlate with the other properties of the exoplanet (incident flux, metallicity, etc)? Are these albedos associated with the presence of clouds or hazes?

- What are the dominant, inert molecules present in exoplanetary atmospheres? What are the mean molecular weights?

- When are clouds present in exoplanetary atmospheres? What is the diversity of the cloud properties (particle size, reflectivity, etc)?

- For small exoplanets (of about 2 Earth radii in size), what are the best targets for follow-up, atmospheric spectroscopy? Here, small planets at intermediate orbital separations are of particular interest.

## 2.11 Characterizing stellar-exoplanet environments

Transit observations of exoplanets around bright host stars together with advanced numerical modelling techniques and known astrophysical parameters, such as the host-star age and radiation environment, offer a unique tool for understanding the exoplanet upper atmosphere-magnetosphere interaction with the star. Hubble Space telescope (HST) UV transmission spectroscopy and Spitzer secondary eclipse measurements of known bright exoplanetary systems have been used to study a number of issues related to the upper atmospheres of planets including space weather events (Lammer et al. 2011a; Lecavelier des Etangs et al. 2012), to infer properties such as the thermospheric structure (e.g., Koskinen et al. 2012; Vidal-Madjar et al. 2011), the exosphere-magnetosphere-stellar plasma environment (Holmström et al. 2008; Ekenbäck et al. 2010; Llama et al. 2011), outflow of planetary gas including atomic hydrogen (Vidal-Madjar et al. 2003; Ben-Jaffel 2007; 2010; Ehrenreich et al. 2012, Haswell et al. 2012), and heavy species such as carbon, oxygen and metals (Vidal-Madjar et al. 2004; Linsky et al. 2010; Fossati et al. 2010, Haswell et a. 2012).

Moreover, the detection of transiting Earth-size or super-Earth-type exoplanets with PLATO 2.0 orbiting bright M-stars can be used as a proxy for early Solar System planets like young Venus, Earth and Mars, which faced a much harsher UV radiation environment than today, closer to that of active M dwarf stars (Lammer et al. 2011b; Lammer et al. 2012). PLATO 2.0 detections of such terrestrial planets will permit study of EUV heated and extended upper atmospheres around Earth-type exoplanets by UV follow up observations. These observations are essential for testing e.g., early terrestrial atmosphere evolution hypotheses (Erkaev et al. 2013; Kislyakova et al. 2013).

To really examine the complex physics of the interaction of close-in exoplanets with their host stars, bright nearby systems are required. PLATO2.0 will make a huge contribution here. UV transmission spectroscopy, particularly examining Lyman alpha is key to examing mass loss from hot Jupiers, but the faintness of the known sample has limited this work to HD 209458, HD 189733, and 55 Cancri. WASP-12 is a particularly interesting and extreme hot Jupier, but is too distant to be studied at Lyman alpha with HST. By using the near-UV where the host star is much brighter the mass loss from WASP-12b was detected (Fossati et al 2010, Haswell et al 2012) but without Lyman alpha data, quantitative comparison of the mass loss rate with the models is uncertain. PLATO 2.0 is expected to find WASP-12b analogues orbiting closer, brighter stars, and follow-up of these discoveries will produce a step-change in our ability to probe the processes governing the catastrophic end-point of hot Jupiter evolution. Near future space observatories such as the World Space Observatory-UV (WSO-UV) (Shustov et al. 2009; 2011; Gómez de Castro et al. 2011), or other future UV observing facilities will be able to take advantage of this PLATO 2.0 legacy.

PLATO 2.0 is also able to detect small planets, about 2-4 $R_{Earth}$, around A-stars. This means that the so-called mini Neptune's that are found in great numbers amongst the G-type stars today, could be detected with PLATO 2.0 for A-type stars. However, it could well be that mini Neptune's do not exist at small distances for these stars, if the XUV-radiation of intermediate-mass stars is strong enough to erode the gaseous envelope. The minimum distance at which mini Neptune's can exist therefore constrains the erosion of planetary atmospheres for such type of stars with extreme environment.

Among the surprising recent findings from the *Kepler* data is the existence of a number of extremely close-in rocky bodies orbiting their host stars at periods of less than a day. Kepler-78b is an Earth-sized planet in a 8.5 hour orbit (Sanchís-Ojeda et al 2013b); Kepler-42c is a sub-Earth sized planet in an 11 hour orbit (Muirhead et al 2012); and KIC 12557548b appears to be a disintegrating mercury-like object in a 16 hour orbit (Rappaport et al 2012). These objects are fascinating from an evolutionary point of view, and may be remnant cores of mass-losing hot Jupiters analogous to WASP-12. Alternatively, they may have been rocky bodies throughout their evolution. In either case these objects, in particular KIC 12557548b with its prodigious mass loss, give an unprecedented opportunity to study the composition of exo-rocks through transmission spectroscopy. The *Kepler* discoveries are distant and hence the signal-to-noise of any follow-up observations will limit the scope of the inferences we can draw from them. PLATO2.0 will find plentiful similar systems around nearby bright targets (see Section 6.3).

## *2.12 Detection of rings, moons, Trojans, exo-comets*

From modulations in the transit light curve planetary rings and large moons can also be detected (Sartoretti & Schneider 1999, Barnes & Fortney 2004, Ohta et al. 2009, Kipping 2009a,b, Tusnski & Valio 2011, Schneider et al. 2014). One of the main drivers of the search for moons is that they might share the orbits of Jupiter-sized planets in the habitable zone, and therefore be interesting targets for atmospheric characterization (Heller & Barnes 2012). There are well-developed projects searching for moons around transiting extrasolar planets in the *Kepler* mission, but so far the search has proven to be elusive (Kipping et al. 2012, Simon et al. 2012).

Moons produce two types of observable effects: photometric transits are superimposed on the planetary transits, and the timing and length of the transits of the host planet is perturbed. Unfortunately, for typical regular Solar System satellites, such as Ganymede around Jupiter, the amplitude of the timing perturbations is extremely small: in the order of several seconds. This is well below current detection limits. Furthermore, the photometric transit of a moon, when superimposed onto planetary transits, can be easily confused with the patterns produced by spot crossing (Silva-Valio & Lanza 2011; Sanchís-Ojeda et al. 2012) or instrumental systematics. On the other hand, moons are not thought to be stable for orbital periods of the host planet below 0.1 au (Namouni 2010). This means that we can only aim at finding moons around planets with large orbital periods, which reduces the number of transit events for a given length of the observations. The scarcity of the observations and the fact that the orbital phase of the moon is sampled at the orbital period of the planet, or below the Nyquist frequency of the moon's orbit, makes the characterization of these systems extremely challenging. Nevertheless, even if the exomoon orbital period cannot be inferred from TTVs, its radius can be measured for large moons, by the depth of its transit superposed to the planet transit. And for transiting moons, their atmosphere can be detected by further transit spectroscopy (Kipping et al. 2009c).

A more favorable scenario is the possibility of detecting binary planetary systems, or systems close to binary such as Pluto-Charon or the Earth-Moon system. In these cases, the combined signal of the planet and the moon is clearly distinguishable in the photometry and the TTVs can be much larger, up to some minutes in the case of the Moon orbiting the Earth. Such binary systems have not been found yet and determining their frequency remains a science case for PLATO.

Trojan-planets moving close to the Lagrange points L4 and L5 in 1:1 mean-motion resonance with planets are thought to be in very stable configurations, even if they reach the size of super-Earth planets. In our Solar System there are multiple examples of bodies in such orbits, albeit with sizes comparable with asteroids, so planetary objects in Trojan orbits would be a new class of system. PLATO 2.0 will have the precision to detect Trojan-planets as small as Earth. However, so far such systems have not been detected by any other survey (Ford & Holmann 2007, Cabrera 2010b and references therein).

Finally, exo-cometary tails lead to transit curves which can be as deep as Earth-sized planets, but with a different shape (Lecavelier des Etangs et al. 1999). Exo-cometary tails detected around nearby stars might be detectable by future direct imaging (Jura 2005). Exo-comets can be a source of interstellar comets transporting organics from one planetary system to another. Also, giant planets can develop cometary-like tails (Schneider et al. 1998). Indications for such tails have already been found in *Kepler* data (Budaj et al. 2013).

# 3 Science Goals II: Probing stellar structure and evolution by Asteroseismology

Asteroseismology is the study of the global oscillations of stars (see, e.g., Aerts et al. 2010 for a monograph on the subject). The frequencies of these oscillations, which can be either trapped acoustic waves (also called p modes) or internal trapped gravity waves (also called g modes) or a mixture of the two, depending on the radially varying density and internal sound speed of the star. Thus, measurements of oscillation frequencies can be used to infer both the internal structure of stars and their bulk properties (see Appendix A2 for details on the method). The precision of stellar bulk parameters determines the related parameter precision for its orbiting planets. Asteroseismology of planet host stars is therefore of key importance to derive accurate planet parameters (Gizon et al. 2013; Van Eylen et al. 2014). Furthermore, it is the only method that allows us to accurately date planetary systems for the first time. Asteroseismology is therefore related to the core science of PLATO 2.0, which will be the first mission to make systematic use of asteroseismology to characterize planet host stars due to its bright target sample.

Beyond the characterization of planet hosts, asteroseismology of stars with PLATO 2.0 will drastically improve our understanding of stellar evolution beyond what has been achieved with previous missions and significantly enhance current stellar evolution models. PLATO 2.0 will measure the oscillation frequencies of over 80,000 dwarf and subgiant stars with magnitudes less than 11. In total,1,000,000 stellar photometric light curves will be obtained for stars ≤13 mag over the course of the full mission. It will thus be a powerful new tool for the characterization and study of the evolution of star-planet systems.

## *3.1 Stellar parameters as key to exoplanet parameter accuracy*

The main focus of the asteroseismology programme of PLATO 2.0 will be to support exoplanet science by providing:

- Stellar masses with an accuracy of better than 10%,

- Stellar radii to 1-2%, and

- Ages to 10%

Gaia will provide the distances to the stars via direct, geometrical measurements, and hence the true absolute luminosity of the star can be derived with high accuracy. Combining the luminosity with the effective surface temperature of the star obtained from (ground-based) high-resolution spectroscopy, we will obtain the radius of the star with 1-2% accuracy. Also, luminosities from Gaia can be used in cases where $T_{eff}$ has not been measured. Notice that Gaia will be complete down to V~20 magnitude, while Plato will observe stars between V=4-16 magnitude, so all PLATO 2.0 targets will also be observed by Gaia (Perryman et al. 2001). In case Gaia data should not become available in future, the stellar radius can also be directly determined by using asteroseimic scaling relations together with the effective temperature (for technical details see Appendix 2.). As of Feb 2014 - Gaia has been succcessfully launched, and is currently commissioning, with the early indications being positive for a successful mission.

In the past, performing asteroseismology was far from straightforward, i.e., even for stars very similar to our Sun. However, real breakthroughs in asteroseismology have recently been achieved through the space missions MOST, CoRoT and *Kepler*. Asteroseismology will provide the mean density of the star, e.g., via the scaling relationships or inversion techniques as outlined in the Appendix. These scaling relations based on solar values have already been tested and validated on *Kepler* targets by comparing the asteroseismic radii and distances with interferometric observations and Hipparcos parallaxes (Huber et al. 2012; Silva Aguirre et al. 2012). By combining the very precise mean density values of PLATO 2.0's asteroseismic analysis and the stellar radii from Gaia we will obtain accurate stellar masses.

The asteroseismic age-determination is more complex and requires invoking models of stellar evolution. Age estimates will be made by comparing grids of stellar models computed for different initial parameters (mass, metallicities, helium abundances, convection parameters) to the combined non-asteroseismic and asteroseismic observational constraints as outlined in the Appendix. The models will themselves be improved using the asteroseismology of PLATO 2.0 observations (Section 3.2). Several publications in recent years have shown that ages can indeed be determined from asteroseismology with high precision, e.g. Metcalfe et al. 2009, 2010 and recently for 22 *Kepler* targets by Mathur et al. 2012 and the bright stars 16 Cyg A and B by Metcalfe et al. 2012. These examples show that even higher precisions than 10% can be achieved.

Other examples come from the CoRoT satellite which has observed several solar-type stars in its asteroseismic programme. One of the cool stars observed is the G0V type star (*$m_V$ = 6.3*) HD52265, a planet hosting star, which was observed with CoRoT for 117 days (Ballot et al. 2011). About 31 oscillation modes were present with sufficient S/N in the power spectrum of the light curve (Figure 3.1). A grid of stellar models was computed (Escobar et al. 2012) and further analysis of convection and rotation performed (Lebreton & Goupil, 2012; Gizon et al. 2013). A seismic radius of 1.34 ± 0.02 $R_{Sun}$ and a seismic mass of 1.27 ± 0.03 $M_{Sun}$ were derived. The age was determined as 2.37 ± 0.29 Gyr. More solar-type stars have been observed by CoRoT in this fashion and several of them are known to have (large) planets.

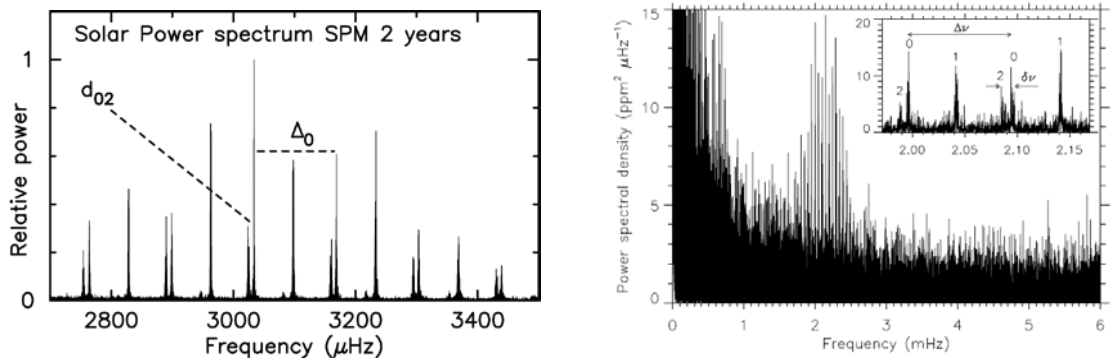

**Figure 3.1** *Left:* Solar power spectrum from 2 years of SPM photometric data. *Right:* Power spectrum of HD 52265 from 117 days of observation with CoRoT (Gizon et al. 2013; ESA/SRE(2009)4).

The *Kepler* mission allows for asteroseismology down to about at least $m_V = 12$. *Kepler* has carried out asteroseismic observations on a large number of stars, including many with transiting planets. Recent examples of such planetary systems containing icy/rocky planets are Kepler-36b (Carter et al., (2012), Kepler-68 (Gilliland et al. 2013) and the smallest planet detected so far, Kepler-37b (Barclay et al. 2013). Based on the asteroseismic analysis of 66 *Kepler* planet host stars, Huber et al. (2013a) claim typical uncertainties of 3% and 7% in radius and mass, respectively, from the analysis of global asteroseismic parameters. PLATO 2.0 will provide similar performances in thousands of stars with planets.

CoRoT and *Kepler* results clearly demonstrate the feasibility of achieving highly accurate star and planet parameters. It should be noted that measurements of effective temperature to within 1% will be achievable through dedicated high-resolution, high signal-to-noise spectroscopic observations obtained as part of the ground-based follow-up program. The determination of the chemical abundances and effective temperature will be based on state-of-the-art techniques and model atmospheres taking 3D and non-LTE effects into account (Bergemann et al. 2012; Magic et al. 2013). Taken together with the luminosities expected from ESA's Gaia mission the effective temperature will lead to stellar radii with a relative precision within 2% for un-reddened stars, as is the case for most of the PLATO 2.0 targets, as illustrated above.

With the ultra-precise, long-term photometry of PLATO 2.0, a number of possibilities open up to measure the logg of the host stars with unprecedented accuracy. These methods include the asteroseismic determination, by acquiring the density as a function of depth from the detection of p-modes, and integrating this to the stellar radii R (as determined by Gaia), and, through calibration of the semi-empirical correlation between the observed time-dependent granulation variations and surface gravity (also only possible through PLATO 2.0's accurate, time-resolved photometry). Comparing the results of these measurements with the exquisite high-resolution spectroscopic follow-up measurements (Section 7) will also improve 'classical' stellar modeling (also through the very precise determination of abundances and stellar surface rotation velocities).

Finally, the frequency analysis can also provide information about stellar interior rotation (e.g., Beck et al. 2012; Deheuvels et al. 2012). Furthermore, the relative amplitudes of the split components depend on the inclination of the rotation axis relative to the line of sight and hence may reveal a possible misalignment between the stellar equator and the orbital planes of transiting exoplanets (Chaplin et al. 2013; Huber et al. 2013b).

## 3.2 Stellar models and evolution

With sufficiently good data the asteroseismic determination of mass and radius is essentially independent of stellar models. For other quantities, particularly the age, the inferences involve fitting models to the observables and their accuracy depends on our ability to model stellar evolution. Thus the asteroseismic investigation of stellar structure and evolution is an essential part of the characterization of planet hosts and to put the discovered planetary systems into an evolutionary context. Asteroseismic investigation of a large number of stars of various masses and ages is a necessary tool to constrain models of stellar interiors, identify missing physics, and thereby improve our understanding of stellar evolution.

One of the main sources of uncertainty affecting age determination is the presence and efficiency of transport mechanisms in radiative zones (Zahn 1992, Maeder 2009). While these mechanisms can have a significant impact on the main-sequence lifetime, they are still poorly understood and crudely modelled.

Rotational mixing is one of such processes that is not yet well understood. Angular momentum and chemical elements can be transported in the radiative zones of rotating stars through meridional circulation and hydrodynamical instabilities. This results in a change of the global and asteroseismic properties of stars when rotational effects are taken into account, and in particular to an increase of the main-sequence lifetime due to the transport of fresh hydrogen fuel in the stellar core (e.g. Eggenberger et al. 2010). These changes depend on the poorly-known efficiency of rotational mixing, which can be constrained by obtaining information about the internal rotation profiles in stellar radiative zones. Radial differential rotation can be inferred by asteroseismology for stars that have mixed modes (e.g., Suarez et al. 2006). These modes have a g-mode character in the core and a p-mode character in the envelope. They are therefore sensitive to the core, while having amplitudes large enough to be detected at the surface. Mixed modes are present in subgiant and red-giant stars (e.g., Beck et al. 2011), and differential rotation has already been detected using *Kepler* data (e.g., (e.g., Beck et al. 2012, Deheuvels et al. 2012).

A thorough investigation of stellar evolution requires a large number of stars which sample all relevant stellar parameters (mass, age, rotation, chemical composition, environment…). The PLATO 2.0 mission will, for the first time, provide such necessary data in order to:

- Improve understanding of internal stellar structure, including the identification of missing physics.

- Better understand the pulsation content and its interaction with the physics of the star, in particular with respect to rotation.

- Improve our understanding of stellar evolution.

# 4 Science Goals III: Complementary and Legacy Science

In addition to its focus on relatively bright stars, one major and crucial advantage of PLATO 2.0 over the CoRoT and *Kepler* space missions is its ability to observe in many directions of the sky. This will enable us to sample a much wider variety of time-variable phenomena in various populations of the Galaxy than hitherto. Moreover, PLATO 2.0's asteroseismic characterization of stellar ensembles, binaries, clusters and populations will be a significant addition to the Gaia data for about 50% or more of the sky. This capability will obviously give rise to a very rich legacy for stellar and galactic physics, promising major breakthroughs in a variety of subjects, some of which are discussed in this section.

## *4.1 Stellar structure and evolution*

### 4.1.1 Low- and intermediate mass red giants

Red giants are an important source of information for testing stellar models. An important legacy from the CoRoT and *Kepler* missions has been the discovery of solar-like oscillations in thousands of G-K giants (De Ridder et al. 2009; Bedding et al. 2010; Hekker et al. 2011). The occurrence of non-radial modes was only unambiguously proven from CoRoT observations (De Ridder et al. 2009). This opened up the field of asteroseismology of low-mass evolved stars.

Thanks to the discovery of gravity-dominated mixed modes from more than 300 days of continuous *Kepler* data of red giants (Beck et al. 2011, Bedding et al. 2011), the promise of asteroseismology being able to discriminate between different nuclear burning phases was delivered. Indeed, the oscillation period spacings of dipole mixed modes probe the properties of the core structure of red giants and reveal if they are already in the helium core burning stage or are still climbing up the red giant branch while burning hydrogen in a shell, despite having the same position in the HRD (Bedding et al. 2011, Mosser et al. 2012). PLATO 2.0 will be able to separate these two kinds of stars, because they have different positions in the frequency spacing-diagrams. Additionally, the capability of performing seismology analysis in red giants can constrain mixing processes in main sequence stars. Red giants in the transition between low and intermediate mass (2-2.5 $M_{Sun}$) will provide information on the extension of the central mixing during their main sequence phase, allowing the study of transport processes in mass range and evolution phase where solar like oscillations are not expected. On the other hand the seismology of low-mass red giants will constrain the extension of the central mixed region during the central He burning phase (Montalban et al. 2013). The description of transport processes as well as the size of the mixed region are matter of strong debate, and have important consequences on stellar population studies (e.g. Straniero et al. 2003; Chiosi 2007). Finally, PLATO 2.0 will also improve the period-luminosity relationships of these kinds of bright objects, which helps to use them as galactic or even extragalactic distance-indicators with higher precision than than currently the case. PLATO 2.0 will improve our understanding of the internal structure of red-giant stars by providing accurate oscillation frequencies for an unprecedented number of targets in different directions of the galaxy.

### 4.1.2 Hot B subdwarf (sdB) stars

Hot B subdwarfs are core He-burning stars with an extremely thin H-rich envelope (Heber 2009). They exhibit pulsation instabilities driving both acoustic modes of a few minutes and gravity modes with 1-4h periods. While the asteroseismic exploitation of the p-mode pulsators started a decade ago (Brassard et al. 2001), it is only recently, with CoRoT and *Kepler*, that data of sufficiently high quality could be obtained for the g mode sdB pulsators (Charpinet et al. 2010; Østensen et al. 2010). Asteroseismic modelling of sdB stars provides measurements of their global parameters such as the mass and radius with a precision of typically 1% (Van Grootel et al. 2013). The mass distribution of sdB stars (Fontaine et al. 2012), is consistent with the idea that sdB stars are post-RGB stars that went through the He-flash and that have lost most of their envelope through binary interaction. While about half of sdB stars reside in binaries with a stellar companion, the recent discoveries of planets around single sdB stars (Charpinet et al. 2011; see Section 2.6) also support the idea that planets could influence the evolution of their host star, by triggering the mass loss necessary for the formation of an sdB star.

PLATO 2.0 will be the only space-based facility permitting to develop further the deep seismic probing of sdB stars and all the related outcomes briefly mentioned previously. It will provide the high-quality data on g mode pulsations in these stars that cannot be obtained from the ground, as well as very high precision data on p mode pulsations. Thereby, PLATO 2.0 will increase the number of sdB stars that can be modeled by asteroseismology. It will also discover new planets around sdB stars, permitting to disentangle the question of the origin of such stars and explore star-planet interactions in the advanced stages of stellar evolution.

### 4.1.3 White dwarfs (WDs)

White dwarfs are the endpoint of the evolution of the vast majority (~95%) of stars in the Universe. They no longer undergo fusion reactions but gradually evolve along the cooling sequence, where several classes of g modes pulsators allow asteroseismic probing of the final stages of stellar evolution (Fontaine & Brassard 2008). Firstly, white dwarfs can be used to constrain the ages of the various populations of evolved stars in the Galaxy, a field called white dwarf cosmochronology (Fontaine et al. 2001, Liebert et al. 2013). The cooling tracks are very sensitive to the exact core composition and envelope layering, two parameters that are inaccessible from direct observations and poorly constrained from theory, but that can be determined from asteroseismology (Giammichele et al. 2013). White dwarf cosmochronology is currently of high interest, especially in the coming era of Gaia to add accurate age estimates to the 3D mapping of the Galaxy. Secondly, internal dynamics can also be probed by asteroseismology, allowing the study of the rotation and angular momentum evolution to the white dwarf stage (Charpinet et al. 2009). Finally, "exotic" physics due to the extreme compact nature of white dwarfs can be calibrated: neutrino production rates, conductive opacities, interior liquid/solid equations of state, crystallization physics at the end of the cooling of white dwarfs. White dwarfs may also become interesting targets for planet search campaigns (Algol et al. 2011).

Simulations show that, assuming V ≤ 16, ~10 pulsating WDs should be observable with a sufficient quality in the long-monitoring fields and ~50 pulsating WDs in the step-and-stare fields. These numbers are in good agreement with the three WD pulsators discovered in the *Kepler* field, while none have been observed by CoRoT.

PLATO 2.0 will be the very first mission to bring WD seismology in the space era, allowing for significant improvements in the asteroseismic probing of the final stages of stellar evolution.

### 4.1.4 Massive stars

Despite their scarcity compared to low-mass stars, stars massive enough to end their lives in core-collapse supernovae dominate the chemical enrichment of galaxies and the Universe as a whole. Most of the heavy elements (by mass fraction) are created by stars with birth masses above about 9 $M_{Sun}$. For such stars, the effects of internal rotational mixing remain largely uncertain, despite being crucial to predict their evolution as blue supergiants. Interestingly, gravity-mode oscillations have been discovered in such evolved massive supergiants (e.g., Saio et al. 2006, Lefever et al. 2007). Such modes hold similar potential to probe the stellar core as the gravity-dominated mixed modes found in red giants (Moravveji et al. 2012a). PLATO 2.0 can provide a homogeneous sample of blue supergiants studied by asteroseismology with a broad range of pulsation periods.

From five months of CoRoT data, Degroote et al. (2010) measured a periodic deviation of amplitude superimposed on a constant period spacing for the high-order gravity modes of a star of about 8 $M_{Sun}$. This allowed deducing that this star passed 60 % of its core-hydrogen burning lifetime. This allowed as well as the determination of the detailed shape of the near-core chemical composition gradient. The only missing ingredient to apply the same type of diagnostic to blue supergiant pulsators is a large number of suitable high-precision uninterrupted light curves of such stars, as can be provided by PLATO 2.0.

### 4.1.5 Probing angular momentum transport using gravity modes

A major missing input for stellar models, as opposed to the solar model, is a measurement of the internal differential rotation as a function of evolutionary stage. Such a measurement is necessary to estimate the amount of rotational mixing and angular momentum transport, which are crucial aspects for the outcome of stellar evolution but which remain essentially unconstrained by experiment so far. It required two years of continuous *Kepler* data to make the first steps towards such input for evolved low-mass stars (Beck et al. 2012; Mosser et al. 2012; Deheuvels et al. 2012). These first results imply that an important angular momentum coupling between the core and the envelope of evolved stars is missing in current models (Eggenberger et al. 2012, Marques et al. 2013, Ceillier et al. 2013). An unknown physical process which transports angular momentum much more efficiently than hitherto assumed during the stellar life is clearly needed. Very recently, the internal gravity waves (IGW) were proposed to leave observable surface light fluctuations at a level of hundreds of micromagnitudes by Shiode et al. (2013) and Rogers et al. (2013). Rogers et al. show that IGWs are very efficient in transporting angular momentum in stars and, in particular, can be responsible for spinning up or/and slowing down their outer layers. The authors suggest that IGW angular momentum transport may explain many observational mysteries, such as: the misalignment of hot Jupiters around hot stars, the Be class of stars, nitrogen enrichment anomalies in massive stars, and the non-synchronous orbits of interacting binaries. Thanks to its high-precision photometric data and long time-base observations, PLATO 2.0 can observationally explore the theory of excited IGWs. **PLATO 2.0 has the potential to characterize this major missing ingredient in stellar evolution theory, by deriving internal rotational profiles from inversion of rotationally split oscillation frequencies for a carefully selected sample of target stars covering entire evolutionary paths.**

### 4.1.6 Early stellar evolution – the pre-main sequence phase

During the stars' early evolution from their births in molecular clouds to the onset of hydrogen core burning, complex physical processes are acting which challenge current theory and observing techniques. Young stars and their photospheres are directly connected to the moving hydrodynamic circumstellar material that is still being accreted. In the gas and dust disks surrounding the protostars, it is assumed that planetary systems – similar to our Solar System - are formed. All these phenomena make young stars interesting objects that allow us to investigate, among other things, how our Sun formed and evolved to its present state and to study how stellar evolution depends on the initial conditions. Due to its large sky access, PLATO 2.0 will be able to target pre-main sequence stars, to study their different types of variability and to reveal groundbreaking insights into early stellar evolution.

## *4.2 Asteroseismology of globular and young open clusters*

Testing stellar evolution theory through asteroseismology will be most successful if applied to the extremes of evolutionary stages within a cluster.This should include both young open clusters with (pre-)main-sequence and pre-supernova supergiant pulsators on the one hand, and  old globular clusters of various metallicities that contain main-sequence, horizontal branch, and white dwarf stars.

Current asteroseismic studies involved, for example, the study of solar-like oscillations of the red giant members (Stello et al. 2011, Hekker et al. 2012) and led to the first seismic cluster constraints on age, metallicity, and mass-loss rates on the red giant branch (Basu et al. 2011, Miglio et al. 2012a,

Corsaro et al. 2012). Unfortunately, only clusters in a relatively narrow range of ages, from 0.4 Gyr for the youngest to ~8 Gyr for the oldest, were studied.

Due to the pointing restrictions of *Kepler* and CoRoT, no young clusters (i.e., with ages younger than a few tens of million years) can be observed by *Kepler*, and only one young cluster, NGC 2264, could be observed in two short runs by CoRoT. Recent asteroseismic results from the NGC 2264 observations include, e.g., the discovery of the first two pre-main sequence ϒ Doradus pulsators (Zwintz et al. 2013) and a homogeneous study of the relation between pulsations and stellar evolution from the early stages to the main sequence phase (Zwintz et al. 2013).

PLATO 2.0 will lead to major breakthroughs in this area, thanks to its large-sky accessibility and its step-and-stare phase. No other astronomical experiment with the capability to investigate stellar evolution at the level of full cluster asteroseismology is presently on the horizon.

## *4.3 Probing the structure and evolution of the Milky Way*

The chemical enrichment of the Universe is one of the main thrusts of modern astrophysics and the Milky Way (MW) can be seen as the Rosetta stone of this evolution. The origin and evolution of the MW is encoded in the motion and chemical composition of stars of different ages. In particular, the MW halo contains the oldest and most metal-poor stars observable, which were born at times, or equivalently redshifts, still out of reach for the deepest surveys of primordial galaxies. These stars retain the memory of the unique nucleosynthesis in the First Stars, as revealed by their striking abundance patterns observed at very low metallicities (Chiappini et al. 2006). A serious obstacle to discriminate between different scenarios of formation and evolution of the Galaxy components (halo, thin and thick disk and bulge) is the difficulty of measuring distances and more importantly ages for individual field stars. Crucial ingredients to study evolutionary processes in the disk are, e.g., the age-metallicity and age-velocity dispersion relations for different directions and at different galactic radii and heights from the plane.

Even if not completely free from stellar modeling, the mass of a red giant star, given its evolution rate, is a good proxy of its age. In addition, oscillation spectra also allow one to distinguish between H-shell burning and central He-burning phases (Bedding et al. 2011, Mosser et al. 2011). So, once the chemical composition is known, asteroseismology can provide stellar ages within a 15% uncertainty, while classical methods such as isochrones may be uncertain by a factor two. Using seismic data from CoRoT and *Kepler*, Miglio et al. (2012b, 2013) showed that pulsating red giants can efficiently be used to map and date the Galactic disc in the regions probed by the observations (Figure 4.1). Note that given the high intrinsic luminosity of red giants compared to dwarfs, these data allow us to see quite far in the Galaxy, up to about 10kpc, whereas Hipparcos precise parallaxes are available only up to 100pc. The capability of seismic data to derive individual stellar ages indicates a clear vertical gradient in the ages of disc red giants. These results show the enormous potential of red giant seismology with PLATO 2.0, which will not be limited to pencil-beam surveys as is the case for CoRoT and *Kepler*.

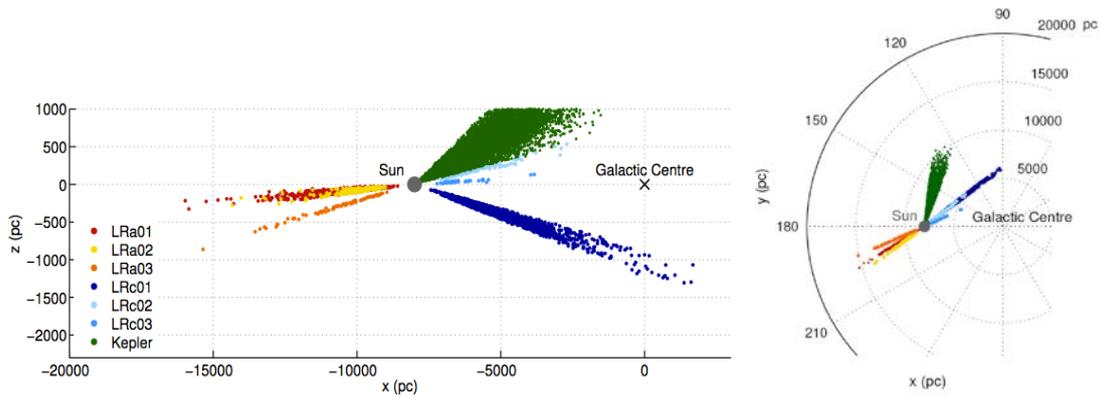

**Figure 4.1** Distribution on the galactic plane of the red giants with asteroseismic characterization from the light curves obtained in the CoRoT exofield for six long runs: (LRa01 (red), LRa02 (yellow), LRa03 (orange), LRc01 (blue), LRc02 (green), LRc03 (cyan)). Reproduced from Miglio et al. (2011)

The European Gaia satellite will create a 3-D map of stars throughout our Galaxy, hence providing an observational test bench to theoretical predictions on the origin, structure and evolutionary history of our Galaxy. Additional crucial information both, on velocities and chemical abundances, will come even earlier (already in 2011-2012) from several ongoing/planed spectroscopic surveys such as SEGUE-2, APOGEE and the Gaia-ESO surveys. The combination of chemical compositions from spectroscopic surveys with distances from Gaia and ages from seismic data as provided by PLATO 2.0 for large samples of stars will allow us to comprehensively study chemical gradients and their time evolution in different directions. It will provide information on the metallicity distribution of thick and thin disk stars at different positions in the galaxy, and their time evolution. In addition, the evolution of the stellar velocity dispersions in the disk can be studied. All of these crucial constraints will allow us to quantify the importance of stellar radial migration in the formation of the MW, otherwise difficult to quantify from first principles. This will represent invaluable information not only for the formation of the MW, but also for the formation of spiral galaxies in general.

## 4.4 Stellar activity

Starspots dim the star when they transit across the surface, allowing a determination of the surface rotation rate and even the surface differential rotation. While the fixed pointing of CoRoT and *Kepler* limited their stellar diversity, PLATO 2.0 will give a full picture of the evolution of angular momentum loss among different populations of stars. One important application will be the calibration of gyrochronology (e.g., Barnes 2007). Gyrochronology is a strict age-rotation relation which is calibrated using precise ages and rotation periods. PLATO 2.0 will be able to provide both: precise ages from asteroseismology and rotation periods from the analysis of light curves.

Magnetohydrodynamics (MHD)-processes in stars have many open issues, e.g., length of activity cycles, location of spots, mass flows around the umbra of spots (Zhao et al. 2001, Kosovichev 2002, Strassmeier 2009), or the structure and evolution of magnetic flux tubes (Parker 1979, Rempel et al. 2009). Some of these issues can be constrained by recording the decay time of starspots, because decay time constrains the magnetic diffusivity, a key factor of the models. CoRoT already provided evidence of a magnetic activity cycle (Garcia et al. 2010) and constraints on stellar dynamo models under conditions different from those of the Sun (e.g., Mathur et al. 2013). This study was extended in terms of number of stars and in precision by Chaplin et al. (2011) who used *Kepler* observations of a sample of solar-type pulsators and found a strong correlation between the strength of the activity and the level of inhibition of stochastically excited solar-like oscillations. Stellar activity has also been used as another indicator of stellar age (Mamajek & Hillenbrand 2008). PLATO 2.0's long time series will furthermore allow the study of magnetic activity cycles and spot decay time for various types of stars. The magnetism has a great impact on the evolution of the stars, and it is poorly known on solar-type stars. The observation of starspots evolution, number, activity cycle and distribution of spots on the surface will allow to better understand the physics at the origin of activity phenomena, and to give constraints to the dynamo theories.

Starspots cannot be observed with the same detail as sunspots and consequently only the extreme types of starspots are known. However, the occultation of starspots by transiting planets has revealed spot sizes similar to those seen on the Sun (e.g., Silva 2003). The occultation of starspots by transiting planets produces anomalies in the transit light curves that may lead to an inaccurate estimation of the transit duration, depth, and timing (Czesla et al. 2009, Oshagh et al. 2013). These inaccuracies can for instance affect the precise derivation of the planet radius, and consequently affect the planet density estimation. Thus, having an estimation on the size and position of starspots would help to overcome this issue when determining the planet parameters. Furthermore, repeated starspot occultations can reveal the stellar rotation period (Silva-Valio 2008) and even differential rotation (Silva-Valio & Lanza 2011) as well as constrain the angle between the spin axis of the star and the orbit of its planets, i.e., the stellar obliquity (e.g., Sanchís-Ojeda et al. 2013a).

While *Kepler* has provided long time series of a large sample of active cool stars (e.g., Wells et al. 2013), including exoplanet hosts (e.g., Bonomo & Lanza 2012), most of its targets are too faint to combine the white-light space photometry with ground-based, ultra-fast, high-resolution Doppler tomography. This will be feasible, however, for a large sample of PLATO 2.0 targets, given the mission's normal cadence of 25 seconds and its relatively bright targets compared to CoRoT and *Kepler*. This combination will allow us to investigate the time-dependence of activity phenomena, differential surface rotation and activity cycles in single and binary stars over a wide range of mass and convection zone depth. The future high-resolution, ultra-stable ESPRESSO spectrograph to be installed at the VLT (planned for 2016) is ideally suited to do such time-resolved spectroscopy for the most interesting part of the PLATO 2.0 active star sample.

PLATO observations of sun-like stars will help to understand long-term changes in activity and brightness of the Sun and sun-like stars. Solar irradiance has been accurately measured for almost 40 years, and its cyclic variability is well established. However, our understanding of irradiance changes during exceptionally low activity periods, like the Maunder minimum, is very limited. In particular, the magnitude of the longer-term (centennial and longer) changes is heavily debated and remains one of the critical factors thwarting reliable assessment of solar influence on Earth's climate (Solanki et al. 2013). The assets of PLATO in this hunt are (i) the very large observed stellar sample, which will be necessary to get stars in such rare activity states, and (ii) the seismic constraints on the stars, which will bring new and robust physical parameters (age, mass, surface rotation, internal rotation) to finally identify stars that clearly deviate from the general activity trends and the physical ingredients that control this temporary deviation.

In addition, PLATO 2.0 will allow us to perform stellar coronal seismology, which was so far mainly restricted to the solar corona. STEREO and SDO detected the so-called transfer loop oscillations in the solar corona (periods are 2-20 minutes and damping times roughly twice that long) which were interpreted in terms of MHD theory (e.g., West et al. 2011, Goossens et al. 2012). Similar oscillations have been detected at optical wavelengths and in X-ray during flaring, with periods of seconds to minutes (e.g., Contadakis et al. 2010, 2012). PLATO 2.0 will be capable to perform coronal and flare seismology, particularly via its fast telescope performance with a cadence of 2.5 seconds. Following coronal seismology of the Solar Orbiter mission (to be launched in 2017), PLATO 2.0's detections of stellar coronal oscillations will allow us to understand the Sun's corona as part of the stellar population, by deducing local plasma properties outside the solar regime. This is important to unravel and understand the overall coronal heating mechanism across the entire spectral range of types A to M.

In summary, the study of stellar activity by PLATO 2.0 will allow to:

- calibrate the stellar age-rotation relationship (gyrochronology)
- study magnetic activity cycles and constrain stellar dynamo models
- perform stellar coronal seismology

## 4.5 Accretion physics near compact objects

While the *Kepler* mission is currently producing an extensive legacy in the area of eclipsing binaries (e.g., Prša et al. 2011, Slawson et al. 2011, Matijevič et al. 2012), including the detection and characterization of circumbinary planets (Doyle et al. 2011, Winn et al. 2011, Welsh et al. 2012), ultra-short periodic phenomena in binaries are hard to catch due to the 59 seconds sampling cadence and the single pointing of the telescope. Accretion phenomena in compact binaries, such as cataclysmic variables (CVs) or X-ray binaries (XRBs), display variability on a range of timescales, involving both the orbital and spin periods of the components. In such systems, the secondary transfers material to the primary, which is either a white dwarf, a neutron star or a black hole. XRBs show variability due to accretion ranging from milliseconds to hours, while the time scales for CVs are in the range from minutes to days. PLATO 2.0's all-sky accessibility, optical photometry and cadence of 25 and 2.5 seconds for normal and fast telescopes, respectively, is well suited to shed new light on the physical processes involved in disc accretion of compact objects, by studying a sample of carefully selected optically bright CVs and XRBs.

Importantly, phase- and time-lags of a few % in radians and 2 to 15 seconds, respectively, have recently been discovered in fast multi-colour optical photometry with ULTRACAM for the two CVs MV Lyr and LU Cam (Scaringi et al. 2013). Similar lags have also been observed for X-rays compared with optical measurements for XRBs and in Active Galactic Nuclei (AGN). PLATO 2.0's fast telescopes hold the potential to unravel the physical origin of these lags by studying a carefully chosen modest sample of bright CVs in two colours.

## 4.6 Classical pulsators

PLATO 2.0 will obtain high-precision photometric light curves for classical pulsators, such as β Cep stars, slowly pulsating B stars (SPBs), δ Sct stars, ϒ Dor stars, as well as distance indicators such as RR Lyrae stars, high-amplitude δ Sct stars, and Cepheids (e.g., Chapter 2 of Aerts et al. 2010). The mean densities of those stars can now be determined with a precision of about 6% for stars observed by CoRoT (e.g., Suárez, J. C. et al. 2014, García Hernández et al. 2009).

CoRoT and *Kepler* photometry have revolutionized our knowledge of such classical pulsators through several new discoveries, e.g., (i) observed regularities in the frequency spectra (Breger et al. 2011, Zwintz et al. 2013) and the existence of relationships between low and high frequencies in δ Sct stars (Breger et al. 2012), (ii) the high fraction of 23% δ Sct – ϒ Dor hybrid pulsators among the A and F type stars (Uytterhoeven et al. 2011, Tkachenko et al. 2013), (iii) the dense frequency spectra of δ Sct stars below 5 μHz (Poretti et al. 2009, Mantegazza et al. 2012), and (iv) SPB type g mode period spacings and p mode frequency spacings in OB type pulsators (Degroote et al. 2010, 2012). Moreover, B type stars were shown to exhibit a much larger diversity in their variability than expected before from ground-based observations. CoRoT and *Kepler* photometry combined with high-resolution ground-based spectroscopy revealed, e.g., stars with spotted (or at least inhomogeneous) surface configurations (e.g., Degroote et al. 2011, Pápics et al. 2012) pointing towards the presence of a magnetic field (e.g., Briquet et al. 2013), pulsating stars outside and constant stars inside the theoretical instability strips (e.g., Briquet et al. 2011, Pápics et al. 2011) calculated with current stellar evolution models and oscillation codes, stars exhibiting gravito-inertial modes (Pápics et al. 2012, Thoul et al. 2013), and pulsations driven by more rare excitation mechanisms, such as tidal excitation and non-linear resonant excitation (Pápics et al. 2013) in addition to the ε-mechanism in blue supergiants (Moravveji et al. 2012b).

The precision of the PLATO 2.0 data and the expected number of β Cep stars may also be key to understand their pulsational properties by the analysis of the splitting asymmetries, as well as the internal rotation profile. For those stars the convective core was found to rotate faster than the surface (Aerts et al. 2003, Dziembowski & Pamyatnykh 2008, Suárez et al. 2009) but the number of studied stars is at present too limited to make general conclusions to improve stellar evolution theory. Besides making significant progress in all these areas, the large number of observable Cepheid and RR Lyrae targets and the precise space photometry of PLATO 2.0 will facilitate the investigation of the Blazhko-effect (Kolenberg et al. 2011) to test the current theories (e.g., Gillet 2013), its occurrence rate and phenomenology (Le Borgne et al. 2012), the excitation of nonradial modes (Poretti et al. 2010) and other light curve variations, the stability of pulsation periods (Derekas et al. 2012), stellar evolutionary effects and nonlinear dynamics (e.g., Molnar et al. 2012; Kollath et al. 2011, Szabó et al. 2010) as well as its appearance in the light curves of high-amplitude δ Sct stars (Poretti et al. 2011). The availability of accurate asteroseismological measurements and radial mode pulsational period estimates, combined with a detailed evolutionary framework could be of pivotal importance in order to shed light on the well-known discrepancy between theory and observations about the pulsational period change rate: observed period change rates are an order of magnitude larger than those predicted by Horizontal Branch models (Kunder et al. 2011 and references therein).

The study of Cepheids and RR Lyrae stars will benefit greatly from the large number of PLATO 2.0 targets. A rough estimate gives 550 (730) Cepheids (of both classical and Type II) down to 13th (15th) magnitude compared to about a half dozen observed with CoRoT, and only one well documented case in the *Kepler* field (Szabó et al. 2011). The improvement is similarly large for RR Lyrae stars: the current design and observing strategy of PLATO 2.0 promises the observation of at least 800 (3600) of such stars down to the 13th (15th) magnitude limit, as opposed to ~30 and ~50 found in CoRoT and *Kepler* fields respectively. These calculations used the GCVS catalogue (Samus et al. 2012) and neglected the results of recent all-sky surveys, and therefore these numbers should be regarded as lower limits.

In summary, PLATO 2.0 will reveal significantly more features of classical pulsators that will lead to a better understanding of the underlying physical processes and their influences on stellar evolution.

### *4.7 Classical eclipsing binaries, beaming binaries and low-mass stellar and substellar companions*

PLATO 2.0 will provide the opportunity to significantly increase the samples of binaries and sub-stellar companions studied in the following areas:
- Classical eclipsing binaries allow us to measure the masses of the components via Kepler's third law in a model-independent way, when high quality photometry and a radial velocity curve of the two-lined binaries are available. However, at present good quality mass, radius and luminosity data for such systems are available only for about 100 such systems (Torres et al. 2010).
- Low-mass stellar companions can be detected via the so-called beaming effect. This relativistic effect causes a small light curve modulation with the period of the orbital period of the companion, and allows us to determine the companion mass without radial velocity measurements (e.g., Zucker et al. 2007, Faigler et al. 2013, Mazeh et al. 2010).
- The gravity darkening effect can be used to probe the internal heat-distribution of stars via radial and meridional circulations (Rafert & Twigg 1980).

- Observations of contact binaries will permit the studies of the formation process, internal structure, acitvity and especially the final evolutionary stage of binary systems (Eggleton 2012; Csizmadia et al. 2004; Tran et al. 2013).

## *4.8 Additional complementary science themes*

Apart from the above short, non-exhaustive list of themes in stellar and galactic physics that PLATO 2.0 will address, various additional subjects are within reach. Examples from stellar physics are common-envelope and Roche-Lobe overflow evolution of close binaries, tidal asteroseismology, mass-loss and structure of stars rotating at critical velocity. In some favorable cases, PLATO 2.0 could observe the microlensing amplification of massive objects eclipsing bright companions (Maeder 1973, Muirhead et al. 2013). On top of this, PLATO 2.0 can address a number of science topics in different areas of planetary, stellar and galactic physics. Topics discussed for further investigations using PLATO 2.0 include phenomena such as super-novae, GRBs, and even microlensing searches for black holes (Griest et al. 2013), as well as Kuiper-belt and Oort clouds objects in our Solar System.

## *4.9 PLATO 2.0's long-term legacy*

The PLATO 2.0 catalogue of thousands of characterized planets and of about 1,000,000 stellar light curves will provide the basis for a huge long-lasting legacy programme for the science community. Planets, around bright stars, detected and characterized by PLATO 2.0 will be a rich input catalogue for spectroscopic studies to investigate their atmospheres and link them with the planetary bulk properties. Observing further transits of large planets around suitably bright objects from the ground over long periods, well beyond the mission lifetime, will allow searching for planets or exomoons by TTVs and Transit Duration Variations (TDVs) over a very long time baseline. During the PLATO 2.0 mission lifetime, RV follow-up to determine planet masses will focus on the scientifically most interesting targets. However, science interests develop with time and there is always room for surprising discoveries. Planet candidates detected by PLATO 2.0, but not confirmed by RV within the mission lifetime, will provide a wealth of targets for future mass determinations by the science community, resulting in thousands of further characterized planets.

The PLATO 2.0 catalogue of about 85,000 stars with known ages and of about 1,000,000 highly accurate photometric stellar light curves complements the results of the Gaia mission and will provide a huge legacy for stellar and galactic science which will be explored by the community in the years to come after the PLATO 2.0 mission.

# 5 Mission and instrument concept

## *5.1 Mission and instrument*

The PLATO concept has been investigated in two independent industrial studies while its payload was studied by the PLATO consortium during the M1/M2 selection process of ESA. After this definition phase the mission designs have been described in a report (ESA/SRE(2011)13) which forms the basis for the mission and instrument details described here as PLATO 2.0. Previous descriptions of the mission can also be found in e.g., Catala (2009), Catala & Appourchaux (2011), Rauer & Catala (2011).

The PLATO 2.0 mission consists of a spacecraft module and a payload module including the telescopes and cameras. PLATO 2.0 can be launched on a Soyuz-Fregat rocket for injection into a Lissajous Orbit around the L2 Lagrangian point. This allows a nominal lifetime of 6 years after commissioning. Optional is an extended science operation phase, lasting up to 2 years. To protect the instrument from solar light, it has to rotate by 90° around the Line-of-Sight (LoS) every 3 months.

After launch and an early orbit phase, the spacecraft will enter a transfer phase to attain the operational orbit around L2. Commissioning starts during the transfer phase and will be completed two months after arrival in the operational orbit. Commissioning includes checking the spacecraft and payload calibration. The nominal science operation phase includes long- and short-duration phases, as described in Section 5.2.

The key scientific requirement to detect and characterize a large number of terrestrial planets around bright stars determined the design of the PLATO 2.0 payload module. It provides a wide field-of-view (FoV) to maximize the number of the sparsely distributed bright stars in the sky with one pointing and allows coverage of a large part of the sky in a step-and-stare mode. In addition, it provides the required photometric accuracy to detect Earth-sized planets and a high photometric dynamic range, allowing us to observe bright stars (≥4 mag) as well as fainter stars down to 16 mag. This performance is achieved via a multi-telescope instrument concept (Figure 5.1, left), which is novel for a space telescope.

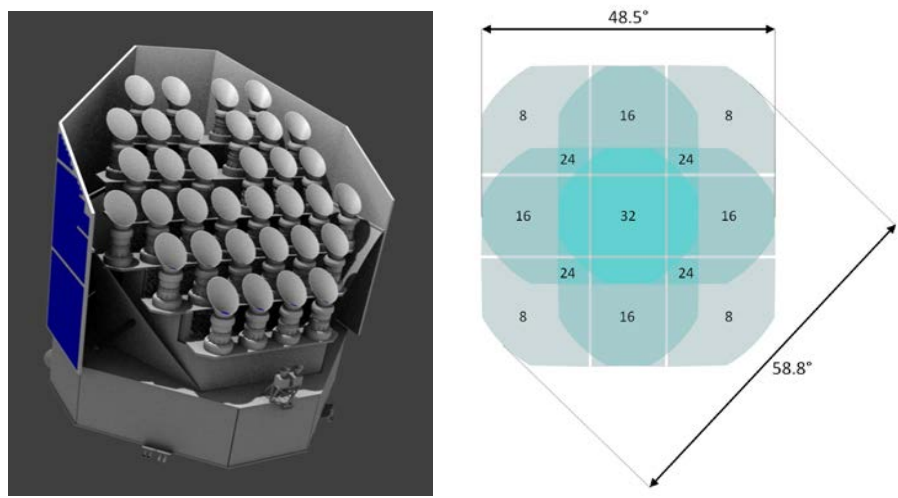

**Figure 5.1** *Left:* Schematic of the multi-telescope design of PLATO 2.0. *Right:* Schematic for the overlapping field-of-view of the four groups of eight 'normal' telescopes (ESA/SRE(2011)13).

The instrument consists of an ensemble of 32 so-called 'normal' cameras and two additional 'fast' ones, thus 34 telescopes in total mounted on an optical bench. Each telescope has a very wide, 1100 deg$^2$, FoV and a pupil diameter of 120 mm. The telescopes are based on a fully dioptric design with 6 lenses each, one of which is aspherical, mounted on an ALBeMet tube. Each telescope is equipped with a focal plane array of 4 CCDs, each with 4510² of 18 µm pixels. The 'normal' camera telescopes are read out in full frame mode with a cadence of 25 seconds and will monitor stars with m>8 mag. The two additional 'fast' telescopes are read-out in frame transfer mode with 2.5 second cadence and are used for stars with m=4-8 mag. While the normal cameras work in a wide bandpass, from 500 to 1050 nm, the two 'fast' cameras will provide colour information in two passbands for stellar analysis.

**Table 5-1 PLATO 2.0 Instrument characteristics in comparison to *Kepler* and CoRoT**

| | PLATO 2.0 | *Kepler* | CoRoT | Ref. |
|---|---|---|---|---|
| **Magnitude range** | - normal cameras: $8 \leq m_V \leq 16$ mag<br>- fast cameras: 4-8 mag | $7 \leq K_p \leq 17$ mag | - Exoplanet channel: $11.5 < m_V < 16$ mag<br>- Asteroseismology channel: $5.4 < m_V \leq 9.2$ mag | 1, 2, 3 |
| **Aperture size** | - **32×12cm** normal cameras<br>- **2×12cm** fast cameras | 99 cm | 27 cm | 1, 2, 3 |
| **FoV** | **2232 deg²** total (48.5°x48.5°)<br>- normal cameras: ~1100 deg²<br>- fast cameras: ~550 deg² | 105 deg² 16° diameter | 2.7°×1.5° (2.7°×3.05° until 03/2009) | 1, 2, 3 |
| **CCDs** | - normal cameras: 4 CCD per camera 4519×4510px, 18 µm square, full frame, 15 arcsec/px<br>- fast cameras: 4519x2255px, 18 µm square, frame transfer | 42 CCDs 1024×2200px | 2 CCDs 2048x4096px (4 CCDs until 03/2009) | 1, 2, 3 |
| **Time sampling of data points** (readout cadence) | - normal cameras: **25 sec** (~22 sec exp. time)<br>- fast cameras: **2.5 sec** (~2.3 sec exp. time) | - LC windows: **1766 sec**<br>- SC windows: **59 sec** | - Exoplanet channel: **512 sec** (normal) **32 sec** (optional)<br>- Asteroseismology channel: **32 sec** | 1, 2, 3 |
| *Spectral range* | - 500-1000 nm (normal cameras)<br>- one broad band for each fast telescope | 423-897 nm | 400-900 nm | 1, 2, 3 |
| **No. of target fields** | Step-and-stare and 1-2 long pointings | 1 | 26 | 3, 5 |
| **Observing period per target field** | 20 days – 3 years | 4 years | 20 - 150 days | 1, 2, 9 |
| **No. of dwarf target stars per pointing** | ~150,000* | 170,000 | - Exoplanet channel: ~6,000 (~12,000 until 03/2009)<br>- Asteroseismology channel: 5 (10 until 03/2009) | 1, 2, 3 |
| **Total no. of target stars over mission** | >1,000,000* | 170,000 | - Exoplanet channel: ~170,000<br>- Asteroseismology channel: ~150 | 2, 3, 5 |
| **No. of bright targets ≤11 mag** | ~85,000 stars total* | ~6,000 stars | ~370 | 2, 4, 5 |
| **No. of dwarf star asteroseismology targets** | ~85,000 stars total* | >512 stars | ~150 | 2, 3, 4, 5 |

1: Auvergne et al. 2009, 2: ESA/SRE(2011)13, 3: Koch et al. 2010, 4: from released *Kepler* data, 5: J. Cabrera, pers. comm.
\* for baseline observing strategy (see Section 5.2.1)

The 'normal' cameras are arranged in four groups of 8 cameras each which are aligned in their pointing direction. The four groups, however, are offset in their pointing by 9.2° from the payload z-axis, thereby increasing the total field surveyed to ~2250 square degrees per pointing (Figure 5.1, right). This strategy was chosen to optimize the dynamic photometric range of the instrument (4-16 mag) as well as to provide a large number of targets observed at a given noise level. As a result, however, the sensitivity of the payload varies over the field, depending on how many telescopes point at a target region. The sky fraction covered in each pointing is:

- The centre of the field is seen by 32 cameras and offers a FoV of 301 deg²
- A second zone, seen by 24 cameras, offers an intermediate FoV of 247 deg²
- A third zone, seen by 16 cameras, offers a FoV of 735 deg²
- A fourth zone in each of the 4 corners of the FoV is seen by only 8 cameras each, and with a total FoV of 949 deg²

In the case that an interesting target is detected in a field with reduced number of telescopes, it can be placed into the centre field at a later mission stage in a short-term pointing. Table 5-1 tabulates the PLATO 2.0 instrument characteristics in comparison to CoRoT and *Kepler*.

Due to its large FoV, the data volume is too large to transmit time series of full frames to ground (this would correspond to 189 Terabits, compared to ~109 Gbit download capacity per day). Therefore, data have to be processed on board to produce light curves per star and telescope. In this way, the light curves from all individual telescopes can be transferred to ground, and no on-board averaging on this level is needed.

Therefore, selected target stars are assigned a window from which the light curve is computed on board. A typical window is about 6x6 pixels (9x9 for the fast cameras) and includes the whole image of the target star. Only a limited number of windows can be transferred to ground, others are processed on board. On-board and ground-based processing includes weighted mask photometry, with frequent updates of the masks, correction for satellite jitter, outlier rejection, PSF fitting on several thousand reference stars for position measurements and instrument PSF control and sky background modelling.

Always two cameras share a common Digital Processing Unit (DPU) which performs the basic photometry. The resulting light curves, windows and centroid data are sent to a common Instrument Control Unit (ICU) and then transmitted to ground. On the ground, data from all 34 cameras are received for further reduction and data analysis. The two fast cameras also provide the required Attitude and Orbit Control System (AOCS) data for the high pointing stability of the satellite.

## *5.2 Observing strategy and sky coverage*

### 5.2.1 Observing strategy

Detecting planets by transit surveys requires long, continuous and uninterrupted observations. It is an advantage of space missions that they can provide the necessary high duty cycle. However, different strategies concerning observational pointings can be selected.

The two recent space exoplanet surveys, CoRoT and *Kepler*, differ significantly in their observing strategy. Due to its low-Earth orbit, the CoRoT satellite could point at one target field in a predefined and fixed viewing zone for about 6 months (Baglin et al. 2006). Therefore, CoRoT is well-adapted to detections of planets on relatively short-period orbits, typically less than ~ 100 days. CoRoT partially compensates the disadvantage of a limited observing duration per target field by flexibility of its pointing. The satellite has observed 26 target fields, located in its visibility zones towards and opposite the galactic centre direction (Figure 5.2). Interestingly, the detection yield per target field varies. We do not understand yet whether this hints towards differences in planet population in the sky, or other issues concerning these fields and their data analysis. But it nevertheless raises the interesting question whether planets are homogeneously distributed in the sky, or not. *Kepler* on the other hand aimed at planets on Earth-like, long-period orbits. It therefore stares at the same field over its whole mission duration, finally for about 4 years in total. TESS (NASA) will follow a similar strategy to PLATO 2.0, covering bright stars over a wide part of the sky. TESS will, however, concentrate on short period planets, up to 10 or 20 days, except for a limited region of the sky (approximately 2%). CHEOPS (ESA) is not a survey mission, but performs pointed follow-up observations, one target at a time, and therefore cannot be compared with the others in this section.

PLATO 2.0 has a more flexible observing approach. Two observing strategies, long continuous pointings versus shorter coverage of different fields, complement each other and allow a wide range of different science cases to be addressed. Long pointings will be devoted to surveys for small planets out to the Habitable Zone of solar-like stars. Short pointings will be devoted to shorter-period planet detections and will address a number of different science cases.

In its nominal science operation phase, PLATO 2.0's current baseline observing strategy combines:

- Long-duration Observation Phases, consisting of continuous observations for two sky pointings, lasting a minimum of 2 years with a maximum of 3 years for the first pointing, and 2 years coverage for the second pointing.

- Step-and-Stare Operation Phases, consisting of shorter-period observations of several sky fields which will last 1-2 years total, depending on the duration of the long duration phases. Sky fields in this phase will be observed for at least 2 months, up to a maximum of 5 months.

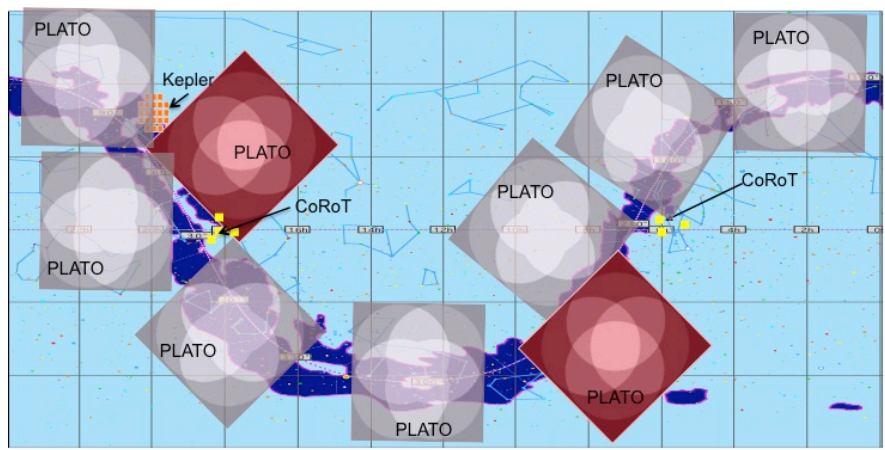

**Figure 5.2** Schematic comparison of observing approaches. Yellow squares: CoRoT target fields in the galactic centre and anti-centre direction. Upper left corner: the *Kepler* target field. Large squares: size of the PLATO 2.0 field. A combination of short and long (red) duration pointings is able to cover a very large part of the sky. (ESA/SRE(2011)13). Note that the final locations of long and step-and-stare fields will be defined after mission selection and are drawn here for illustration only. See Figure 5.2 for preliminary location of long-duration fields.

The proposed observing strategy aims at covering a large fraction of the sky, thereby maximizing the number of well characterized planets and planetary systems, in combination with wide-angle long pointings that will significantly increase the number of accurately known terrestrial planets at intermediate distances up to 1 au. The latter detection range will be unique to PLATO 2.0 and is not covered by any other planned transit survey mission nor can it be achieved for a large number of RV detections in any feasible observing time.

In view of the exceptionally fast development of exoplanet science, the order of long and short runs can be re-investigated after mission selection and adapted to the needs of the community by 2022/24, e.g., to investigate interesting sky regions and targets earlier in the mission with a step-and-stare run. The PLATO 2.0 observing concept offers sufficient flexibility.

### 5.2.2 PLATO 2.0 Input Catalogue (PIC) and long duration pointing field selection

Telemetry limitations impose the pre-selection of PLATO 2.0 targets for the detection of planets. The optimal field selection is closely related to the target selection. The success of the mission is related to our ability to select fields that maximize the number of F5 or later spectral type dwarfs and sub-giants. We need to prepare a PLATO 2.0 input catalogue (PIC) which includes the targets in the priority magnitude range (4-11 mag), and provide their main parameters. A limited number of additional targets may be added to the PIC, to monitor special objects (e.g., in star formation regions or star clusters within the long monitoring fields) for the main and complementary science cases. Finally, the PIC will help us to assess the nature of the detected transiting bodies: a good knowledge of the central star will help us to exclude false alarms and will trigger the most appropriate follow up strategy. It will also allow us to get a first estimate of the size of the planet.

The PIC will serve to: 1) finally select the optimal PLATO 2.0 fields; 2) select all appropriate >F5 dwarf and sub-giants within them; 3) characterize as much as possible the selected targets, i.e., estimate their temperature, gravity, variability, metallicity, binarity, chromospheric activity; 4) provide a list of neighbours that contaminate the target star flux; 5) give a first estimate of the transit object radius; 6) optimize the follow-up strategy.

The building of the PIC will require the assembly of information from very different input catalogues on a wide range of targets (from mid-F to M-dwarfs and subgiants). The main source for the PIC will be the Gaia catalogue. A complementary survey of available photometric, spectroscopic catalogues and other data bases for the assessment of stellar activity will be carried out in addition. This survey, in addition to the option of dedicated surveys for further characterization, can also be used as back up for the PIC target selection and characterizations in the case of delays in the publication of Gaia catalogues. First results of a statistical analysis of available stellar catalogues, and a contamination analysis to minimize the number of targets for which follow-up is needed to eliminate false alarms, are described in ESA/SRE(2011)13.

The two long-duration PLATO 2.0 fields form the core of the mission. Their centres must stay within two regions imposed by observability constraints. These "allowed regions" are spherical caps defined by an ecliptic latitude $|\beta|>63$, and are located respectively in the southern and northern hemispheres, mostly at high declinations ($|\delta|>40$). The choice of the long-duration fields is driven by the need to find a trade-off between the number of priority targets and a minimum rate of false-alarms due to crowding.

A proposed conservative choice (to minimize contaminants, still satisfying the scientific requirements in terms of target numbers) for the field centres is ($l$=253, $b$=-30) for a Southern sky field and ($l$=65, $b$=30) for a Northern sky field (see Figure 5.3). These fields are centred approximately on the Pictor (South) and Lyra/Hercules (North) constellations. The northern field includes the *Kepler* field on a corner. An additional, thorough study of the contaminant problem will allow us to verify whether the field centre can be moved to lower Galactic latitudes ($|b|$~25), thus increasing the number of targets.

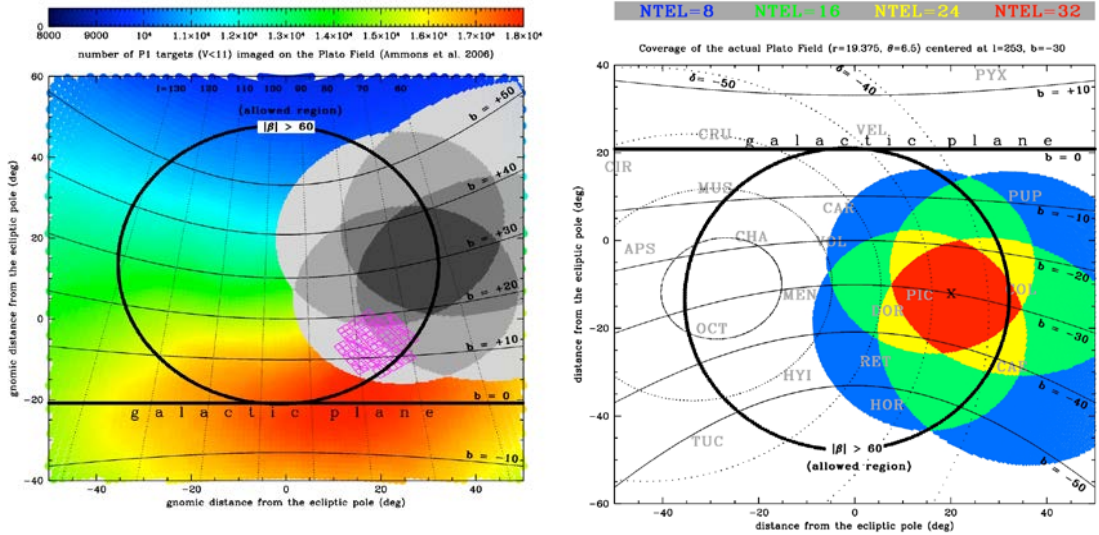

**Figure 5.3:** *Left:* Density of targets ≤11 mag for the northern region, averaged over the area of the PLATO 2.0 field, following Ammons et al. (2006). The preliminary long-duration PLATO 2.0 Field is shown in gray. The *Kepler* field is indicated in pink colours. *Right:* The preliminary long-duration PLATO 2.0 field chosen for the southern allowed region, with the number of telescopes covering the single sub-regions indicated by different colours.

# 6 PLATO 2.0 planet detection performance

PLATO 2.0 detects planets primarily by searching for periodic transit events. However, planets can also be detected by other methods, e.g., reflected stellar light variations or astrometry. In addition, planetary rings and moons of transiting planets can be detected. In this section, we discuss PLATO 2.0's performance for these detection methods, as well as the means to constrain masses of transiting planets by transit timing variations and 'Blender' analyses.

The transit detection performance can be characterized in two ways, which are both addressed below:
- Via the total number of stars that will be monitored down to a given magnitude. This performance indicator depends on the global FoV of the instrument and on the number of fields observed.
- Via the total number of stars that will be monitored down to a given photometric noise level. This performance depends on a combination of the pupil size and FoV of each camera, the configuration of the cameras in the overlapping LoS concept, and on the number of fields observed.

## 6.1 Instrument performance

The PLATO 2.0 end-to-end simulator (Zima et al. 2010) was used to generate simulated light curves for various sets of stars representing realistic parts of the fields to observe. Known sources of noise were introduced in these simulations, including photon noise, readout noise, jitter noise, background noise, etc., as well as a standard on-board and on-ground data treatment system. See ESA/SRE(2011)13 for details. Figure 5.4 provides an overview of the resulting signal-to-noise performance of the instrument. In the central part of the FoV all 32 cameras overlap and 27 ppm in one hour can be reached for an 11 mag star, degrading to 60 ppm in one hour at the edge of the FoV where only 8 cameras observe. Across most of the FoV 34 ppm over one hour can be obtained down to ~11mag, reaching the photon limit. This is sufficient for detection of an Earth-sized planet as well as asteroseismology analysis of its host star. The fast cameras reach the same photon limited noise level for ~7.5 mag. Much better performance is reached for both camera types for bright stars.

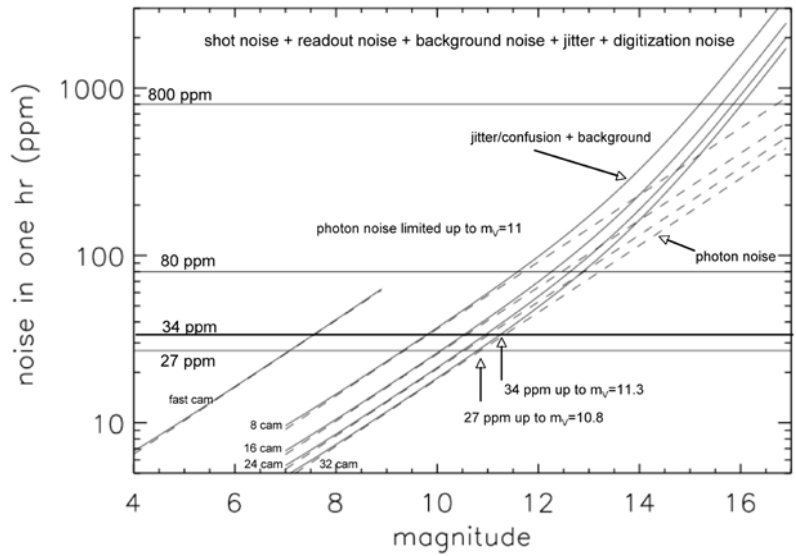

**Figure 5.4** PLATO 2.0 instrument noise performance (Catala 2009; ESA/SRE(2011)13).

## *6.2 Observed number of stars*

The final number of stars monitored with PLATO 2.0 will of course depend on the final observing strategy chosen. However, to obtain an estimate of the expected mission performance, we present the number of observed stars for the current baseline observing strategy (Section 5.2.1) as an example.

In order to derive the number of stars observable during the mission as a function of magnitude and achievable noise level we used two galactic models, Besancon (Robin et al. 2003) and TRILEGAL (Girardi et al. 2005), and all-sky stellar classifications based on astrometric and photometric catalogues (Ammons et al. 2006, among others). Based on the expected instrument performance (Figure 5.4) the yield of stars monitored by PLATO 2.0 up to a given noise level or stellar magnitude has been computed assuming two long runs of 2 years each, and a 2 year step-and-stare phase including the following successive runs: 3 x 5 months, 1 x 4 months, 1 x 3 months, 1 x 2 months.

**Table 5-2** Expected number of monitored cool dwarf and sub-giant stars with PLATO 2.0 in comparison to *Kepler*. See text for assumed PLATO 2.0 observing strategy.

|  |  | PLATO 2.0 |  | *Kepler* |
|---|---|---|---|---|
| **Noise level (ppm in one hr)** | $m_v$ | 2 long pointings | 2 long pointings + step-and-stare | Fixed *Kepler* field |
| **8** | 8 | >1000 | >3000 | 30 |
| **34** | 11 | 22000 | 85000 | 1300 |
| **80** | 13 | 267000 | 1000000 | 25000 |

Table 5-2 shows the expected PLATO 2.0 performance in terms of number of monitored stars in comparison to the *Kepler* mission. Concentrating on the main PLATO 2.0 magnitude range $m_V<11$ mag, we find that PLATO 2.0 outnumbers the *Kepler* performance by a factor of about 50 in the long-pointing phases and by a factor of about 140 over the total mission duration for the typical observing scenario assumed here.

## 6.3 Expected number of characterized super-Earths (≤2 $R_{Earth}$)

The *Kepler* mission has detected about a hundred planets with known radii and masses, including planets with hosts characterized by asteroseismology, and thousands of planet candidates. The TESS mission is expected to detect about 1000 small planets, including hundreds of Earths to super-Earth's. Here, we study the impact of the PLATO 2.0 mission on the bulk characterization of super-Earth planets around bright host stars in comparison to these transit survey missions.

In this section, we define 'super-Earths' as planets with radii ≤ 2 $R_{Earth}$. Bulk characterization requires RV follow-up spectroscopy, and asteroseismology of the hosts. RV follow-up to determine planet masses with reasonable telescope resources for a large number of targets is limited to about 11 mag. The *Kepler* mission performed asteroseismology for stars up to about 12 mag (Huber et al. 2013a), and PLATO 2.0 will do the same for stars up to 11 mag. For the all-sky survey by TESS, little has been published on its asteroseismology performance yet. However, from the much smaller aperture per star (10 cm) it is expected that asteroseismology should be limited to stars brighter than about 7.5 mag. Thus, for *Kepler* and PLATO 2.0 full characterization is limited by RV follow-up to stars ≤11 mag. For TESS asteroseismology limits fully characterized planets to host stars brighter than about 7.5 mag.

With these magnitude constraints for fully characterized planets, we can estimate the number of suitable target stars within the fields surveyed. The *Kepler* field is well-known, and for PLATO 2.0 we use the current baseline observing strategy described above. Stellar catalogues are used to estimate the number of sufficiently bright stars for PLATO 2.0 (≤11 mag) and in the all-sky survey TESS (≤7 mag). We take into account that, accordings to its current, preliminary observing strategy, most fields are observed by TESS for 27 days only, but about 2% of the sky at the equatorial poles is observed for 1 year.

To convert from the observed number of stars to an expected planet detection yield, we take into account the transit geometrical probability and the expected planet frequency per star when known. For planets with orbital periods up to 50 days we apply the published (Fressin et al. 2013) rate of super-Earth planets per star based on *Kepler* planet candidates. Figure 5.5 shows that we expect for PLATO 2.0 transit signals from about 1000 super-Earth planets around stars ≤11 mag, in the range where full characterization is in principle possible. We note that an orbital period of 50 days (~0.4 au) includes the HZ of cool (M-K) stars (see Figure 2.2).

For planets with periods longer than 90 days, no certain planet frequency of super-Earths is known to date, with frequency ranging from few up to 64% (Catanzarite & Shao 2011; Traub, 2012; Gaidos 2013; Petigura et al. 2013). We therefore assume 40% of the stars have a super-Earth in the two distance bins considered (0.4-0.8 au and 0.8-1.2 au) as a typical, mean value. With this assumption, we expect transit signals of about 40-70 super-Earths in the HZ of G-type stars. For these planets, bulk characterization for RV masses and asteroseismology of the hosts can be performed.

Figure 5.5 shows the expected yield of transits from super-Earths which can be fully characterized in comparison to *Kepler* and TESS. PLATO 2.0 will increase the yield of characterized Earth to super-Earth planets (1-2 $R_{Earth}$) by a factor ~10 above *Kepler* and ~1000 above TESS for planets with 90 d < P < 500 d periods and by a factor of ~10-20 above *Kepler* and TESS for planets up to 50 d orbital period.

We emphasize that the total super-Earth detection of *Kepler*, but also in particular of TESS, is expected to be much larger, as discussed at the beginning of this section. Here, we restrict our comparison to those targets which fall into the prime science goals of PLATO 2.0, hence full bulk characterization in particular in the HZ of solar-like stars. It is obvious that PLATO 2.0 will outnumber any foreseen mission within this decade, and bulk characterization of super-Earths in the HZ of solar-like stars remains basically unique to PLATO 2.0. Among these planets some will be around the brightest stars, which will form prime targets for future spectroscopic follow-up to study their atmospheres.

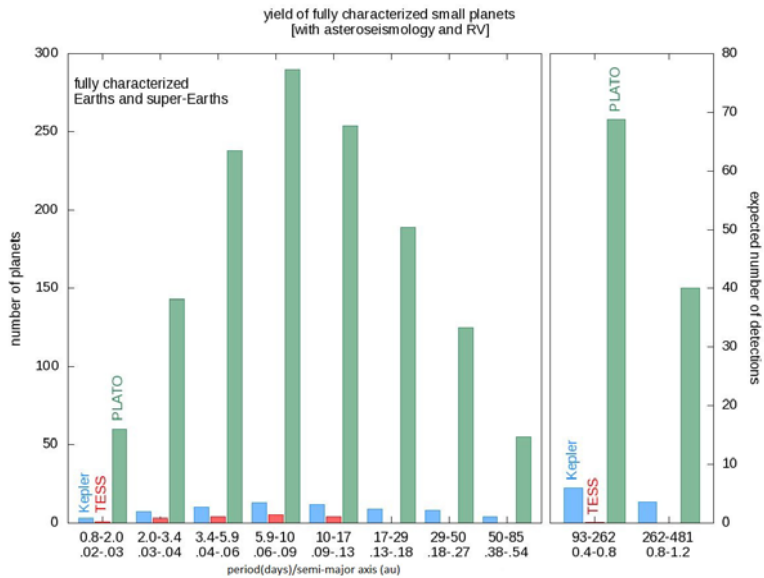

**Figure 5.5** PLATO 2.0 transit signal detection performance for super-Earth planets (≤ 2 $R_{Earth}$) for stars ≤11 mag, hence with RV follow-up and host star asteroseismology possible. For comparison, *Kepler* results are shown (Fressin et al. 2013) and expected yields for TESS assuming 27 day observing coverage per field and 2% of the sky observed for 1 year.

In addition to the expected transit detections for terrestrial, super-Earth planets discussed here, PLATO 2.0 will characterize thousands of mini-Neptunes and gas giants. From the large number of expected transit detections, it is evident that the limitation for the final catalogue will come from RV follow-up resources (see Section 7). We however point out that planet candidates not followed-up by RV within the mission lifetime will become part of PLATO 2.0 legacy science (Section 4.9) and will provide a wealth of interesting targets for future RV measurements within the community.

## 6.4 Detections and Validation via Transit Timing Variations

Deviations from strictly periodic transit events, so-called Transit Timing Variations (TTVs), provide further information about planetary systems because additional planets are sources for TTVs. These planets may also produce observable transits if their orbital plane is sufficiently aligned (e.g., the co-planar Kepler 11 System with 6 transiting planets, Lissauer et al. 2011), or they may remain unseen and only appear through their gravitational interaction with the observed transiting planet and the host star (e.g., Kepler-19c, Ballard et al. 2011). TTVs can be as large as 24h in some cases (e.g. Kepler-90 with 7 transiting planets, Cabrera et al. 2014).

Data from the *Kepler* satellite suggest that multi-planetary systems are relatively common (20-23%, Batalha et al. 2013; Fressin et al. 2013; Burke et al. 2014) and that about 13% of the multiple systems show significant TTVs (Mazeh et al. 2013). Therefore, they provide a valuable extension of the transit method for planet detections. Planets with TTVs in principle also allow us to determine planetary masses, independently of radial-velocity follow-up observations. This is particularly useful for multi-transiting co-planar systems where masses can be relatively well constrained, and for planets in near-resonant orbits where TTVs are large (e.g., Carter et al. 2012, Fabrycky et al. 2012).

TTVs are a useful method to confirm the nature of multiple planetary systems and can be highly complementary to RVs (e.g., Steffen et al. 2012). In particular, a TTV analysis allows one to estimate or constrain the mutual inclination of non-transiting planets (Meschiari & Laughlin 2010; Nesvorný et al. 2013). Concerning mass determination, however, TTVs reach the precision and accuracy obtained with RV observations only in exceptional cases, since e.g., possible systematics, in particular when the planetary orbits are eccentric or have large relative inclinations, can significantly affect the result. Therefore, RV remains the preferred method when high accuracy masses are needed.

The photometric precision and time-resolution of PLATO 2.0 will allow for the detection of TTV perturbations with amplitudes of a few seconds, which is a performance at least two times better than *Kepler* and four times better than CoRoT. Therefore the accuracy will be significantly improved and future interesting discoveries are expected, e.g., large bodies on Trojan orbits (see, for example, Cabrera 2010a and references therein, and Szabó et al. 2013), or exomoons from the detection of Transit Duration Variations (TDV; Kipping 2009).

The TTV-method can also be an important tool for characterizing close-in planets. When the planet has a relatively short (ca. <1-2 weeks) and eccentric orbit, then the semi-major axis of the orbit rotates around the stellar centre (apsidal motion). This causes observable TTVs. The rate of the apsidal motion is a function of the $k_2$ Love-number that describes the internal mass distribution of the planet (e.g., Mardling et al. 2010). Therefore this kind of TTV can be used to obtain additional information on the interior structure of planets.

We highlight that re-observing the *Kepler* field with PLATO 2.0 will provide a time baseline for TTV observations from 2009-2013 to 2024-2029, hence providing an ideal set-up to study accurate long-term TTVs which otherwise are extremely difficult to detect.

In summary, although TTVs do not replace the need for accurate RV mass measurements, TTVs will significantly expand the detection range of PLATO 2.0 beyond what can be reached with this method by existing and near-future space missions. They will be particularly important for faint targets (<11mag) where direct RV for low-mass planets is difficult and will significantly expand the scientific results of PLATO for these targets.

Finally we point out that TTVs are very interesting when combined with RV for PLATO's bright targets. Then we can reconstruct the whole 3D orbits and derive masses with very high accuracy (up to 2%) (Borkovits et al., 2013). In such cases, TTV+RV combinations give a mass estimate independent of asteroseismology and RV, which is an interesting and valuable cross-check of independent methods.

## 6.5 Pre-spectroscopic validation of planet candidates from PLATO 2.0

Stellar blends (Brown 2003, Santerne et al. 2013), e.g. eclipsing binary stars within the photometric point-spread function (PSF) of target star, are one of the major sources of false alarms in transit surveys (Almenara et al. 2009) and their rejection is quite expensive in terms of follow-up requirements (e.g. Santerne et al. 2012). Transit surveys use different tests to reject false alarms based on the analysis of the light curve (see Collier-Cameron et al. 2007 for the WASP survey; Tingley et al. 2011 for the CoRoT survey; Batalha et al. 2010 for the *Kepler* survey). The planet validation approach consists in comparing the relative capabilities of these blend scenarios and of the transiting planet scenario to explain the available data (Torres et al. 2011; Diaz et al. 2013). For space-based surveys, which explore the small-size planetary domain, the so-called 'Blender' (Torres et al. 2011) and 'PASTIS' (Diaz et al. 2013) softwares perform this procedure. They combine a detailed analysis of the transit light curve with a statistical study of the stellar background (or foreground) population which may mimic the planetary signal. One of the key pieces of information in this procedure is the measurement of the transit signal at different wavelengths, especially in the infrared which constrains the colour difference between the target and the potential false-positive system. This procedure has been successfully applied to several *Kepler* and CoRoT cases (Torres et al. 2011, Fressin et al. 2012a,b, Borucki et al. 2012, Moutou et al. 2013), but its performance is severely limited if the transit is observed in only one wavelength.

The exquisite photometric precision of PLATO 2.0 and the simultaneous observations with the two fast cameras, which will observe in different photometric bands, will allow for a first-order rejection of potential blend scenarios with the planet-validation analysis, in a similar way as was done in the case of, e.g., CoRoT-7b by Léger al. (2009). Furthermore, the brightness of the main PLATO 2.0 target sample will facilitate the separation from faint background stars in comparison to *Kepler*. Additionally, centroid determinations (see also Section 6.6.1) will help in many cases to exclude false alarms as part of a blender analysis, again as for *Kepler*. Thus, the number of false alarms entering into the RV follow-up list will be significantly reduced for PLATO 2.0 saving telescope time for the most promising small size planet candidates. The planet-validation tools, such as PASTIS that models spectroscopic data (Santerne et al., in prep.), will be used to secure the RV detection of the coolest PLATO 2.0 rocky planets. This software will therefore help to fully exploit the PLATO 2.0 data.

## *6.6 Detection of non-transiting planets*

For large, close-in planets around bright stars, PLATO 2.0 is not limited to transiting planets. Although these cases are only a sub-set of the observed targets, they provide key objects for further follow-up observations for their characterization, e.g., for direct observation with the ELT.

### **6.6.1 Astrometric detections**

The stellar reflex motion induced by a planet's revolution creates an astrometric wobble of the host star. PLATO 2.0 will measure the astrometric position of each star in the surveyed field relative to all other stars in the field, which will provide a very precise reference frame. The resulting precision of PLATO 2.0 is particularly interesting for close-by Jupiter-sized planets at intermediate distances since it is basically independent of the transit geometry. A 1 $M_{Jup}$ exoplanet, orbiting a 1 $M_{Sun}$ star at 1 au and assuming a stellar magnitude of $m_V$ = ~6, would induce a 60 µas wobble. Preliminary simulations have indeed shown that relative centroid measurements will reach sufficient precision down to $m_V$ = 6 after one month of integration (ESA/SRE(2011)13). This will allow us to detect virtually all giant exoplanets with orbits near 1 au orbiting nearby bright stars, irrespective of the inclination angle of the orbital plane with respect to the line of sight. These astrometric measurements, coupled with measurements of reflected stellar light described below, will constitute a powerful tool for identifying exoplanetary systems around nearby stars, out to distances of 15-20 pc. These observations will complement data obtained with the Gaia satellite and e.g., help characterizing unresolved planetary systems with the better time baseline coverage of PLATO 2.0.

### **6.6.2 Detection of reflected stellar light**

The high-precision photometry of PLATO 2.0 will allow the detection of close-in non-transiting planets by the modulation of the flux in the light curve, as performed e.g., for CoRoT-1b (Snellen et al. 2009). Because the monitoring of such targets will cover several hundred planetary orbital periods, such a modulation will be detectable by PLATO 2.0 down to $m_V$ = 9-10 for albedos as small as A = 0.3. For a CoRoT-1b-like target, PLATO 2.0's noise level below 30 ppm/hr on stars down to $m_V$ = 11 will allow us to detect on its main targets reflected light signals at least 7 times weaker than CoRoT, which could correspond to planets 2.5 times smaller in radius or 2.5 times further out, assuming a similar albedo.

Very bright stars, typically with $m_V$ = 6, will be observed with a noise level of approximately 10 ppm/hr. Such a low noise level will enable the detection of reflected stellar light from planets with 0.15 $R_{Jup}$ radii. We will therefore be able to detect super-Earths in close-in orbits and identify a large fraction of nearby stars hosting close-in planets that will become priority targets for further observations, including searching for smaller and further out planets.

In binary systems, reflected light competes in amplitude with the tidal distortion created by the gravitational interaction of massive bodies and with the relativistic beaming effect (Zucker et al. 2007). This method has proved useful to improve the characterization of known planets (Welsh et al. 2010) and to confirm the masses (independent of RV measurements) of transiting bodies discovered by CoRoT (Mazeh & Faigler 2010) and *Kepler* (Mazeh et al. 2012, Shporer et al. 2011). In the case of *Kepler*, it has also allowed the discovery of a few non-eclipsing stellar binary systems (Faigler et al. 2012). It is expected that PLATO 2.0 will outperform *Kepler* for short period, massive planets.

In some cases, high precision photometric measurements allow the determination of the spin-orbit alignment of transiting bodies, providing additional constraints for the theories studying the formation and evolution of planetary and stellar systems (Barnes et al. 2011; Shporer et al. 2012).

# 7 Follow-up observations

The prime goal of PLATO 2.0 is to deliver detected planets with well determined radii and masses. RV follow-up observations are ultimately needed to derive accurate planetary masses. However, since the observational effort for RV spectroscopy is large, it is important to reduce the number of planet candidates entering to the list for RV follow-up observations as much as possible with less resource intensive methods. This is a multi-step process.

The first step of planet detection is to separate false alarms from real planet signals (Figure 5.6). False alarms can be caused by e.g., diluted signals from eclipsing binaries within the large pixel scale of PLATO 2.0. Many causes of false alarms can be identified already from close inspection of the stellar light curve. The light curves undergo several checks, e.g. for out-of-transit photometric variations as found for binary stars, a check for adequate transit and occultation depth and duration consistent with a planet-sized object, and most importantly from a pre-spectroscopic validation procedure providing a reliable probability that the signal is of planetary nature. Many false alarms can already be rejected in this way and upper limits to the planetary masses can be obtained before putting planet candidates on the observational follow-up list. These analysis methods help to separate planetary candidates from binary stars or intrinsic stellar brightness variations, such as the one caused by spots and give confidence in the planetary nature of the detected object even though its final mass is not known yet. These procedures have been very successfully applied to CoRoT and *Kepler* planet candidates.

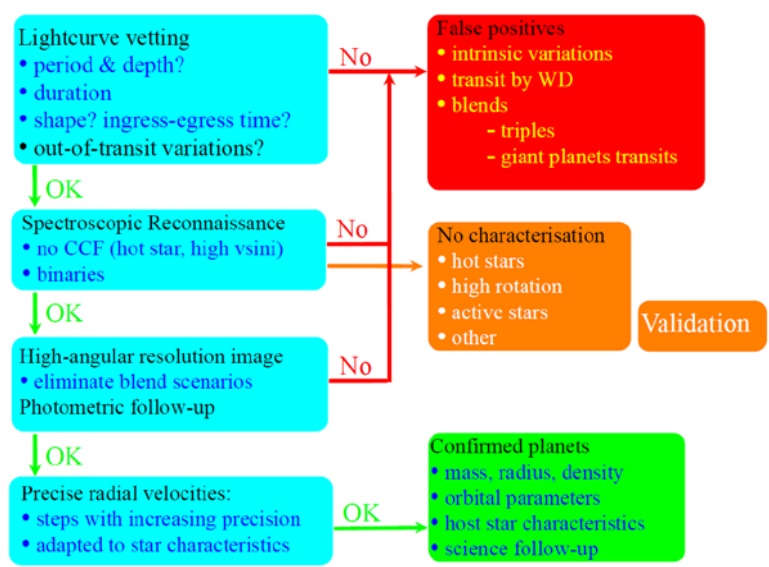

**Figure 5.6** Schematic organization chart for follow-up observations. See text for details.

In addition it is important to obtain high spatial resolution imaging of the planetary transit candidates to exclude contaminating objects in the PLATO 2.0 PSF, and to verify that the observed transit is indeed observed from the target star. This is especially crucial for shallow transits that are candidates for the most interesting, terrestrial planets, that are also the more demanding in RV follow-up time.

Once these tests have been successfully passed by a planet candidate, spectroscopic follow up observations are the next step. They come in a multi-step approach from low- to high-spectral resolution instruments. This approach excludes the remaining binary cases by utilising less cost-intensive instruments on small telescopes. Spectrographs like FEROS on the ESO 2.2-m telescope, CORALIE on the 1.2m Swiss telescope at La Silla, SOPHIE on the 1.93-m telescope at OHP, have already performed such observations in the past, and future instruments with similar performance will be used for PLATO 2.0. For close-in planets with masses down to the super-Earth regime, instruments like HARPS at the ESO 3.6-m telescope, HARPS-N on the TNG (La Palma), or similar instruments in development (Spirou/CFHT, Carmenes/Calar-Alto) will be the main work horses. The most interesting (and demanding) low-mass, longer-period planets will require the very high radial-velocity precision of instruments on larger telescopes, like PEPSI on the LBT (on sky in 2014) and ESPRESSO on the VLT (foreseen on the sky in 2017), and possibly a super stable spectrograph on the E-ELT for the Southern sky.

Even with a rigorous pre-selection of planetary candidates, a significant amount of telescope time will be required for PLATO 2.0 follow-up. Efforts will therefore concentrate on the most interesting prime targets, leaving cases like e.g., 'hot-Jupiter candidates' as a legacy for the community to study over a longer future time period, depending on science interest. In priority it is planned to concentrate on low-mass planets at large orbital separations, but it is also assumed that several hundred/thousands of low-mass planets with short periods will be followed-up with high precision. It is furthermore assumed that 20 observations per planet are adequate to characterize the candidates. With these assumptions for RV follow-up over the 6-year mission lifetime, the required observing time for the RV follow-up is approximately 50 nights/year for several 1-2m and three 4m-class telescopes, and up to 40 nights/year on one 8m-class telescope (ESA/SRE(2011)13). Such follow-up effort would provide on the order of 1500 highly accurate Earths to super-Earths on short to medium period orbits and long period gas giants, and about 100 terrestrial planets out to 1 au.

The real number of confirmed planets with highly accurate RV mass measurements will of course depend on the actual telescope time invested. We only provide a conservative estimate here. Clearly, further developments of data analysis procedures as well as the availability of smaller telescopes will be important to identify potential false alarm scenarios early in the analysis and thereby limit the number of candidates which go to high-precision RV follow-up (see Section 6.5). Also, we strongly emphasize that follow-up for interesting targets will continue as a legacy. Thus if more telescope time is required to cover all interesting targets, this can, and likely will be, performed on a longer timescale after the end of the PLATO 2.0 satellite mission.

# 8 Data products and data policy

The final data products of PLATO 2.0 consist of stellar light curves of about 1,000,000 stars, detected planets and fully characterized stellar and planetary parameters. Specifically, the data products are split into several levels as described in ESA/SRE(2011)13. They range from validated light curves of individual telescopes to averaged light curves, corrected for instrumental effects, up to the final PLATO 2.0 data products which are:

- Planetary transit candidates and their parameters.

- The list of confirmed planetary systems, which will be fully characterized by combining information from the planetary transits, the seismology of the planet-host stars, and the follow-up observations.

- Asteroseismic mode parameters. Stellar rotation periods and stellar activity models inferred from activity-related periodicities in the light curves. Seismically-determined stellar masses and ages of stars.

It is the intention of the PLATO 2.0 mission to make as much data available to the community as fast as possible during the mission. Calibrated lightcurves and centroids (L1 data) will be made public, based on current best knowledge, on time scales ranging from about a few months in the early phases of the mission to days later on. Planet and stellar parameters (L2 data) which also require additional observations will be made publicly available in a timely manner, and no later than the acceptance for publication of the first refereed papers based on them (ESA/SRE(2011)13). Only a small number, i.e., 2000 light curves (out of 1,000,000), will be property of the PLATO 2.0 team and involved ESA scientists for one year. This list of proprietary targets is established at least 6 months prior to each phase of the mission (one phase being defined as one long run or the step & stare phase), as the outcome of a call for proposals aimed at PLATO-involved scientists. PLATO-involved scientists will respond by specifying the scientific use they propose to make of a limited number of targets. The PLATO Science Team, established by ESA, will review the proposals and come up with a final selection of proprietary targets. For details on the propriety target selection see ESA/SRE(2011)13.

Calls for proposals to the general science community will also be made during the mission phases to ask for complementary science programs, not covered in the PLATO 2.0 core program. These proposals add another opportunity for the general scientific community to use PLATO 2.0 for various scientific fields.

# 9  *Summary*

The PLATO 2.0 mission promises to perform the discovery of and first systematical characterization of thousands of Earth-sized or smaller exoplanets, out to 1 au orbital distance from their respective host stars, delivering precise radii and uniquely high precision ages.

The about $10^6$ high precision and long duration light curves will, together with Gaia, be the base of providing a revolution in stellar physics, the impact of which will be felt across essentially all fields of modern astrophysics and cosmology.

Building on the CoRoT, *Kepler*, CHEOPS and TESS missions, PLATO 2.0 will be the only mission dedicated to such detection and characterisations and will be complemented by facilities for spectroscopic follow-up such as E-ELT, JWST or future missions dedicated to exoplanet atmospheric spectroscopy.

# Appendix A

# Methods: Characterizing planets and their host stars

Since the PLATO 2.0 mission addresses science goals in very different scientific communities, we briefly describe the key aspects of the methods used.

## *A1 Planetary Transits, a method to detect planets and determine their parameters*

The transit method photometrically measures the flux of target stars over time, searching for a dimming of stellar flux by an orbiting planet passing through the line-of-sight to Earth. When the planet is in front of the star, it shades an area on the stellar surface proportional to its size. The dimming of stellar flux is therefore proportional to the square of the radius of the planet, $R_{planet}$, relative to the radius of the star, $R_{Star}$: $\Delta F \; \alpha \; (R_{planet}/R_{Star})^2$. Figure A1 shows as an example the transit light curve of Kepler-10b, the smallest known exoplanet with radius and mass measurement so far ($R_{planet}$ = 1.416+/-0.03 $R_{Earth}$, $M_{planet}$=4.6+/-1.2 $M_{Earth}$, Batalha et al. 2011). The round shape during transit is caused by the limb darkening of the host star. The transit method allows us to directly measure a planet's size once the size of the star is known.

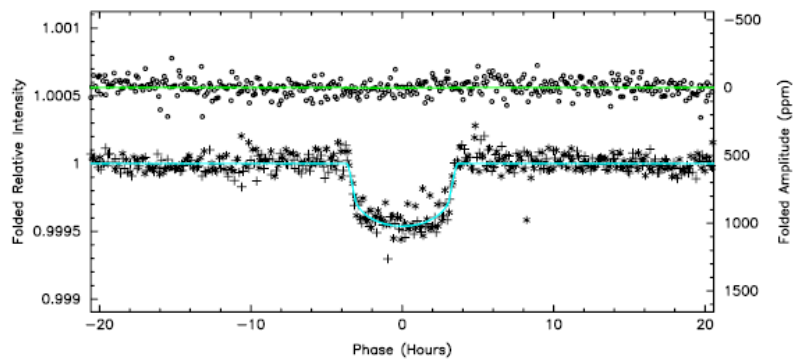

**Figure A1** Transit light curve of Kepler-10b (Batalha et al. 2011). The planet has an orbital period of about 0.8 days and was observed by *Kepler* for a period over 8 months for this data set. The V magnitude of the host star is 10.96 mag.

The mass of a detected transiting planet then has to be determined by other means, for example by spectroscopic radial-velocity follow-up. Once radius and true mass of the detected object are known, its bulk parameters are well determined and the object can be clearly separated from possible false-alarm events also causing periodic dimming of stellar intensity, such as spots or eclipsing binaries. The combination of radius and true mass provides the mean density of the planet. In combination with models of planetary interiors, the inner structures of planets can be constrained.

The periodicity of transit events allows us to derive the orbital period and therefore orbital distance according to Kepler's 3$^{rd}$ law. If the secondary eclipse can be detected, i.e. the planet disappears behind its host star, the orbital eccentricity can also be derived. Furthermore, the combination of transits with spectroscopic radial-velocity follow-up allows us to determine the alignment of the planetary orbital plane with the projected stellar rotation axis and the sense of orbital revolution of the planet around its star by the Rossiter-McLaughlin effect (Rossiter, 1924; McLaughlin, 1924).

High-precision light curves, such as those provided by the PLATO 2.0 mission, will allow us to detect exomoons and possibly even rings of Saturn-like exoplanets.

What makes the transit method a 'gold-mine' for planetary research is the ability to not only detect planets, but also characterize them physically. The prime planet parameters radius, true mass and therefore mean planet density have already been mentioned. Furthermore, photometric measurements of the stellar light reflected on the surface of the orbiting planet allow us to determine the planetary albedo. During secondary eclipse the emitted infrared flux can be derived and the planet's effective temperature determined. Spectroscopic observations during primary transit and during secondary eclipse permit detection of atmospheric absorption by atoms and molecules in the planetary atmosphere. The analysis of the transit ingress and egress can be used to map the planetary atmosphere, at least for close-in hot giants.

In summary, transiting planets allow us to derive the following parameters of a planet:
- Orbit: - Period, semi-major axis, spin-orbit alignment
- Planet parameters: - radius, mass, density, constrain inner structure and composition
  - effective temperature, albedo, atmospheric composition, surface heat distribution and reflectivity variations from phase curves for gas giants in IR and optical
  - exomoons, planetary rings, Trojan objects

The main detection and characterization goal of PLATO 2.0 is focused on small, terrestrial planets, down to Earth-sized and smaller. An Earth-sized planet around a solar-like star causes a transit depth of about 0.008%. It is obvious that high signal-to-noise-ratio (SNR) light curves are needed to detect such small signals and disentangle them from stellar activity. As an example, one can look at CoRoT-7b and Kepler-10b (Léger et al. 2009; Batalha et al. 2011) two planets slightly larger than Earth (1.6+/-0.1 $R_{Earth}$, Hatzes et al. 2011 and 1.416+/-0.03 $R_{Earth}$, Batalha et al. 2011, respectively) on short-period orbits (<1 day period), orbiting stars around 11 mag. Transits of both planets were clearly detected by the satellite telescopes and allowed determination of their radii. However, several transit events were co-added to achieve this precision. It is clear that brighter target stars must be screened when targeting small planets on long orbital periods. Furthermore, the investment of observing time to determine the planetary mass from radial-velocity measurements for such low-mass planets around faint stars is large and restricted to relatively bright host stars. It will be even worse for Earth-mass planets. This is the particular strength of the PLATO 2.0 mission which is designed to detect planets around bright stars in large numbers, allowing for such follow-up investigations and thereby providing statistical information on planet properties.

**Table A1** Examples for transit parameters in our Solar System, for the hot Jupiter planet CoRoT-1b (Barge et al. 2008; Gillon et al. 2009) and the super-Earths CoRoT-7b (Léger et al. 2009).

| Planet | Orbital period [days] | Transit duration [hours] | Flux dimming [%] | Geom. Prob. [%] |
|---|---|---|---|---|
| CoRoT-1b | 1.5 | 2.5 | 2.2 | ~10 |
| CoRoT-7b | 0.85 | 1.3 | 0.035 | ~10 |
| Mercury | 88 | 8 | 0.0012 | 1 |
| Earth | 365 | 13 | 0.008 | 0.5 |
| Jupiter | 4332 | 30 | 1 | 0.1 |

We furthermore note that planetary masses and radii cannot be determined independently from the properties of their host stars. For both of these parameters, the result is expressed in terms of the corresponding stellar parameter. The accuracy of the planetary parameters derived, therefore, is ultimately limited by our knowledge of the star. PLATO 2.0 addresses this key issue by performing asteroseismology analysis of all planet hosting stars in its main magnitude range (4-11 mag), thereby providing radii and masses with unprecedented accuracy. In addition, the asteroseismology analysis allows determining the age of stars, hence planetary systems, as accurate as 10%.

A disadvantage of the transit method is the required transit geometry. The so-called geometrical probability is the probability to see a system edge-on. Table A1 provides some examples. The geometrical probability mainly scales with the orbital distance, a, of the planet as 1/a. It is around 10% for a close-in hot Jupiter and decreases to about 0.5% for a planet on a one year orbit like Earth. The transit event itself is always short compared to the orbital period of planets. Searching for transits therefore requires continuous coverage of planetary orbits not to miss the short transit events. PLATO 2.0 addresses these challenges of the transit method by providing a very large field-of-view (FoV) and many pointings on the sky covering a very large number of stars, combined with a flexible observing strategy.

## *A2 Asteroseismology: a technique for determining highly accurate stellar and planetary parameters*

**Properties of stellar oscillations**

Modes of stellar oscillations can be described by spherical harmonics $Y_{\ell,m}(\theta, \phi)$ as functions of position $(\theta, \phi)$ on the stellar surface. The *eigenfrequencies* $\nu_{n,\ell,m}$ are described by the three "quantum numbers" $(n,\ell,m)$, where n is the radial order, $\ell$ the latitudinal degree, and $m$ azimuthal order of the spherical harmonic. For a spherical star there is no dependence on the azimuthal order *m*; but this degeneracy is broken by rotation and/or magnetic fields. For slow rotation, the frequencies $\nu_{n,\ell,m} = \nu_{n,\ell} + m <\Omega>$, where m belongs to $\{-\ell, \ell\}$, and $<\Omega>$ is a weighted average of the interior rotation depending on the internal structure of the star and the particular eigenmode. This can be used to probe the internal angular velocity of the star. Measurements of modes with $\ell$ values only up to 3 are expected for PLATO 2.0 targets; since the stellar disk cannot be resolved the signal from modes of higher degree is strongly suppressed by averaging over regions with different oscillation phases.

The oscillation frequencies, including the rotational splitting, are found by fitting peaks in the power spectrum of the light curve. Determining frequencies of modes with $\ell$ = 0, 1, 2, 3 with the solar data is quite straightforward giving estimated errors < 0.1µHz, while for the 137 day run on HD 49385, we can extract frequencies with errors ∼ 0.3µHz. The goal with the much longer monitoring to be performed with PLATO 2.0 is to achieve accuracies ∼ 0.1µHz.

The power spectra show characteristic spacings between the peaks. These are usually described in terms of separations such as the large separations $\Delta_\ell = \nu_{n,\ell} - \nu_{n-1,\ell}$ between modes of the same degree $\ell$ and adjacent *n* values and the small separations, *e.g.*, $\delta_{02} = \nu_{n,0} - \nu_{n-1,2}$ between the narrowly separated peaks corresponding to modes $\ell$ = 0, 2. Additionally we have the small separations $\delta_{01} = \nu_{n,0} - (\nu_{n-1,1} + \nu_{n,1})/2$. These are particularly valuable when only modes of degree $\ell$ = 0, 1 can be reliably determined. The separations provide diagnostic information on the stellar internal structure near the core and hence information on the age of the star. The large separations provide a measurement of the star's acoustic radius, i.e., the travel time of a sound wave from the stellar centre to the surface, which is related to the stellar mean density $\sim M/R^3$, while the small separations such as $\delta_{01}$, $\delta_{02}$ give diagnostics of the interior structure. Periodic modulations in the frequencies or separations give diagnostics of the location of the boundaries of convective cores and envelopes, as well as properties of the helium ionization zone (e.g., Ballot et al. 2004, Cunha & Metcalfe 2007,

Houdek & Gough 2007, Roxburgh 2009a, Miglio et al. 2010, Mazumdar et al. 2014). The outer layers of the star are poorly understood and their contribution to the frequencies must be modeled or corrected for in the asteroseismic analysis (Roxburgh & Vorontsov 2003a, Kjeldsen et al. 2008).

**Asteroseismic inferences**

From the frequencies determined from the power spectrum of the light curve we need to extract physical information. There are several techniques for this, the choice depending on the quality of the data and the type of information desired, ranging from overall properties such as mass and radius of the star to detailed information about its internal structure.

We first consider the case where the S/N ratio in the seismic data is insufficient to allow robust extraction of individual p-mode frequencies; here it may still be possible to extract average estimates of the large and perhaps small separations <$\Delta_0$> <$\Delta_1$>, <$d_{01}$>, <$d_{02}$> and their ratios over one or more frequency ranges, owing to their regularity. For very low signal-to-noise data the mean large separation <$\Delta$>, some indication of its variation with frequency, and possibly an average value of the small separation $d_{02}$, can be determined from frequency-windowed autocorrelation of the time series (Roxburgh & Vorontsov 2006, Roxburgh 2009b, Mosser and Appourchaux 2010). These average values provide a set of seismic data well-suited to constraining the exoplanet host star parameters (cf. Christensen-Dalsgaard, 1988). Coupled with classical observations of $L$, $T_{eff}$, [Fe/H], log $g$ delivered by Gaia (or even more precisely by other means) this has considerably better diagnostic power than the classical observables alone.

For most stars measurement of the average large separation should allow the stellar density to be constrained to a precision of several percent from model fitting (e.g., White et al. 2011). With an accurate knowledge of the effective temperature and/or luminosity, a seismic radius can be determined with a similar precision. The use of the average large separation together with the radius from Gaia can also provide a seismic mass with a precision higher than provided by the classical observables alone. For example, scaling relations relate the averaged large separation, <$\Delta\nu$>, and the frequency at maximum power, $\nu_{max}$, to the mass, radius and effective temperature of the star (Kjeldsen & Bedding, 1995), leading to:

$$\frac{R}{R_{sun}} = \left(\frac{135 \mu Hz}{\langle \Delta\nu \rangle}\right)^2 \left(\frac{\nu_{max}}{3050 \mu Hz}\right) \left(\frac{T_{eff}}{5777 K}\right)^{1/2}; \quad \frac{M}{M_{sun}} = \left(\frac{135 \mu Hz}{\langle \Delta\nu \rangle}\right)^4 \left(\frac{\nu_{max}}{3050 \mu Hz}\right)^3 \left(\frac{T_{eff}}{5777 K}\right)^{3/2}$$

where the radius and mass are normalized to the solar values.

Mass and radius determinations that are based on average seismic quantities will also be used to yield a first, very rapid determination of mass and radius for a large sample of stars. These seismic radii and masses will also serve as initial input for the more precise forward and inversion techniques described below.

Averaged oscillation quantities or individual frequencies can be used in *forward model fitting* which has been extensively used. Here one compares an observed data set with the predictions from a grid of evolutionary stellar models in order to find the model that best fits the observables (*e.g.,* Brown et al. 1994, Miglio & Montalbán 2005, Metcalfe et al. 2009, Aerts et al. 2010). The grid is composed of stellar models that are computed under a range of assumptions about the physical processes that govern stellar evolution. The search in the grid is restricted to satisfy the fundamental properties of the star (magnitude, effective temperature, gravity, metallicity, projected rotational velocity ($m_V$, $T_{eff}$, log $g$, [Fe/H], v sin i,..) and the oscillation observables. In practice one seeks to minimize the differences between observed and computed, seismic and non-seismic, parameters. Several methods can be used to carry out such minimization (see for instance Stello et al. 2009, Quirion et al. 2010). The unknown effect of the surface layers on the absolute values of the frequencies can be overcome by different techniques (e.g., Roxburgh & Vorontsov 2003a, Kjeldsen et al. 2008). The best fit model then gives values for the mass, radius, age and internal structure of the stars. If individual frequencies are used the fit is typically overdetermined, and significant differences between the observed and model values may indicate inadequacies in the stellar modelling being used.

The minimum seismic information necessary in the fitting process can be estimated following Metcalfe et al. (2009). The authors found that with half a dozen surface-corrected frequencies available at each of $\ell = 0$ and $\ell = 1$, it becomes possible to constrain the model-dependent masses to within 3%, and the corresponding ages that the star has spent on the main sequence to within 5%, if the heavy-element abundances are known to within a factor of two. Note that this result assumes that the model physics is correct. With the addition of more frequency estimates (i.e., of $\ell = 2$ modes, and of more overtones) further improvement of the parameter uncertainties will be possible. For a main-sequence target observed at $m_V = 11$, we would expect to be able to measure more than ten overtones of its $\ell = 0, 1$ and $2$ frequencies.

Individual frequencies can de determined as in Appourchaux et al. (2012). When this is done, more precise and detailed information about the stars can also be obtained through inversion techniques.

The probably most suitable technique is *model-independent* and seeks to infer the internal density profile inside a star which best fits a set of observed frequencies (see e.g., Roxburgh & Vorontsov 2003b for more details on the technique. Alternative inversion techniques are described in e.g., Reese et al. 2012). Once we know the density profile, the total mass of the star can be simply computed as the integral of the density over the radius of the star. It is assumed that this will be determined from Gaia results. Note that the regions where the density is least well constrained make only a small contribution to the total mass: in the centre the radius *r* is small and in the outer layers the density is small. The resulting density profiles can then be compared with those predicted by stellar evolution models to estimate the evolutionary age of the star. The analysis is iterative, with the mass and radius for the initial model derived for instance from observed average seismic properties, as discussed above. It should also be stressed that the derivation of a model-independent mass requires that the radius *R* of the star is determined by non-seismic means. As mentioned earlier, radii of the PLATO 2.0 exoplanet host stars will be known to an accuracy better than 2% thanks to Gaia, which translates into *a well constrained model-independent exoplanet host star mass with a relative precision better than 2%.*

As a star evolves towards the end of, and beyond, the main sequence it becomes more centrally condensed. As a result, an increasing number of frequencies of oscillation modes behave like g modes in the core and p-modes in the envelope ("mixed modes"). Their frequencies deviate from the regular spacing of asymptotic pure p modes and can therefore be identified. This behavior changes very quickly with stellar age, and the modes therefore yield a strong (though model-dependent) constraint on the age of the star. Both CoRoT and *Kepler* have observed stars presenting such modes (e.g., Metcalfe et al. 2010, Deheuvels & Michel 2011). For the non-seismic parameters, the largest source of observational uncertainty comes from the estimated heavy-element abundances. From the luminosity precision expected from Gaia, it would in principle be possible to constrain the abundances seismically to a precision of about 10% (Metcalfe et al. 2009), thus further improving the accuracy of the star's mass and age.

**Effects of rotation**

As mentioned above, stellar rotation induces a splitting of the frequencies according to the azimuthal order *m* of the mode, by an amount which is essentially a weighted average of the internal rotation rate. The weight function (the so-called rotational kernel) can be determined given the inferred structure of the star. For predominantly acoustic modes most weight is given to the stellar envelope, with little dependence on the mode, and the rotational splitting thus predominantly gives an average of the rotation rate in the outer parts of the star. If in addition the surface rotation rate can be determined from spot modulation, as has been done in several cases from CoRoT and *Kepler* data (e.g., Nielsen et al. 2013), some indication can be obtained of the variation of rotation with depth. In evolved stars, on the other hand, the observation of mixed modes provides information about the rotation in the deep interior (e.g., Beck et al. 2012, Deheuvels et al. 2012, Mosser et al. 2012).

Equally important are the inferences that can be made from the amplitudes of the rotationally split components. Given the stochastic nature of the mode excitation, all components are expected to be excited to the same intrinsic amplitude, on average. However, the observed amplitudes depend on the inclination of the rotation axis relative to the line of sight (Gizon & Solanki 2003). If the rotation axis points towards the observer only the *m* = 0 modes are visible, while if it is in the plane of the sky only modes where ℓ-*m* is even are seen. For intermediate inclination all 2ℓ+1 components are visible, and from their relative amplitudes the inclination can be inferred. This is particularly interesting in the case of stars where planetary systems have been detected using the transit technique, as will be the case for PLATO 2.0; here such observations can test the alignment or otherwise of the orbital planes of the planets with the stellar equator (Gizon et al., 2013; Chaplin et al. 2013).

## Acknowledgements:

The authors acknowledge many fruitful discussions with the ESA study team members, in particular the project scientist and the project manager (Ana Heras and Philippe Gondoin), as well as previous ESA members of the PLATO M1/M2 proposal (Osvaldo Piersanti, Anamarija Stankov). The M3 PLATO 2.0 Mission Consortium also thanks the consortium participants, in particular CNES, France, who prepared the M1/M2 PLATO proposal, which forms the basis on which PLATO 2.0's application as M3 candidate is built on. We also thank Kayser Threde (in particular the study manager Richard Haarmann) for their inputs and fruitful cooperation in the M3 proposal preparation phase. We also thank the referees for their thorough revision and insightful comments, which have led to a significant improvement of this manuscript.

M. Ammler-von Eiff acknowledges support by DLR (Deutsches Zentrum für Luft- und Raumfahrt) under the project 50 OW 0204. M. Bergemann acknowledges the support by the European Research Council/European Community under the FP7 programme through ERC Grant number 320360. I. Boisse acknowledges the support from the Fundacao para a Ciencia e Tecnologia (Portugal) through the grant SFRH/BPD/87857/2012. For J. Christensen-Dalsgard, funding for the Stellar Astrophysics Centre is provided by The Danish National Research Foundation (Grant DNRF106). The research is supported by the ASTERISK project (ASTERoseismic Investigations with SONG and *Kepler*) funded by the European Research Council (Grant agreement no.: 267864). L. Gizon acknowledges support from Deutsche Forschungsgemeinschaft SFB 963 "Astrophysical Flow Instabilities and Turbulence" (Project A18). M. Godolt and J.L. Grenfell have been partly supported by the Helmholtz Gemeinschaft (HGF) through the HGF research alliance "Planetary Evolution and Life". S. Hekker acknowledges financial support from the Netherlands Organisation for Scientific Research (NOW) and the Stellar Ages project funded by the European Research Council (Grant agreement number 338251). K. G. Kislyakova, N. V. Erkaev, M. L. Khodachenko, H. Lammer, M. Güdel acknowledge support by the FWF NFN project S116 "Pathways to Habitability: From Disks to Active Stars, Planets and Life", and subprojects, S116 604-N16 "Radiation & Wind Evolution from T Tauri Phase to ZAMS and Beyond", S116 606-N16 "Magnetospheric Electrodynamics of Exoplanets", S116607-N16 "Particle/Radiative Interactions with Upper Atmospheres of Planetary Bodies Under Extreme Stellar Conditions". F. Kupka is grateful for support through FWF project P25229-N27. M. Mas-Hesse was supported by Spanish MINECO under grant AYA2012-39362-C02-01. L. Noack has been funded by the Interuniversity Attraction Poles Programme initiated by the Belgian Science Policy Office through the Planet Topers alliance. D.R. Reese is supported through a postdoctoral fellowship from the "Subside fèdèral pour la recherche 2012", University of Liège. I.W. Roxburgh gratefully acknowledges support from the Leverhulme Foundation under grant EM-2012-035/4. A. Santerne and N.C. Santos acknowledge the support by the European Research Council/European Community under the FP7 through Starting Grant agreement number 239953. S. G. Sousa acknowledges the support by the European Research Council/European Community under the FP7 through Starting Grant agreement number 239953. Gy.M. Szabó acknowledges the Hungarian OTKA Grants K104607, the HUMAN MB08C 81013 grant, by the City of Szombathely under agreement No. S-11-1027 and the János Bolyai Research Scholarship of the Hungarian Academy of Sciences. R. Szabó was supported by the János Bolyai Research Scholarship, Hungarian OTKA grant K83790, KTIA URKUT_10-1-2011-0019 grant Lendület-2009 Young Researchers' Program and the European Community's Seventh Framework Programme (FP7/2007-2013) undergrant agreements no. 269194 (IRSES/ASK) and no. 312844 (SPACEINN).